\documentclass[useAMS,usenatbib,usegraphicx]{mn2e}
\usepackage{epsf}
\usepackage{graphicx}
\usepackage{subfigure}
\usepackage{psfig}

\title[Brown Dwarfs and Planetary Mass Objects in the Orion Nebula Cluster]{Infrared Spectroscopy and Analysis of Brown Dwarf and Planetary Mass Objects in the Orion Nebula Cluster}
\author[D. J. Weights, P. W. Lucas, P. F. Roche, D. J. Pinfield, F. Riddick]{D. J. Weights$^{1}$\thanks{E-mail:
d.j.weights@herts.ac.uk}, P. W. Lucas$^{1}$, P. F. Roche$^{2}$,
D. J. Pinfield$^{1}$, F. Riddick$^{2}$\\
$^{1}$Dept of physical sciences, University of Hertfordshire, College Lane, Hatfield Al10 9AB, England.\\
$^{2}$Astrophysics Dept, University of Oxford, 1 Keble Road, Oxford,
OX1 3RH, England.}
\begin{document}
\input{epsf}

\pagerange{\pageref{firstpage}--\pageref{lastpage}} \pubyear{2008}

\maketitle

\label{firstpage}

\begin{abstract}
We present near-infrared long slit and multi-slit spectra of low
mass brown dwarf candidates in the Orion Nebula Cluster. The long
slit data were observed in the \emph{H-} \& \emph{K-}bands using
NIRI on the Gemini North Telescope. The multi-object spectroscopic
observations were made using IRIS2 on the Anglo Australian Telescope
at \emph{H-}band. We develop a spectral typing scheme based on
optically calibrated, near infrared spectra of young sources in the
Taurus and IC 348 star forming regions with spectral types M3.0 to
M9.5. We apply our spectral typing scheme to 52 sources, including
previously published UKIRT and GNIRS spectra. 40 objects show strong
water absorption with spectral types of M3 to $>$M9.5. The latest
type objects are provisionally classified as early L types. We plot
our sources on H-R diagrams overlaid with theoretical
pre-main-sequence isochrones. The majority of our objects lie close
to or above the 1 Myr isochrone, leading to an average cluster age
that is $<$1 Myr. We find 38 sources lie at or below the hydrogen
burning limit (0.075 M$_{\odot}$). 10 sources potentially have
masses below the deuterium burning limit (0.012 M$_{\odot}$). We use
a Monte Carlo approach to model the observed luminosity function
with a variety of cluster age and mass distributions. The lowest
$\chi$$^{2}$ values are produced by an age distribution centred at 1
Myr, with a mass function that declines at sub-stellar masses
according to an M$^{\alpha}$ power law in the range $\alpha$=0.3 to
0.6. We find that truncating the mass function at 0.012 M$_{\odot}$
produces luminosity functions that are starved of the faintest
magnitudes, even when using bimodal age populations that contain 10
Myr old sources. The results of these Monte Carlo simulations
therefore support the existence of a planetary mass population in
the ONC.
\end{abstract}


\section{Introduction}
The characterisation of the brown dwarf and planetary mass
populations in star forming regions is important in order to measure
the Initial Mass Function (IMF) and to learn whether the timing and
duration of the formation process changes at low masses. Young
clusters are amongst the best locations to probe the substellar IMF
because young brown dwarfs are several orders of magnitude brighter
than mature field brown dwarfs.

The Orion Nebula Cluster (ONC) is an excellent location to search
for brown dwarfs and planetary mass objects. Extensive research
demonstrates that the cluster is extremely young with an average age
of $\leq$1Myr (e.g. Prosser et al. 1994; Hillenbrand 1997, Palla \&
Stahler 1999; Riddick, Roche \& Lucas (2007, hereafter RRL)). The
cluster is also nearby ($\sim$450pc (Luhman et al. 2000)),
permitting us to probe for sources with masses below the hydrogen
burning limit. The dense background of the OMC-1 cloud and
relatively high galactic latitude ($b=-19^{\circ}$) minimise
contamination from background stars (Hillenbrand \& Hartmann 1998).

In recent years deep near-infrared studies of young clusters have
led to the detection of large numbers of brown dwarf and planetary
mass candidates (Comeron et al. 1993; Williams et al. 1995; Lucas \&
Roche 2000; Lucas, Roche \& Tamura 2005). The luminosity and colour
criteria (see Section \ref{obs} for more details) used for
differentiating between sub-stellar and higher mass cluster members
have proven to be successful, with spectroscopic follow-up
confirming the existence of significant numbers of brown dwarfs in
Orion and other star-forming regions (Lucas et al. 2001; Luhman et
al. 2003; Meeus \& McCaughrean 2005).

Despite spectroscopic confirmation of a large number of brown
dwarfs, a considerable fraction of potentially low-mass cluster
members remain uncharacterised. Spectroscopy has demonstrated that
we cannot rely on photometry alone to determine the nature of
objects. Samples are subject to contamination from both foreground
and reddened background stars. Photometric parameters can be skewed
by scattering of light from dust and circumstellar material. In
addition to this, theoretical pre-main-sequence models become less
certain at young ages. Spectroscopic observations are therefore
critical to accurately calibrate the physical properties of young
objects and determine the true nature of the initial mass function
(IMF) and luminosity function (LF) below the hydrogen burning limit.

Measurements of the low mass IMF are useful to inform and constrain
the theory of star formation as a whole. By characterising the
nature of the IMF in different environments we can establish whether
it is universal. The associated mapping of the spatial distribution
of low and high mass sources which may also improve star formation
theory.

This work extends our previous spectroscopic studies to a larger
sample in order to analyse the substellar mass function and age
distribution. We analyse low resolution near-infrared \emph{H-} and
\emph{K-}band spectroscopic observations of 55 brown dwarf
candidates in the ONC. In Section \ref{obs} we describe the
observations and the data reduction process. We present our
\emph{H-} and \emph{K-}band spectra in Section \ref{specsec}. In
Section \ref{specType} we describe the spectral typing scheme that
was developed for our different data sets and present the results.
H-R diagrams are used to to determine source mass in the following
section. Here we discuss the results from the H-R diagram and
implications on cluster age. In our penultimate section we show
results of Monte Carlo modelling of the LF. We experiment with
different mass functions and age distributions and see how these
parameters effect the LF. Our modelling sheds light on the initial
conditions that are most likely in the ONC and how a population of
planetary mass objects effect the LF.

\section{Observations}
\label{obs}
Near-infrared spectra of Orion brown dwarf candidates
were obtained in classical observing mode at two epochs, each using
a different observational technique. The first data were acquired
using the Near InfraRed Imager (NIRI) on the 8-m Gemini North
Telescope over the period 5-8 December 2003. Objects in this dataset
were observed through a single slit. The second data-set was
obtained with the Anglo Australian Telescope (AAT) using the
Infrared Imager and Spectrograph (IRIS2, see Tinney et al. 2004).
These observations were carried out during 20-25 November 2004. With
the implementation of high astrometric precision masks, the
instrument was operated as a multiple-slit spectrograph.

Candidate sources were selected for observation from UKIRT
photometry of Lucas et al. 2000 \& Lucas et al. 2001 and from Gemini
South/Flamingos photometry of Lucas, Roche and Tamura 2005. All
sources had prospective masses ranging from below the hydrogen
burning limit ($\la$ 75M$_{\emph{Jup}}$) down close to the deuterium
burning limit ($\sim$ 13M$_{\emph{Jup}}$; assuming solar
metalicity). The main selection criteria were identical for both
data sets; however, subtle differences in the selection strategy
were required due to the capabilities and limitations of each
instrument and telescope. The critical requirements were as follows:
(1) Sources were avoided that were close to or embedded in bright
nebulosity. For optimum sensitivity objects were chosen that were
located in regions of low nebula surface brightness. (2) Sources
were only chosen if they possessed fairly low extinction values
(A$_{V}$ $\la$7.5). (3) Candidates that were in a resolved close
binary system were avoided.

In total 16 of the 17 brown dwarf candidate spectra observed with
NIRI were useful. 4 of these objects do not appear in this paper as
they were published in paper 1 (Lucas et al. 2006). A sample of 44
targets were observed with the AAT using IRIS2, 17 of which were
brown dwarf candidates and of sufficient quality to be of scientific
value. 2 of these objects were observed by chance due to
coincidental slit alignment. Table \ref{IntDataTable} lists
photometric properties, total integration times and the observed
waveband(s) for each source. Dereddened \emph{H-}band magnitudes
have been tabulated as they are referred to throughout the paper.
Emphasis has been put on objects that were not on our target list
and were observed by chance.

The observations in the NIRI data set were carried out using the f/6
camera in conjunction with the 1024$\times$1024 pixel ALADDIN InSb
array. This combination yielded a scale of 0.117 arcsec per pixel.
Efficiency was maximised by simultaneously observing closely spaced
candidate sources in pairs. The spectra were observed with a 0.7
arcsec slit with a resolving power of \emph{R}$=$520. A data
acquisition sequence consisting of two nod positions (ABBA pattern)
separated by 3 arcsec was employed. This allowed the removal of the
dark current and near perfect removal of telluric OH emission and
emission from nebulosity. All of the candidates in this data set
were observed with the \emph{K-}band grism. Two pairs were also
observed with the \emph{H-}band grism. Flat fields were taken at the
beginning of each evening and morning using the Gemini Calibration
unit (GCAL). Stellar standards were observed throughout each night
at a similar airmass to each of the sources. The UKIRT list of faint
standards was used to carefully choose F- and G-type stars prior to
the observing run. An Argon lamp was used to create several arc
images each night. The observing conditions were photometric on
three of the four nights. The observations on 6th December were cut
short due to cloud.

The observations with IRIS2 employed cryogenically cooled slit
masks, an observing mode which had not been fully commissioned at
the time. With the implementation of two different masks, a total of
44 brown dwarf candidates were observed. The detector used was a
1024$\times$1024 Rockwell HgCdTe Hawaii array. A scale of 0.4487
arcsec per pixel provided a field of view of 7.7 arcmin$^{2}$. Brown
dwarf candidates were selected for observation that had apparent
magnitudes brighter than mag 17 at \emph{H-} band. Each mask
contained holes for four bright fiducial stars that were used to
precisely align the telescope to the correct field. In an attempt to
remove the telluric OH emission the data were again acquired using
an ABBA nod pattern. The nod distance was experimented with in order
to maximise the number of sources that appeared in their relative
slit in both nod positions. Several candidates in each field only
had a spectrum visible in one nod position. This problem was
encountered for approximately two sources in each mask. Both masks
were observed with a \emph{J-}long (1.10$-$1.33 $\umu$m) and an
\emph{H-}short (1.46$-$1.81 $\umu$m) grism with integration times
ranging between 3 and 4 hours.

Standard stars were observed throughout each night using an
unorthodox method to illuminate all the slits in each mask. The
standard star was defocused until it illuminated two or three
adjacent slits. Then a fast chopping mode of the telescope was used
to further spread out the image profile so that several adjacent
slits were illuminated simultaneously. By repeating this method
three or four times a calibration spectrum could be taken through
every slit in a mask.

Flat fields and Xe arcs were taken at the beginning of each night
through each filter and each mask. Three of the six nights allocated
for observing were abandoned due to bad weather. The data from the
remaining three clear nights had fairly poor seeing (typically $>$ 1.6 arcsec) that was variable on short timescales.\\

\noindent\emph{Erratum}

We would like to point out a naming error in Lucas et al. 2006
(paper 1). The spectrum plotted for 022-115 is actually the
\emph{K-}band spectrum of 023-1939. Both objects are reddened
background stars, with similar spectral profiles. We include the
spectral typing results of 022-115 in this paper.

\begin{table*}
            \caption{Photometric data and total integration times for sources from the IRIS2 and NIRI data sets that are presented in this paper.
                    Objects that could not be spectral typed due to insufficient signal to noise have
                    not been included. Photometric data is from Lucas \& Roche (2000) and Lucas et al.(2001;2005).}
                    \label{IntDataTable}
    \begin{minipage}{17cm}
        \begin{centering}
        \begin{tabular}{@{}lccccccc@{}}
            \hline
            \hline
                Source$^{a}$ & Data Set & J & H & H$_{\emph{dr}}$ & K & A$_{\emph{v}}$ & Integration Time (min)\\
                & & & & & & & \emph{H-} , \emph{K-}\\
            \hline
                010-109 & NIRI & 17.39 & 16.37 & 15.55 & --- & 4.67 & ---~ , 186 \\
                011-027 & IRIS2 & 15.55 & 14.99 & 14.95 & --- & 0.25 & 200 , ~--- \\
                013-306 & NIRI & 16.90 & 15.61 & 14.37 & 14.63 & 7.06 & ---~ , \ 40 \\
                015-319 & NIRI & 18.08 & 17.26 & 16.78 & 16.49 & 4.26 & ---~ , \ 40 \\
                020-1946 & IRIS2 \& NIRI & 16.48 & 15.35 & 14.39 & --- & 5.47 & 110 , 112$^{b}$ \\
                023-1939 & NIRI & 18.87 & 17.98 & 17.46 & --- & 3.01 & ---~ , 112 \\
                024-124$^{c}$ & IRIS2 & 14.32 & 13.33 & 12.70 & --- & 3.60 & 200 , ~--- \\
                030-524 & NIRI & 17.86 & 17.40 & 17.40 & 16.77 & 0.00 & ---~ , \ 60 \\
                031-536 & IRIS2 & 16.72 & 16.23 & 16.23 & 15.52 & 0.00 & 200 , ~--- \\
                044-527 & NIRI & 17.38 & 16.88 & 16.88 & 16.18 & 0.00 & ---~ , \ 60 \\
                047-245 & IRIS2 & 17.27 & 16.62 & 16.41 & 16.10 & 1.23 & 110 , ~--- \\
                047-436 & IRIS2 & 15.61 & 14.37 & 13.28 & 12.64 & 6.22 & 110 , ~--- \\
                053-323 & IRIS2 & 15.98 & 15.43 & 15.39 & 15.15 & 0.25 & 110 , ~--- \\
                056-141 & IRIS2 & 17.34 & 16.34 & 15.54 & --- & 4.54 & 200 , ~--- \\
                057-305 & IRIS2 & 17.31 & 15.61 & 13.75 & 14.81 & 10.66 & 110 , ~--- \\
                067-651 & IRIS2 & 15.78 & 14.95 & 14.46 & --- & 2.78 & 110 , ~--- \\
                077-453 & IRIS2 & 13.90 & 13.10 & 12.76 & 12.79 & 1.93 & 200 , ~--- \\
                084-104 & NIRI & 17.42 & 16.89 & 16.88 & --- & 0.05 & ---~ , \ 32 \\
                087-024 & IRIS2 & 16.60 & 15.56 & 14.74 & --- & 4.67 & 110 , ~--- \\
                095-058 & NIRI & 13.71 & 13.25 & 13.25 & --- & 0.00 & ---~ , \ 32 \\
                121-434 & IRIS2 & 14.51 & 13.43 & 12.66 & --- & 4.44 & 200 , ~--- \\
                127-044$^{c}$ & IRIS2 & 13.67 & 13.19 & 13.19 & --- & 0.00 & 110 , ~--- \\
                130-053 & IRIS2 & 15.77 & 15.20 & 15.12 & --- & 0.44 & 200 , ~--- \\
                186-631 & NIRI & 15.94 & 15.28 & 15.07 & 14.58 & 2.27  & ~52 , 180 \\
                192-723 & NIRI & 17.11 & 16.08 & 15.27 & 15.41 & 4.64 & ~34 , 124 \\
                196-659 & NIRI & 17.85 & 17.00 & 16.44 & 16.34 & 3.20 & ---~ , 130 \\
                235-454 & IRIS2 & 14.84 & 13.82 & 13.11 & --- & 4.06 & 228 , ~--- \\
                255-512 & IRIS2 & 17.11 & 16.10 & 15.31 & --- & 4.50& 228 , ~--- \\
            \hline
        \end{tabular}
            \\
              \end{centering}
              Notes:\\
              $^{a}$  Source names are coordinate based, following O'Dell \& Wong (1996).\\
              $^{b}$  \emph{H-}band data was from IRIS2 and \emph{K-}band was from NIRI.\\
              $^{c}$  Source was observed in the slit by chance.\\
        \end{minipage}
\end{table*}

\subsection{Data Reduction}
\label{datRed} Both single slit and multiple slit data were reduced
using {\sc IRAF} software.

Standard techniques were used to reduce the NIRI long slit data. For
each brown dwarf candidate pair, exposures were separated into their
relative nod positions (A or B). Groups A and B were median combined
separately using the {\sc IMCOMBINE} task in {\sc IRAF}. One
combined image pair was then subtracted from the next, removing the
dark current and most of the background. An argon arc was used to
wavelength calibrate the data. Each image was wavelength calibrated
and corrected for distortion using the {\sc IDENTIFY, REIDENTIFY,
FITCOORDS} and {\sc TRANSFORM} routines in {\sc IRAF}. All spectra
were extracted using APALL. The residual background that remained
after subtraction was removed during the extraction using two
background software apertures adjacent to the target spectrum.
Telluric standards were created from standard stars that had been
divided by a blackbody representing the temperature of the
associated stellar spectral type. Spectra were cleaned in order to
remove spurious noise spikes, residual OH lines, nebula lines and
any remaining residual background features. Extracted pairs from
each nod position were coadded and finally dereddened using the {\sc
DEREDDEN} task in {\sc IRAF}, with an adopted reddening parameter of
3.1.

\subsection{IRIS2 Reduction Techniques}
The initial reduction steps of combining and subtracting pairs was
carried out in the manner described above. Poor observing conditions
meant that the combined images contained a significant amount of
residual background caused by very short timescale variable telluric
OH. Shorter exposure times were experimented with at the telescope
but were inadequate for checking the position of spectra. Due to the
nature of Multi-Object Spectroscopy (MOS) the subtracted images
could not be wavelength calibrated and rectified for distortions in
the manner described above. Instead spectra were extracted
individually with APALL. The trace was recorded and used again to
extract a Xenon arc from the same position in the relative slit.
This was sufficient for wavelength calibration due to the fact that
geometric distortions were insignificant. A master telluric standard
was constructed for each mask rather than a separate standard star
for each individual slit. To achieve full wavelength coverage the
standard stars were observed through the furthermost left and right
slits and a central slit. We checked that this approach was
appropriate by comparing an extracted spectrum from each slit. By
dividing each spectrum by the next we were able to demonstrate that
the throughput did not change from slit to slit. The telluric
standard was created in the same manner described in Section
\ref{datRed}. Each object spectrum was divided by the master
telluric standard. Cleaning, coadding and dereddening procedures
were as described above. Due to the lateral variation of slit
positions in the masks used for multiple object spectroscopy, the
final wavelength coverage of each spectrum is different.

\section{Spectroscopy}
\label{specsec}
\subsection{\emph{H-}Band}
The \emph{H-}band spectra of objects observed with IRIS2 and NIRI,
that show H$_{2}$O absorption and have sufficient signal to noise to
be spectral typed are presented in Figure \ref{hspectraa}. We also
plot the spectrum of 047-436. This object has a red continuum and
shows no H$_{2}$O absorption. All spectra are plotted as F$_\lambda$
with a flux scale normalised at 1.68 $\umu$m. Spectra from our
sample that had a Rayleigh-Jeans type continuum were not plotted.
The distinctive triangular peak that is ubiquitous in young low
surface gravity objects (Lucas et al. 2001), due to strong H$_{2}$O
absorption and pressure sensitive H$_{2}$ absorption, can be seen in
each of the spectra close to 1.675 $\umu$m. None of the ONC objects
reveal obvious narrow features that are unassociated with the pixel
to pixel noise fluctuations. The plots do not demonstrate the
complete wavelength coverage of each spectrum due to the fact that
telluric noise begins to increase dramatically outside this spectral
range for most of our Orion spectra. Many of the IRIS2 spectra have
low signal to noise and have subsequently been smoothed for clarity.
The quality of the AAT spectra drops towards the edges of the
atmospheric window due to the low altitude of the observatory so
absorption longward of 1.72 $\umu$m may be masked by noise. Careful
observation of 047-436 reveals a steep gradient between 1.68 and
1.72 $\umu$m may exist. However, without improved spectral coverage
this remains uncertain. Despite the low signal to noise in the
spectrum of 031-536 the pseudo continuum shows convincing evidence
of strong water absorption. The spectroscopic data becomes
unreliable shortward of 1.59 $\umu$m where the noise significantly
increases. Spectral data is absent at wavelengths shorter than 1.55
$\umu$m for 024-124 due to limited wavelength coverage.

5th order cubic splines have been fitted to each of the spectra. The
continuum fits were created using a fitting procedure within SPLOT,
in IRAF. To ensure a high quality fit could be established, each
spectrum was first box-car smoothed using an 11 pixel size box and
trimmed to give a wavelength range limited between 1.50 and 1.675
$\umu$m. The cubic spline fits represent the pseudo continuum
satisfactorily, avoiding spurious narrow structures in the spectra
that are a due to noise. The cubic spline fits were constructed to
provide us with a robust reference for spectral typing (see section
\ref{specType}).

The IRIS2 \emph{H-}band source 047-436 has a red continuum that has
no strong H$_{2}$O absorption. 047-436 is a deeply embedded cluster
member which displays strong H$_2$ emission lines in its
\emph{K-}band spectrum and strong CaII emission lines in its optical
spectrum (Lucas et al. 2001; RRL). We have plotted the spectrum of
this object as F$_{\nu}$ so that the reader can make a direct
comparison of the two spectra. Between 1.58 and 1.79 $\umu$m we
measure a rise in relative flux that is marginally less than
observed by Lucas et al. 2001. The difference in this slope is
probably due to the fact that we dereddened our spectrum based on a
larger A$_{\emph{v}}$ estimate. Shortward of 1.58 $\umu$m our
spectrum is approximately flat. This contrasts with the previously
published \emph{H-}band spectrum where the gradient of the pseudo
continuum is constant over the entire wavelength range. We believe
that this difference may be attributed to the combination of the low
signal to noise present in our spectrum and the fact that the
shortest wavelength region is at the edge of the array for this
object. The spectral profile of 047-436 changes significantly when
converted to an F$_{\lambda}$ flux scale, presenting a pseudo
continuum that gently declines between 1.50 and 1.60 $\umu$m and
then proceeds to remain flat until 1.80 $\umu$m. The spectrum is not
characteristic of a young, low mass cluster member. However, Lucas
et al. 2001 demonstrated that this object showed signs of youth at
\emph{K-}band and is a definite cluster member. The photometry of
this object reveals an anomalously blue (\emph{I$-$J}) colour
((\emph{I$-$J})$=$3.21) suggesting the source is associated with
circumstellar matter. The slope of the dereddened spectrum may
therefore be unreliable due to the fact that scattered light is
causing us to overestimate the flux at shorter wavelengths. Due to
potential water absorption being veiled by extinction we are unable
to determine whether this object is a low mass star or a brown
dwarf. This object may be a proto-brown dwarf or a background star
that has been dereddened inaccurately.

We have not plotted the \emph{H-}band spectrum of 057-305 despite
reasonable signal to noise. The spectrum of this object is similar
to that of 047-436, showing a red continuum when displayed as
F$_{\nu}$ and a flat spectrum when plotted as F$_{\lambda}$. The
extinction to this this object is relatively high
(A$_{\emph{v}}$=10.662) thus no \emph{I-}band detection has been
attained. Due to the fact that there is no \emph{I-}band photometry
for this source and we do not have a \emph{K-}band spectrum, we
cannot determine whether this object is a cluster member or
background star.

\begin{figure*}
\centering{\ } 
  \hbox{
    \psfig{file=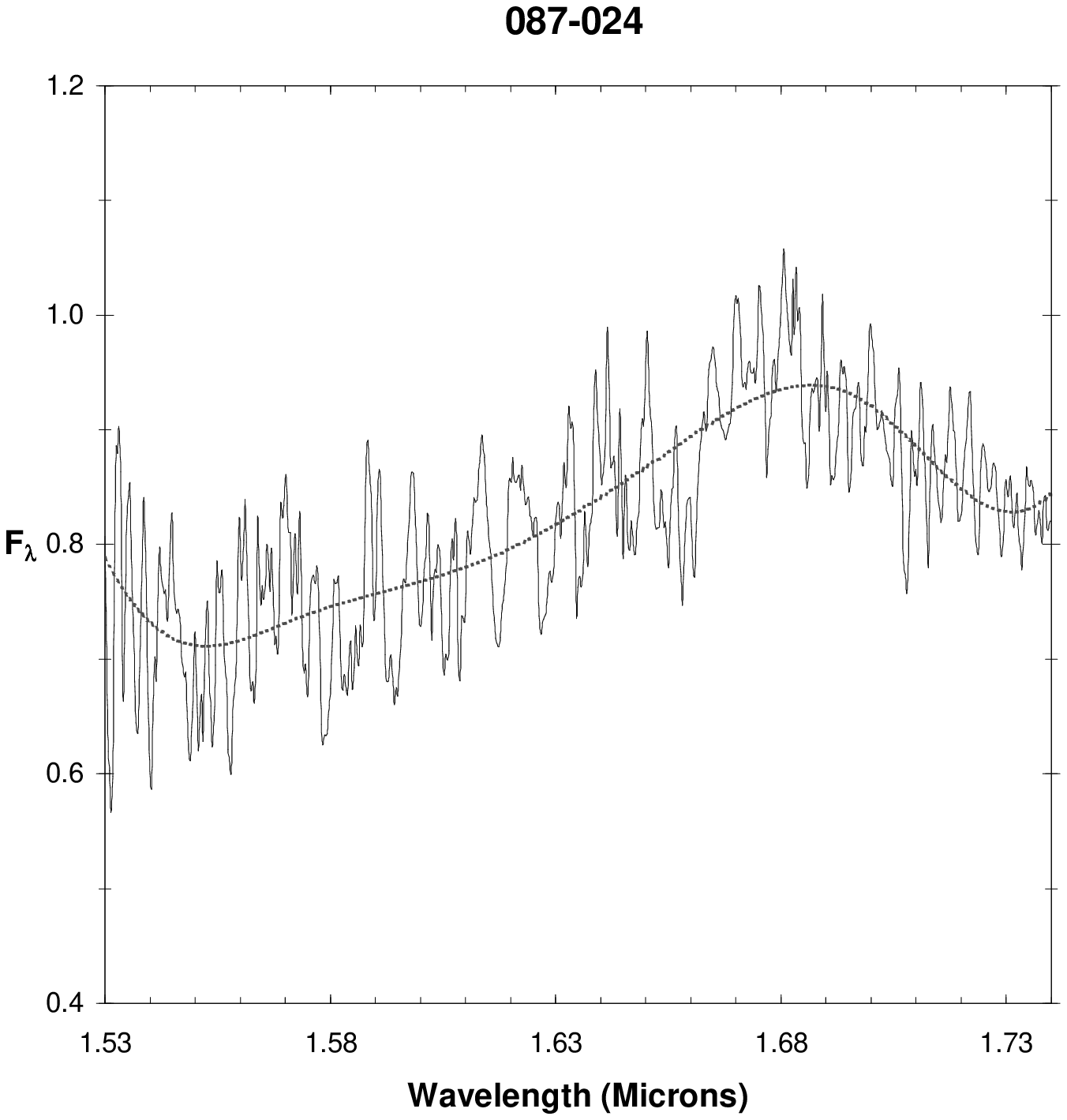,width=7.5cm,angle=-90}
    \psfig{file=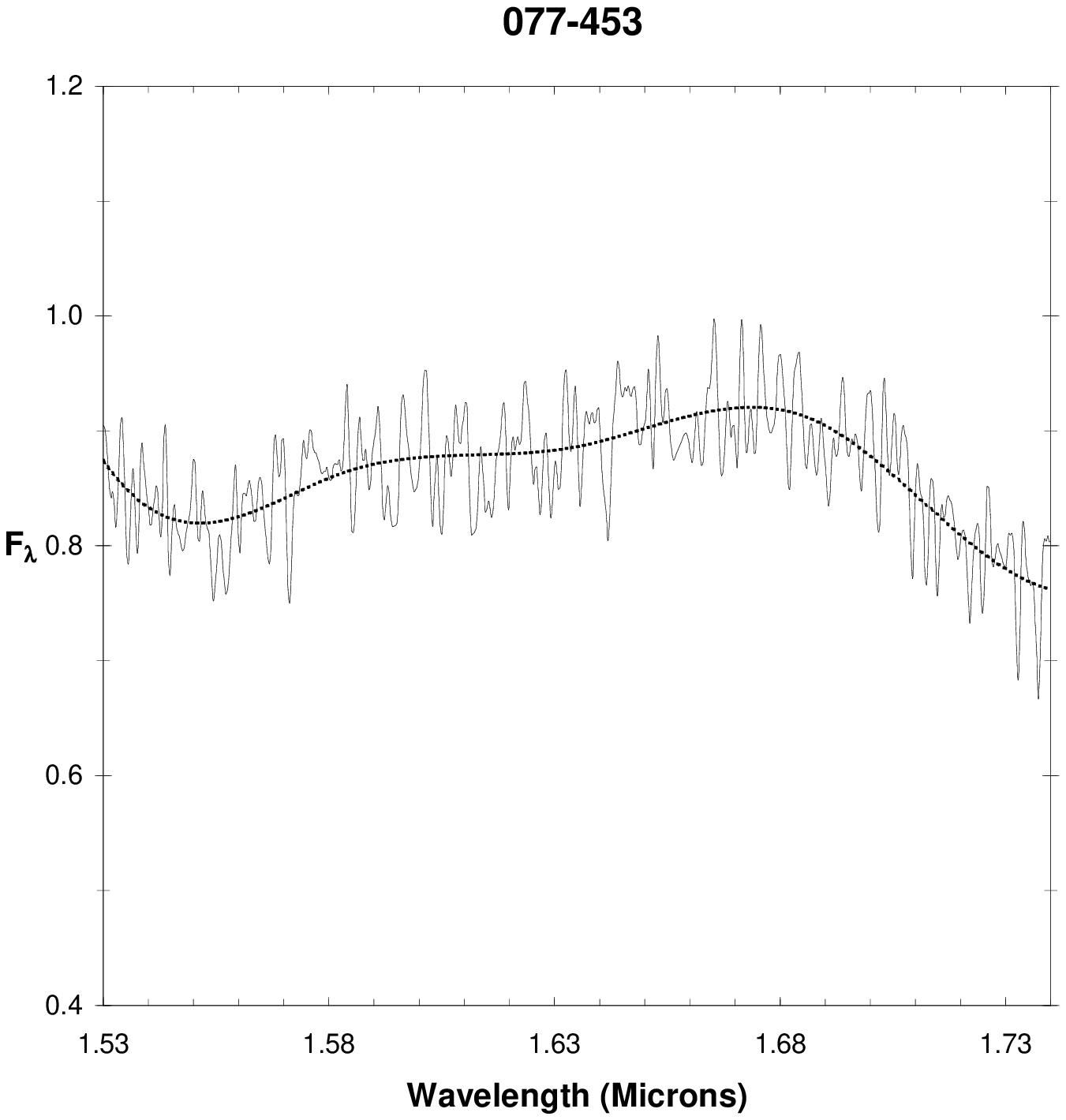,width=7.5cm,angle=-90}}
  \hbox{
    \psfig{file=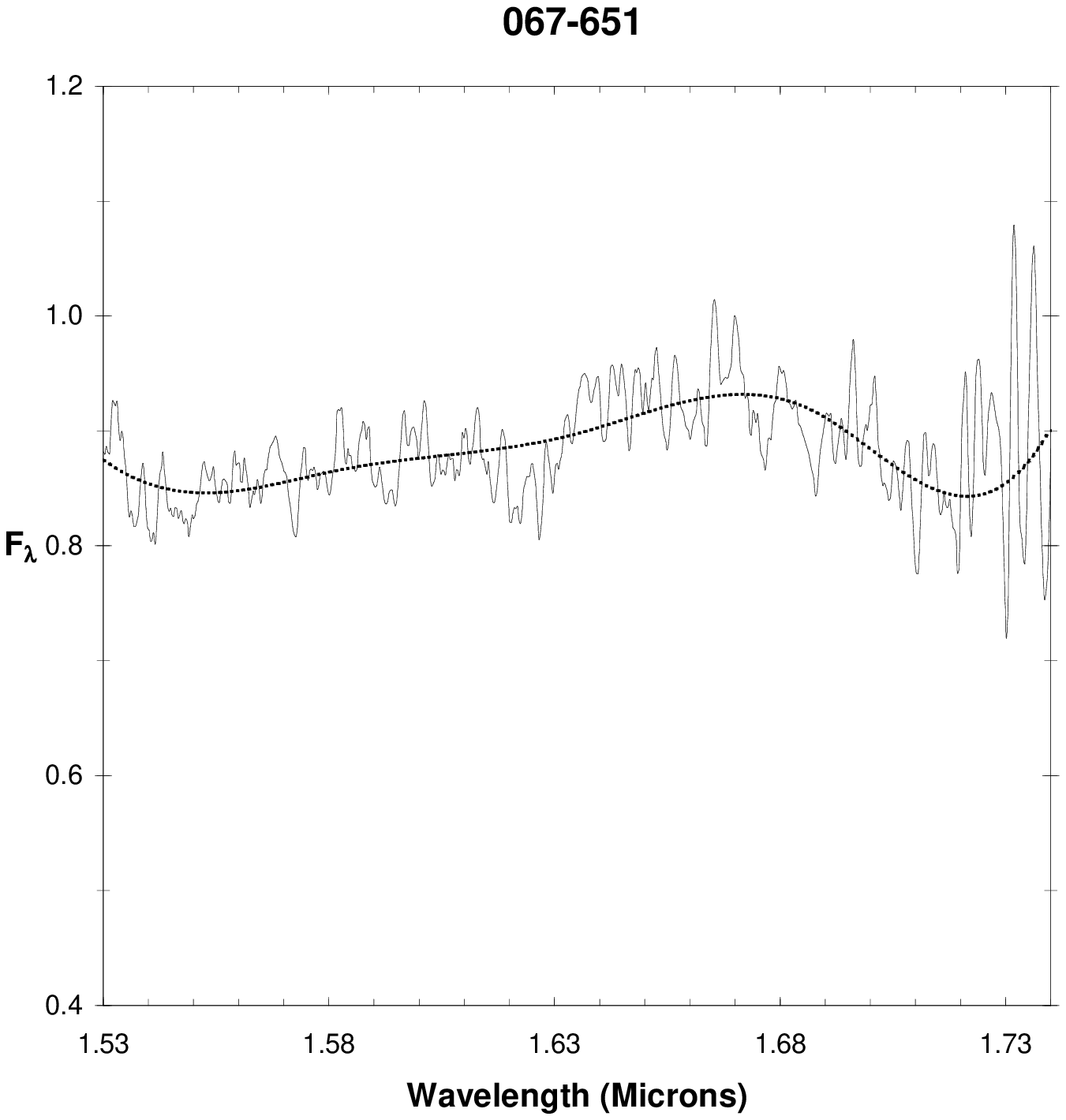,width=7.5cm,angle=-90}
    \psfig{file=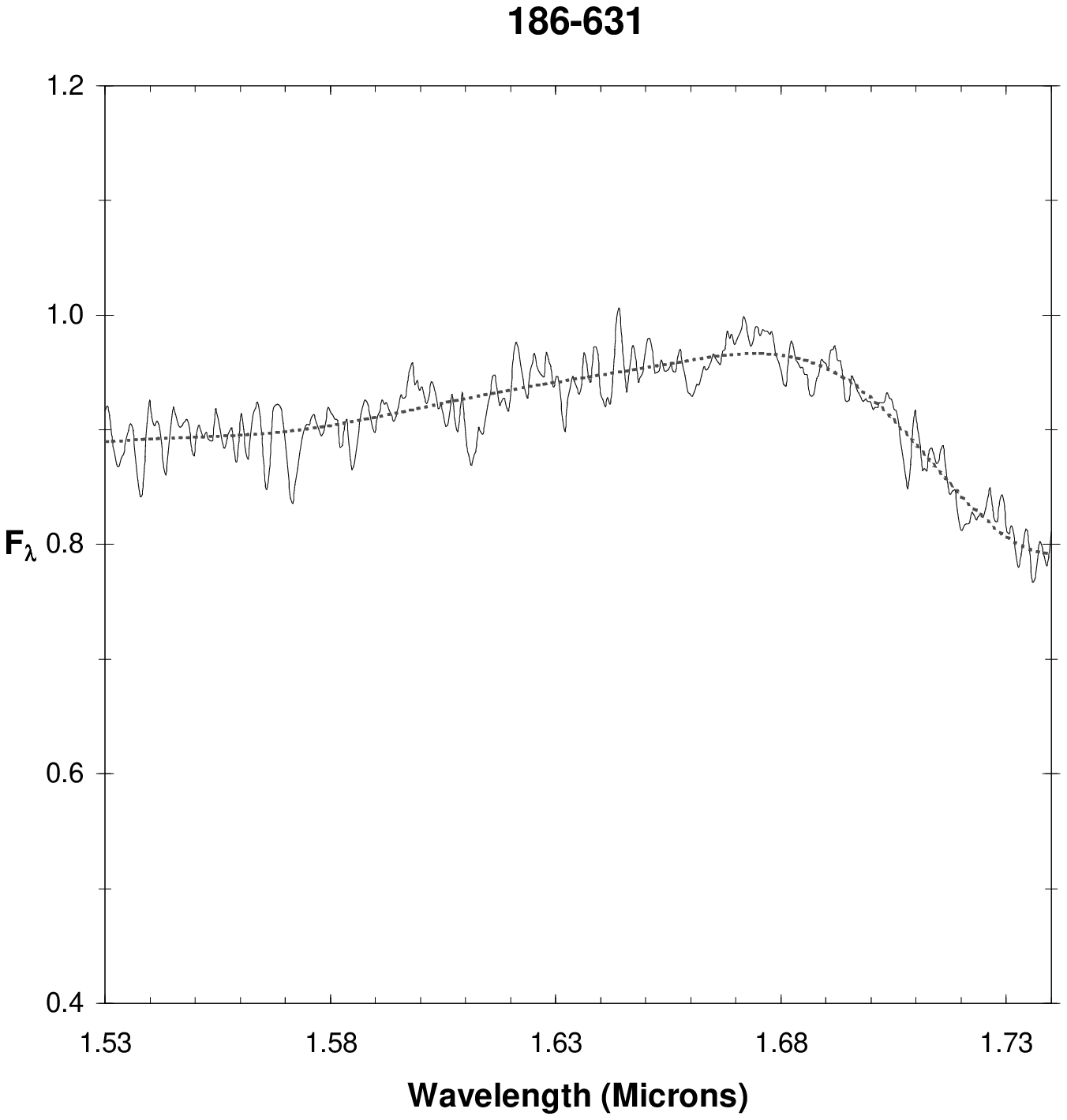,width=7.5cm,angle=-90}}
  \hbox{
    \psfig{file=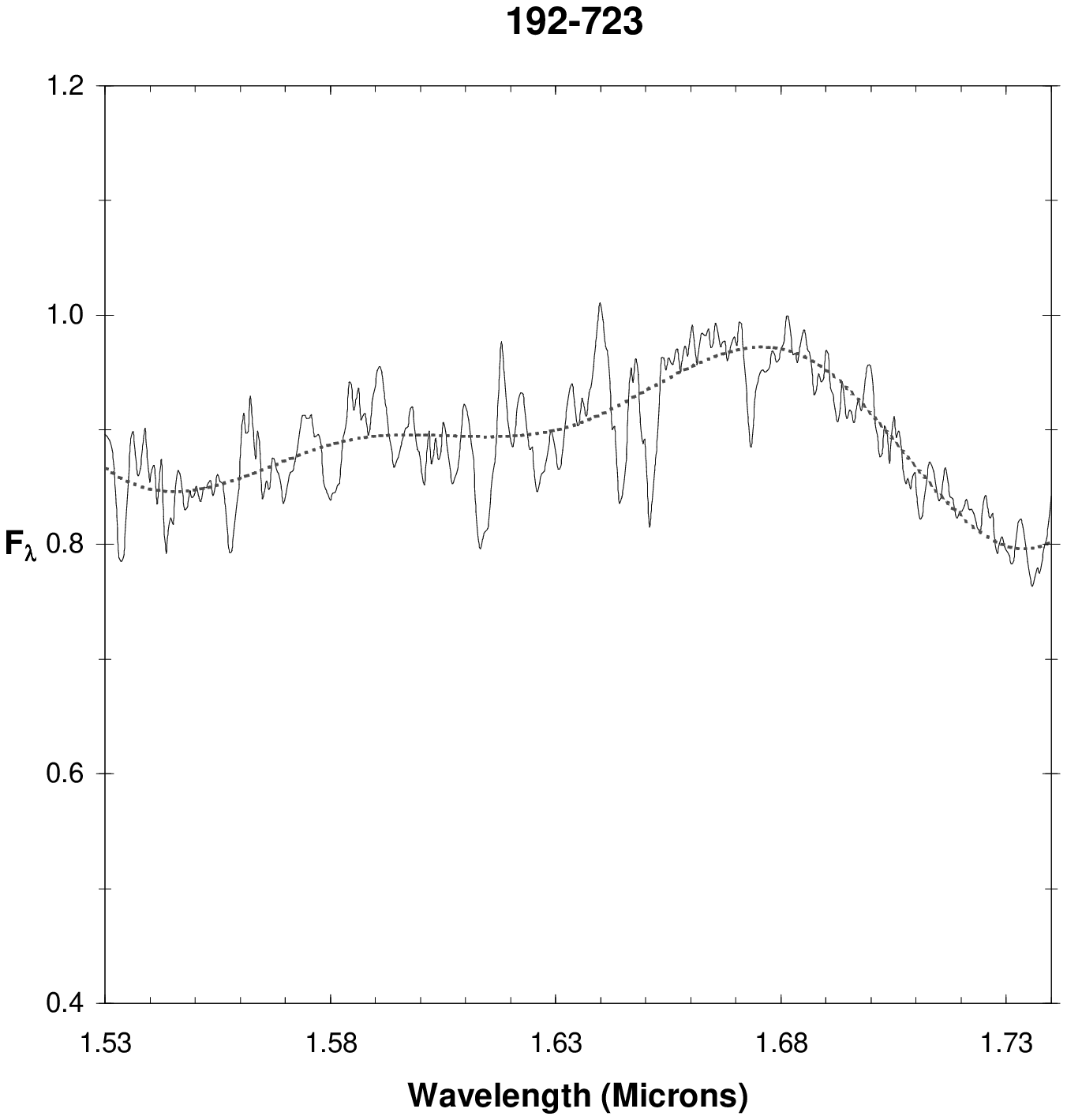,width=7.5cm,angle=-90}
    \psfig{file=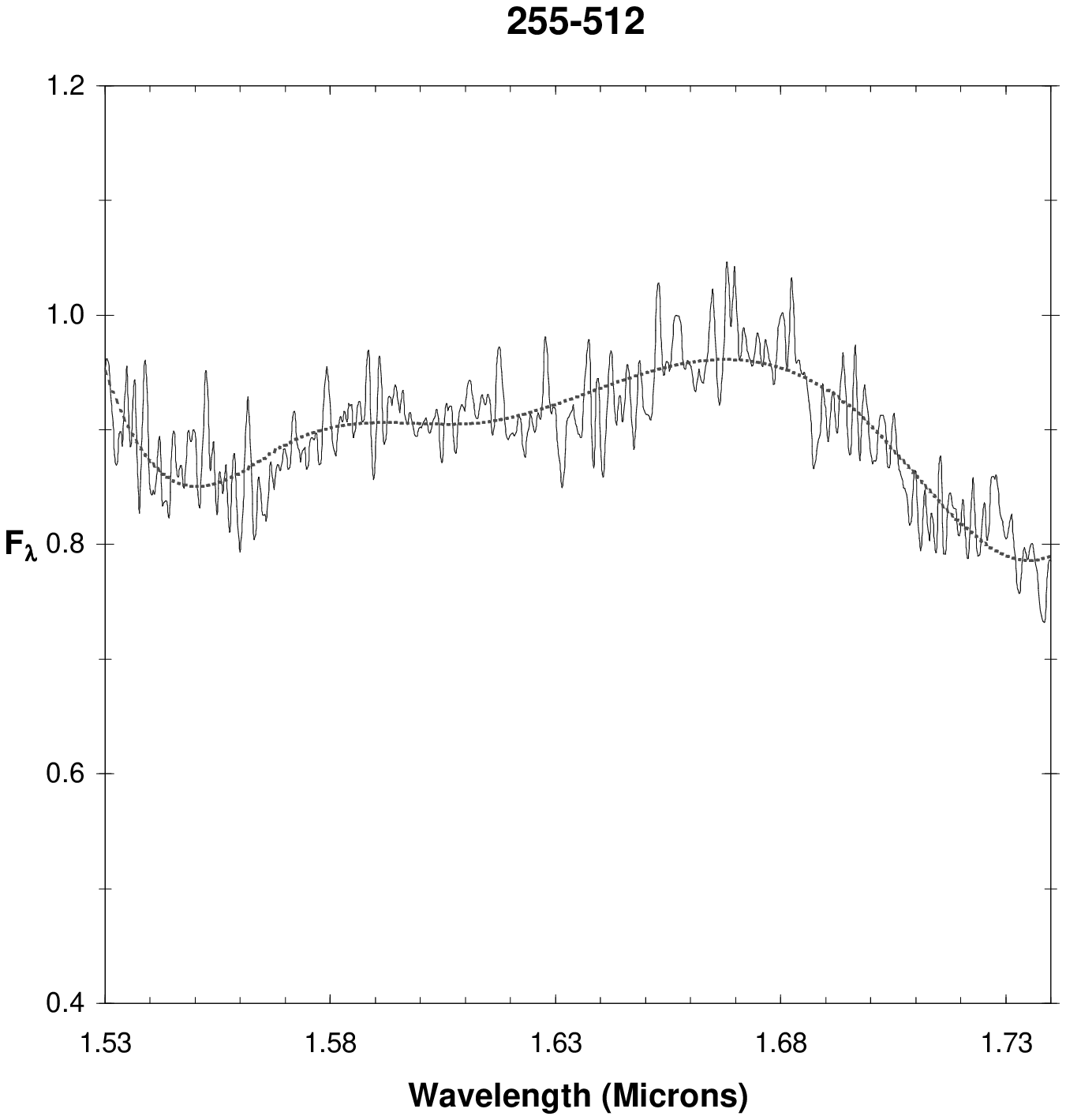,width=7.5cm,angle=-90}}
\caption{Normalised, dereddened \emph{H-}band IRIS2 and NIRI
spectra. Each source shows strong H${_2}$O absorption, resulting in
a triangular spectral profile. This provides strong evidence that
the objects are young and have low surface gravity. The spectral
resolution is too low to resolve narrow features. Therefore the
noise is indicated by the pixel to pixel variations. A 5th order
cubic spline fit is overplotted onto each spectrum. Note that the
spectrum of 047-436 is plotted as F$_{\nu}$} \label{hspectraa}
\end{figure*}

\addtocounter{figure}{-1}
\begin{figure*}
\centering{\ } 
  \hbox{
    \psfig{file=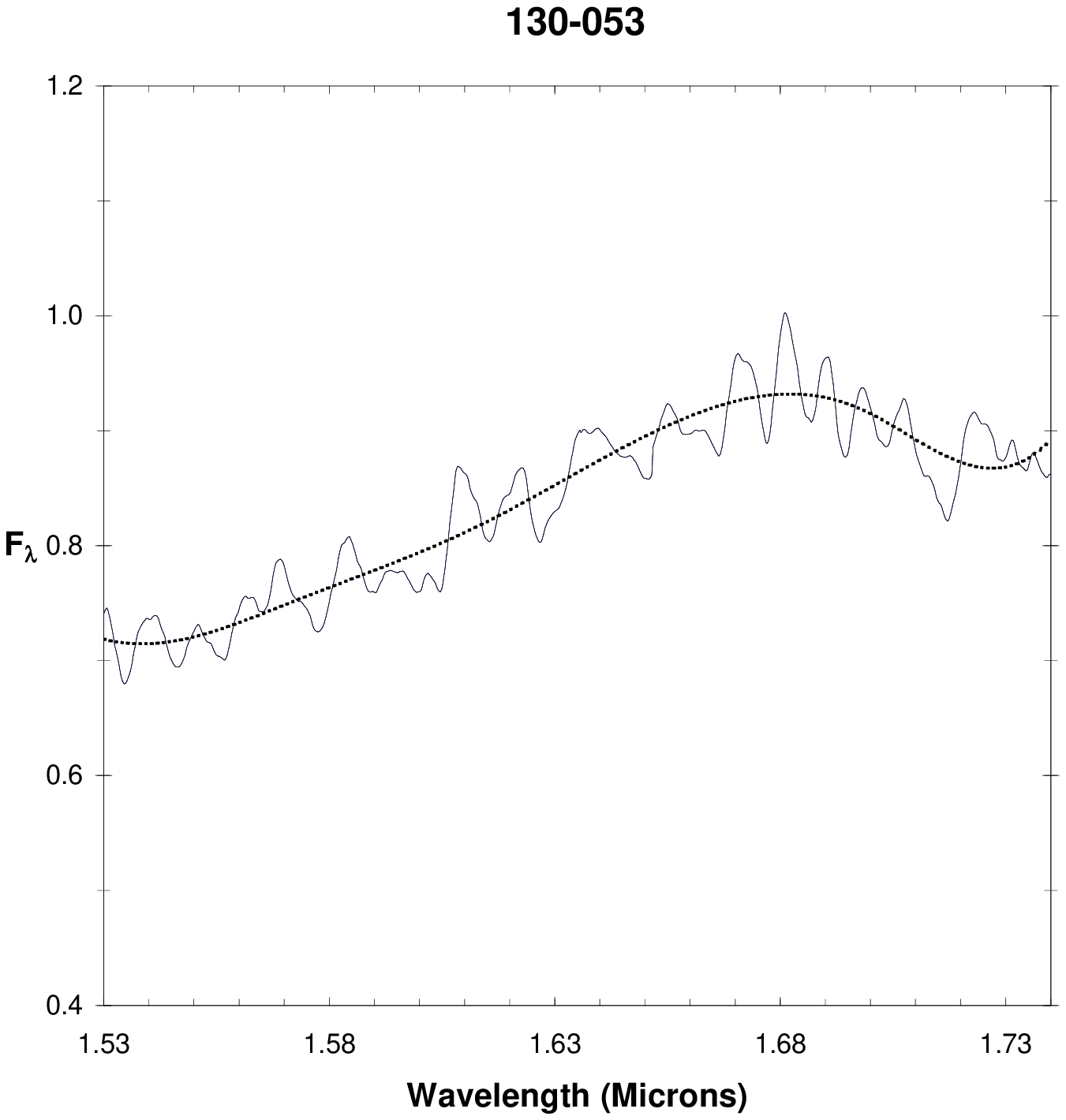,width=7.5cm,angle=-90}
    \psfig{file=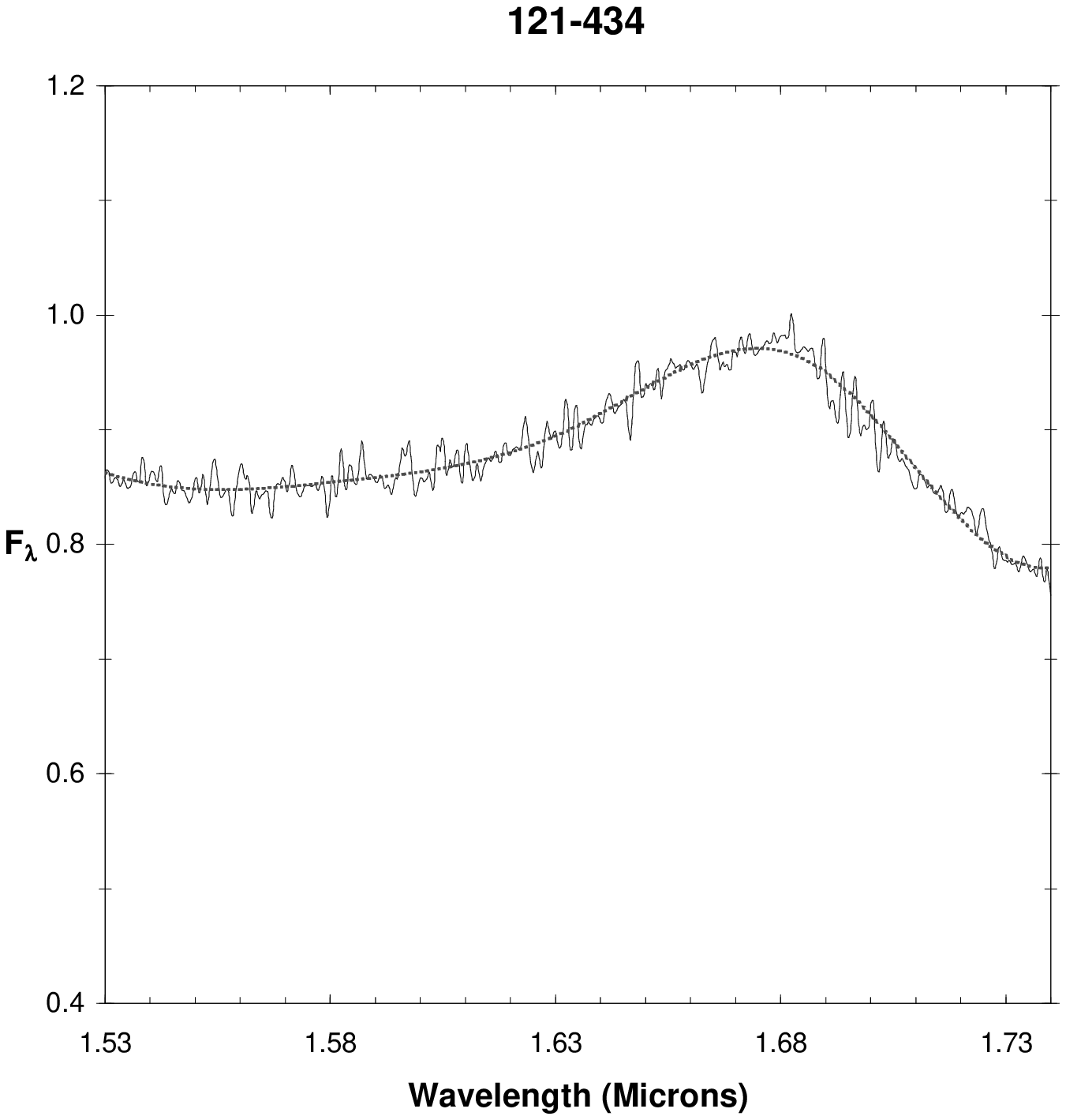,width=7.5cm,angle=-90}}
  \hbox{
    \psfig{file=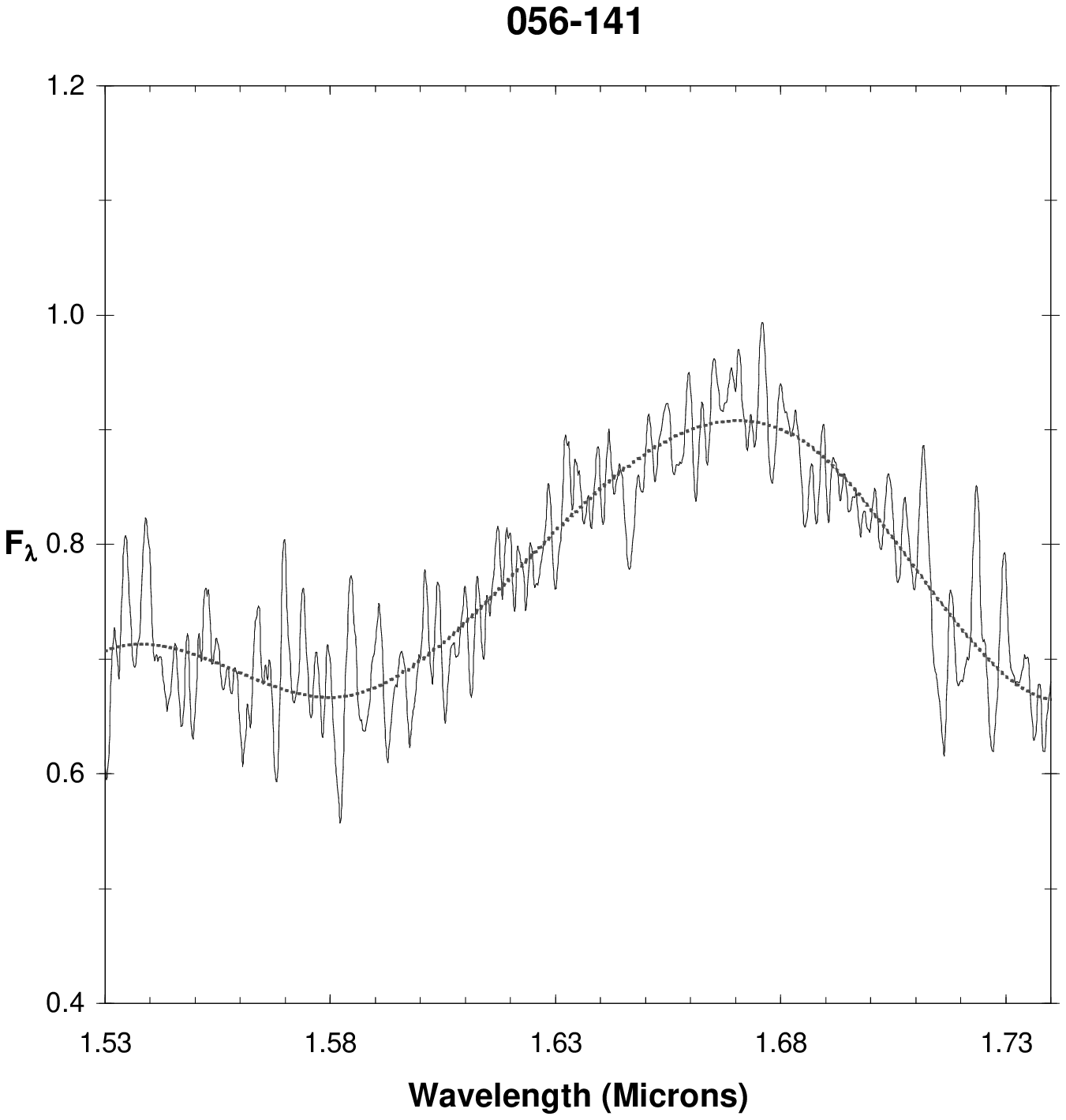,width=7.5cm,angle=-90}
    \psfig{file=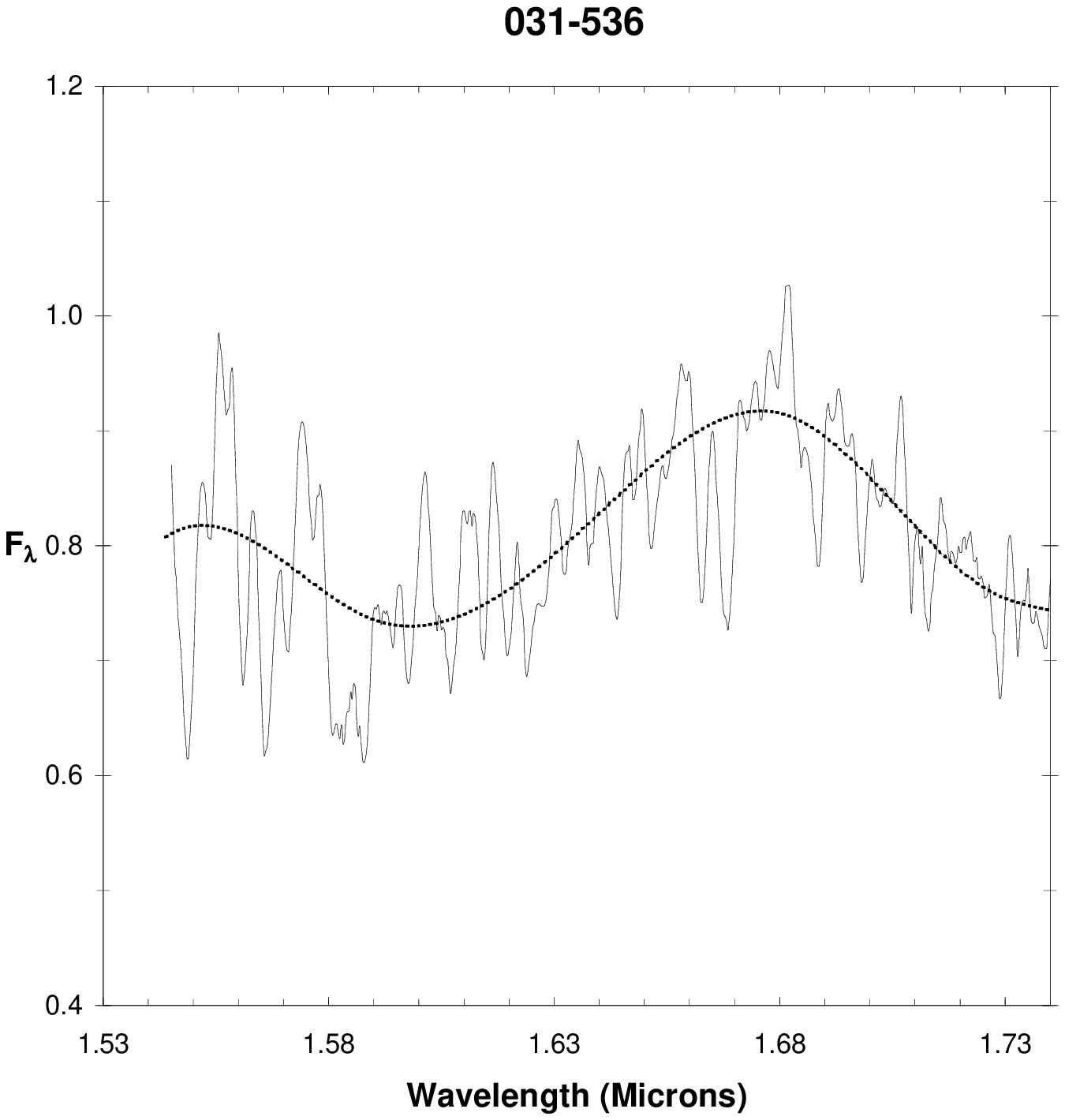,width=7.5cm,angle=-90}}
  \hbox{
    \psfig{file=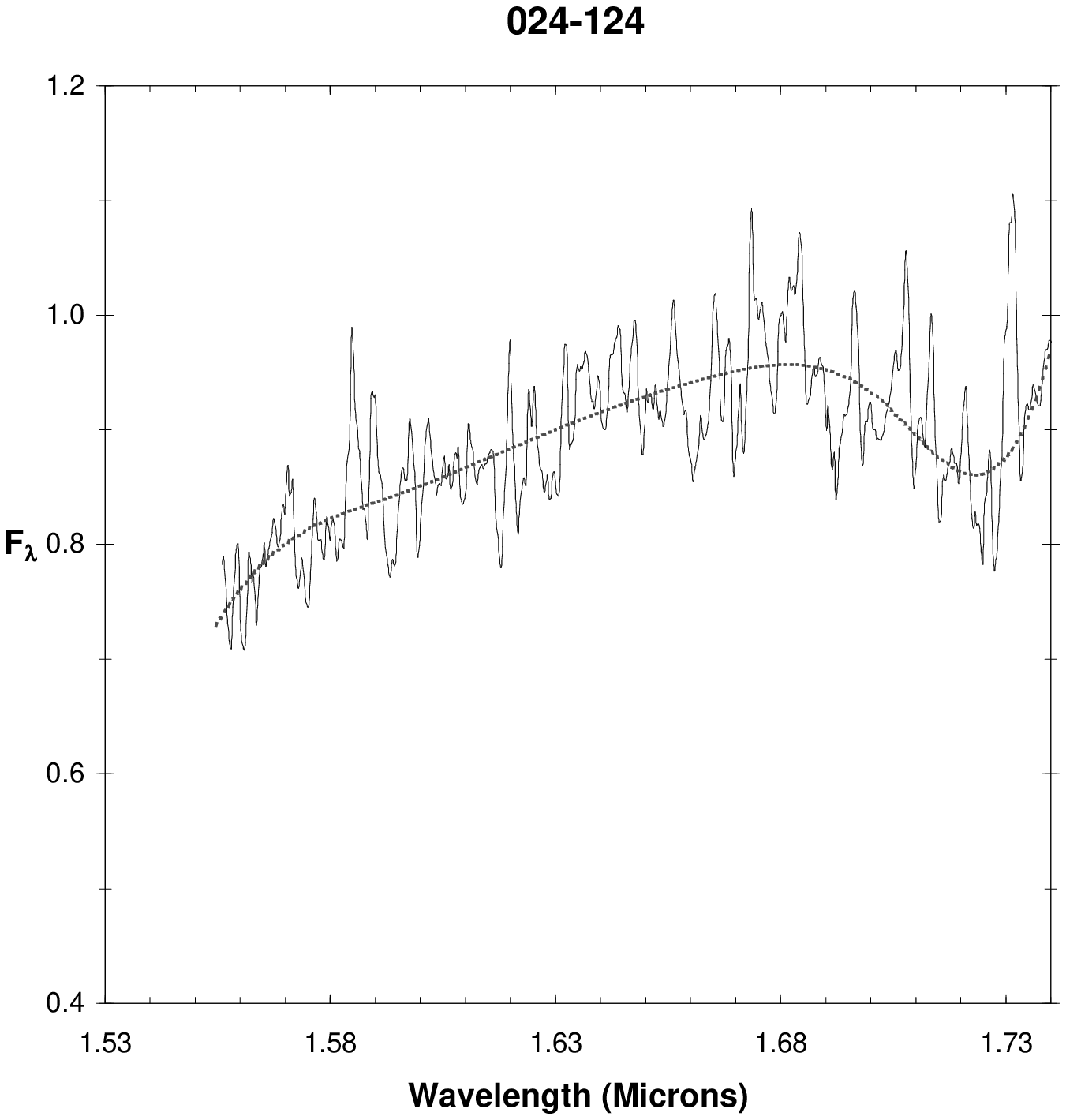,width=7.5cm,angle=-90}
    \psfig{file=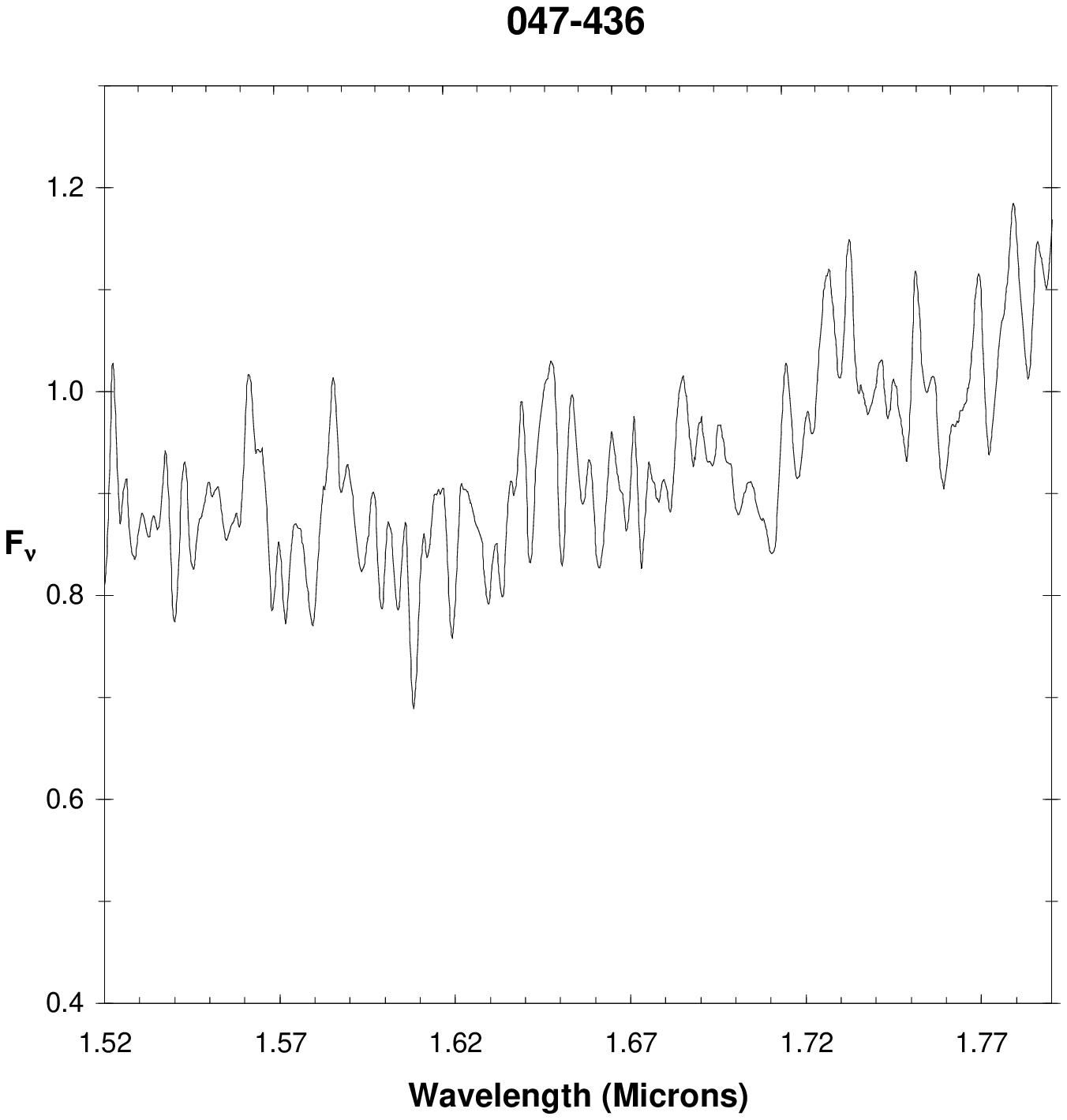,width=7.5cm,angle=-90}}
\caption{continued} \label{hspectrab}
\end{figure*}


\subsection{\emph{K-}Band}
\label{kbandsection} NIRI \emph{K-}band spectra are presented in
Figure \ref{kspectra}. Only objects that were of high enough quality
to be spectral typed have been shown. Spectra that had a
Rayleigh-Jeans type continuum were not plotted. For consistency the
wavelength coverage in the plots ranges between 1.96 and 2.36
$\umu$m. This does not represent the entire wavelength coverage and
has been done due to some spectra having very poor signal to noise
beyond this range. The spectra are plotted using a normalised
F$_{\lambda}$ flux scale. The pseudo continuum of each spectrum has
been fit by a 6th order cubic spline that has been over-plotted in
all cases. The continuum fits were created using the same procedure
described for the \emph{H-}band spectra. The spectra were smoothed
with an 11 pixel boxcar and trimmed to give a wavelength range of
1.96 to 2.36 $\umu$m. Not every spectrum in the sample required this
much smoothing or trimming. However, the same parameters were used
on each object for consistency. Smoothing by this amount may dilute
narrow lines such as CO and NaI. However, these are not required for
the spectral typing procedure. The final fits followed the
pseudo-continuum very well and were used for spectral typing.

The shape of the water absorption profile can be seen in a similar
configuration as observed at \emph{H-}band, with a flux peak close
to 2.20 $\umu$m. The flux maxima are flatter than seen at
\emph{H-}band, producing a spectral profile that is less obviously
triangular in shape. All spectra show strong absorption beyond this
wavelength suggesting that water absorption is not significantly
veiled by hot circumstellar dust. Three of the ten spectra presented
at \emph{K-}band (010-109, 095-058 and 186-631) show evidence of the
first ($ \nu$ $=$ 2$-$0) vibration-rotation bandhead of CO at 2.295
$\umu$m, a feature often seen to be weak in young objects. The
spectra of 095-058 and 186-631 show clear evidence of multiple CO
bandheads. Figure \ref{186-631abs} shows 186-631 with extended
wavelength coverage out to 2.46 $\umu$m. Multiple CO bandheads can
be seen at 2.32, 2.35 and 2.38 $\umu$m. All of these
vibration-rotation absorption features can be seen in the spectrum
095-058. A further three spectra show weak evidence of the 2.295
$\umu$m CO absorption feature; 013-306, 015-319 and 084-104. It is
possible that the tentative CO absorption feature seen in 084-104 at
2.295 $\umu$m is attributable to noise. The spectra of 186-631 and
095-058 appear to be somewhat different to the other \emph{K-}band
spectra. The spectra of 095-058 and 186-631 peak at a longer
wavelength than the 2.2 $\umu$m peaks of the other young objects, at
a wavelength of $\sim$2.26 $\umu$m. This is not an artifact
introduced by noise as the signal to noise for these spectra are
relatively high. Empirical evidence in numerous publications
suggests this is not abnormal for young objects and is in fact
commonly observed; e.g. Luhman, Peterson \& Megeath 2004, Luhman et
al. 2005, Luhman et al. 2006. Both 044-527 and 030-524 have low
signal to noise and have been smoothed. Despite low signal to noise
the spectral profile of 044-527 appears to be consistent with that
of a young, low gravity object. The cubic spline fit represents the
pseudo-continuum for this object well and is therefore suitable for
spectral typing. The \emph{K-}band spectrum of 030-524 is less
convincing due to the low signal to noise. However, this object is a
probable low mass cluster member. RRL have obtained a high quality
spectrum for this object at optical wavelengths. From the data they
derive a spectral type of M7.5$\pm$1.5. This indicates that the
broad H${2}$O absorption bands depicted by the cubic spline fit are
likely to be genuine.


\begin{figure*}
\centering{\ } 
  \hbox{
    \psfig{file=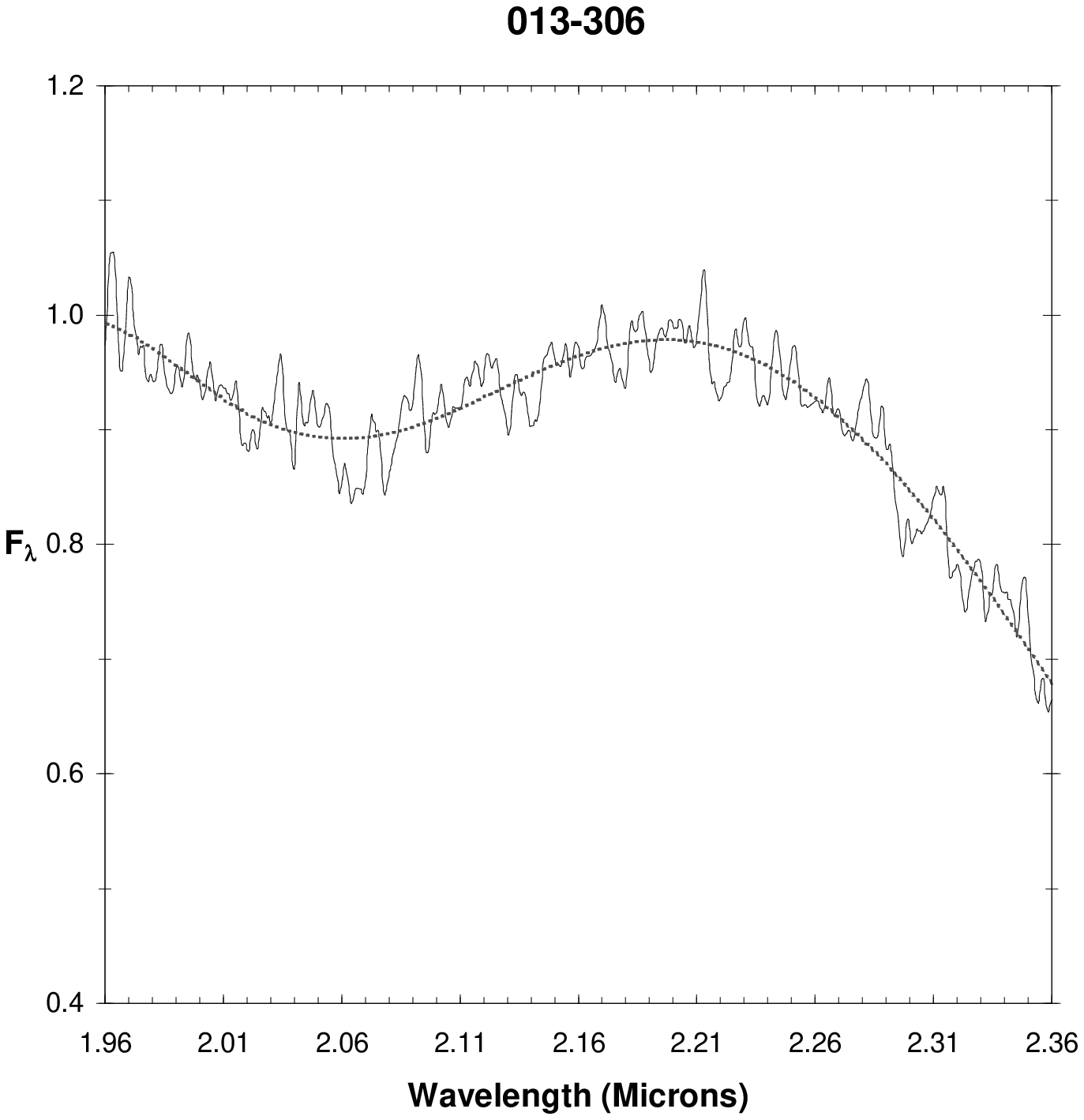,width=7.5cm,angle=-90}
    \psfig{file=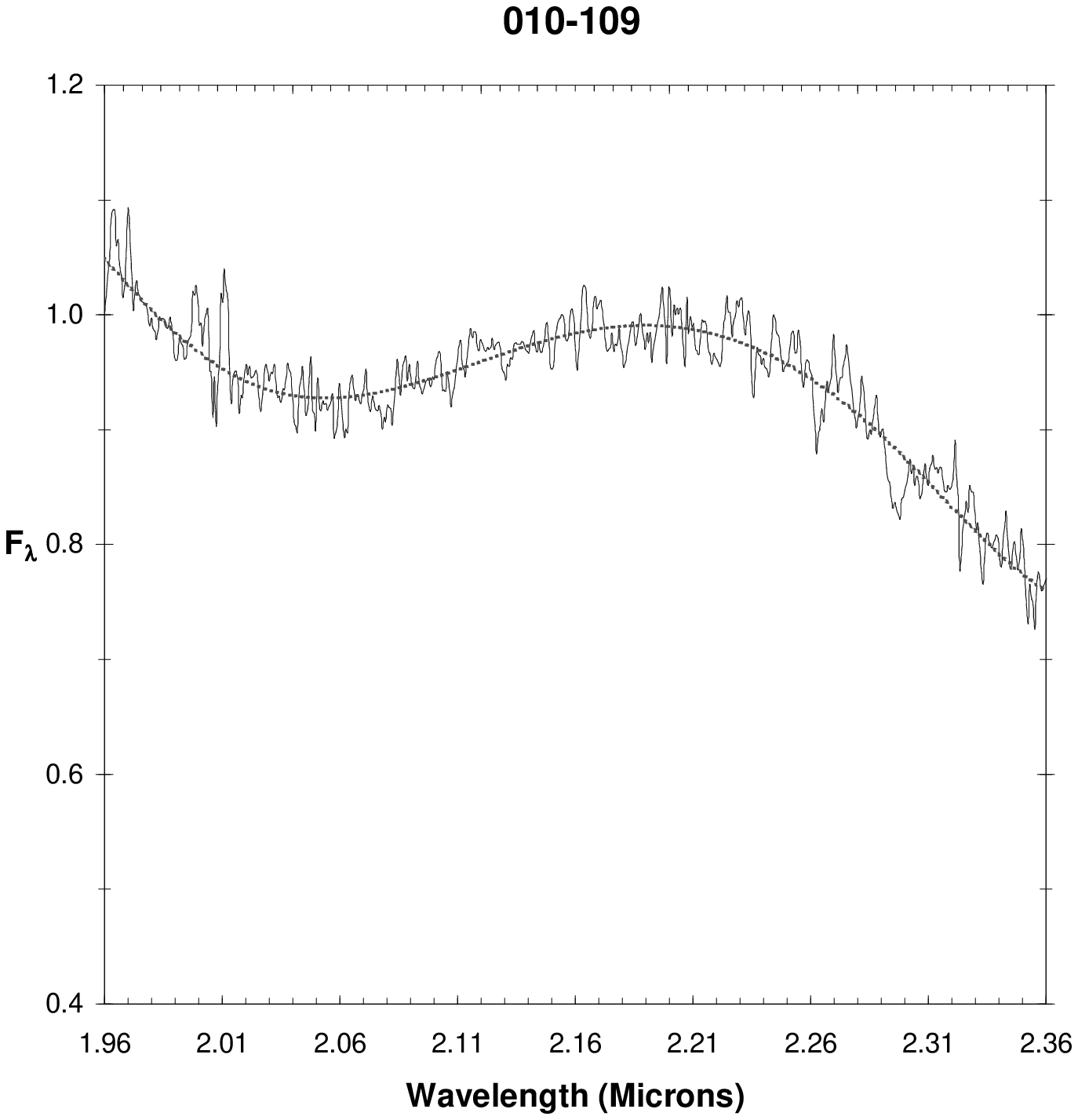,width=7.5cm,angle=-90}}
  \hbox{
    \psfig{file=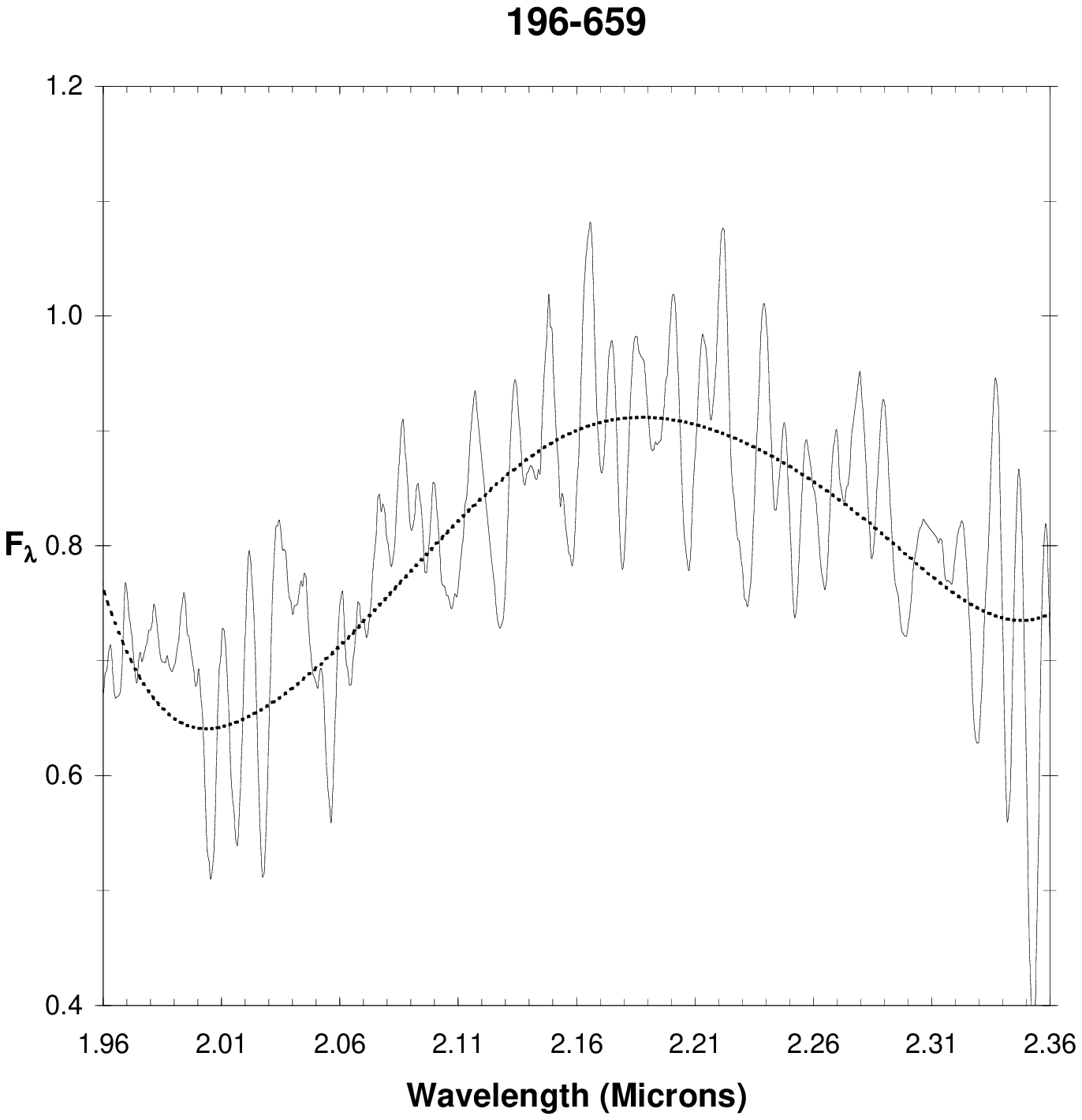,width=7.5cm,angle=-90}
    \psfig{file=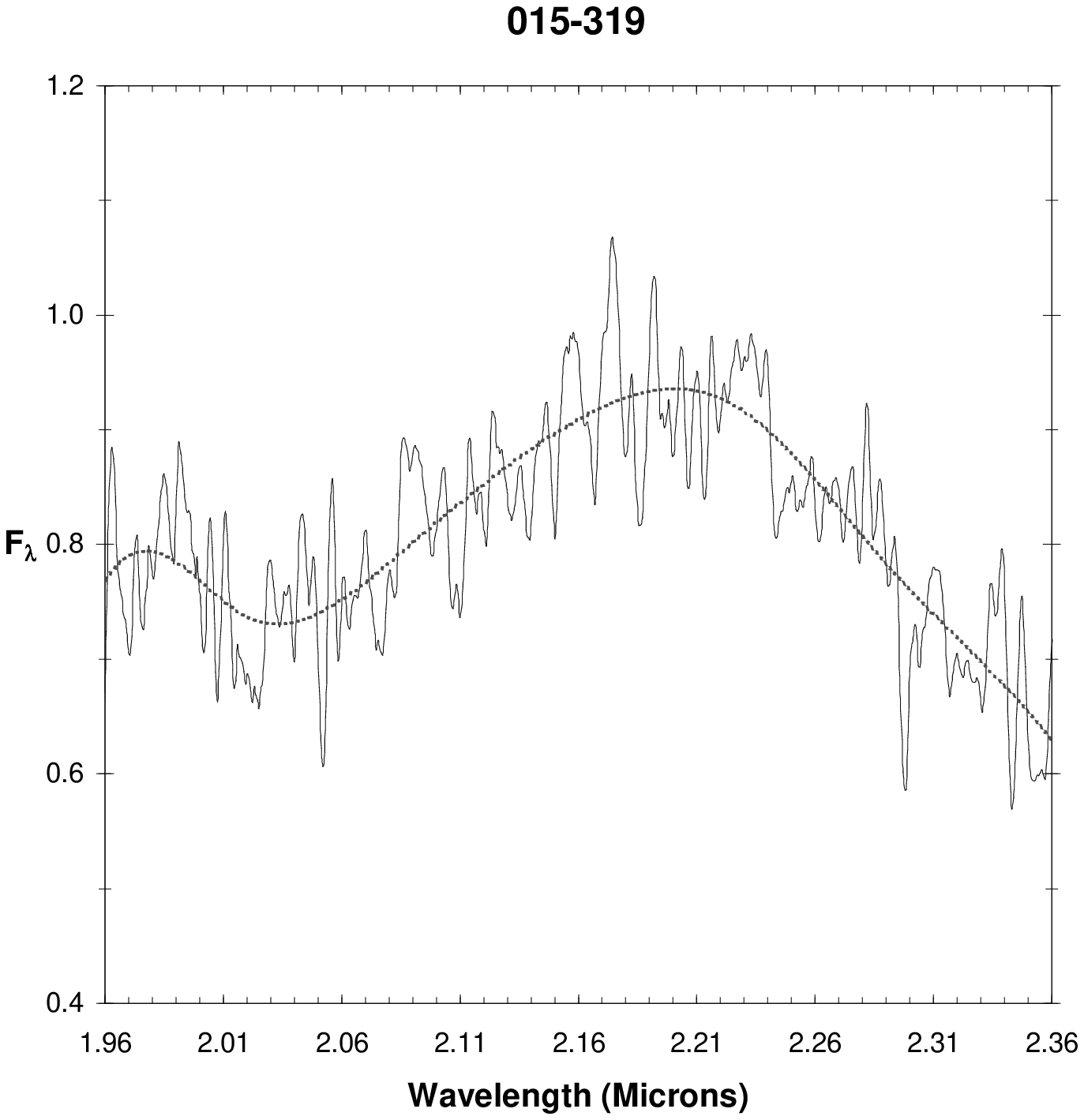,width=7.5cm,angle=-90}}
  \hbox{
    \psfig{file=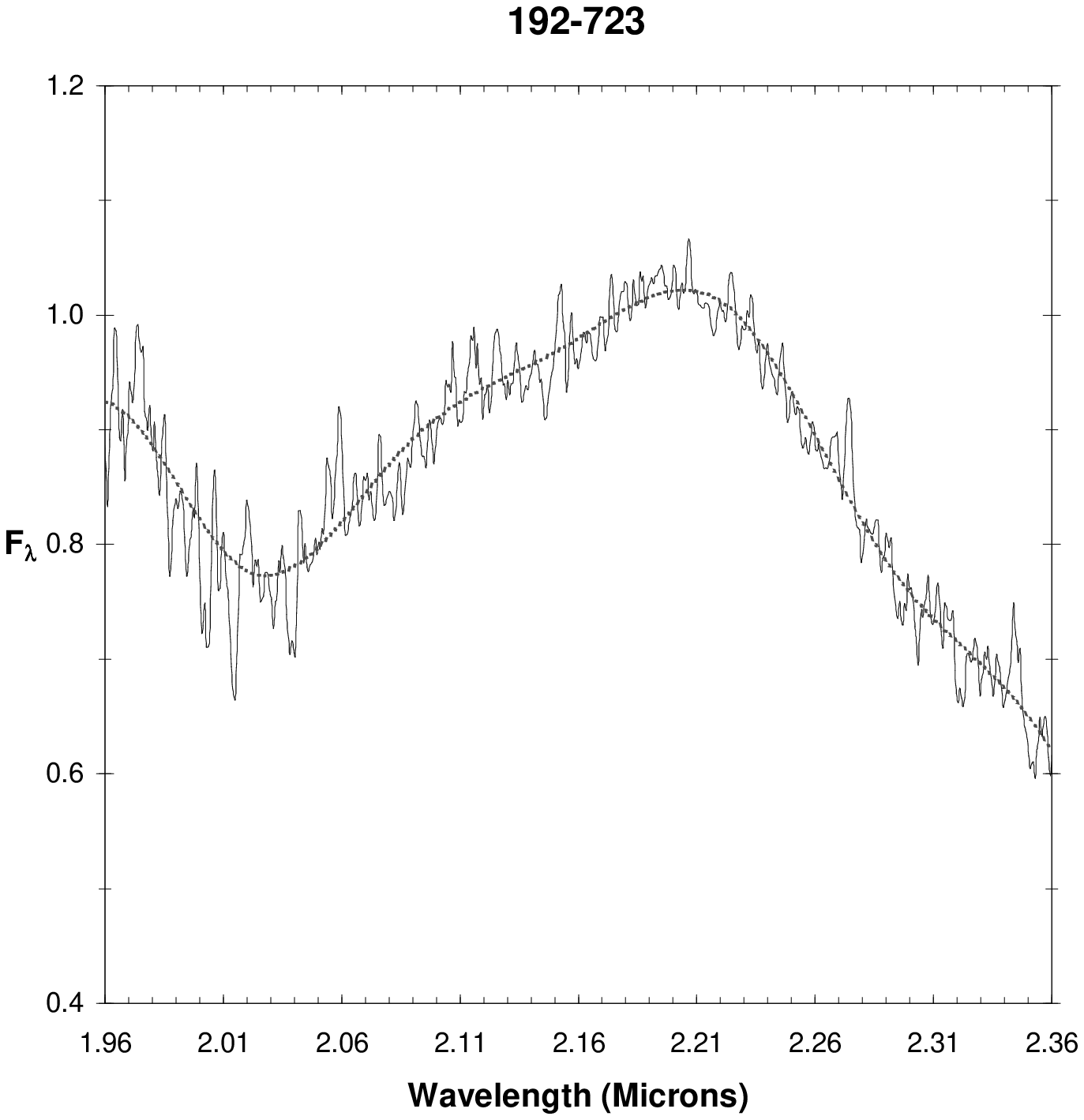,width=7.5cm,angle=-90}
    \psfig{file=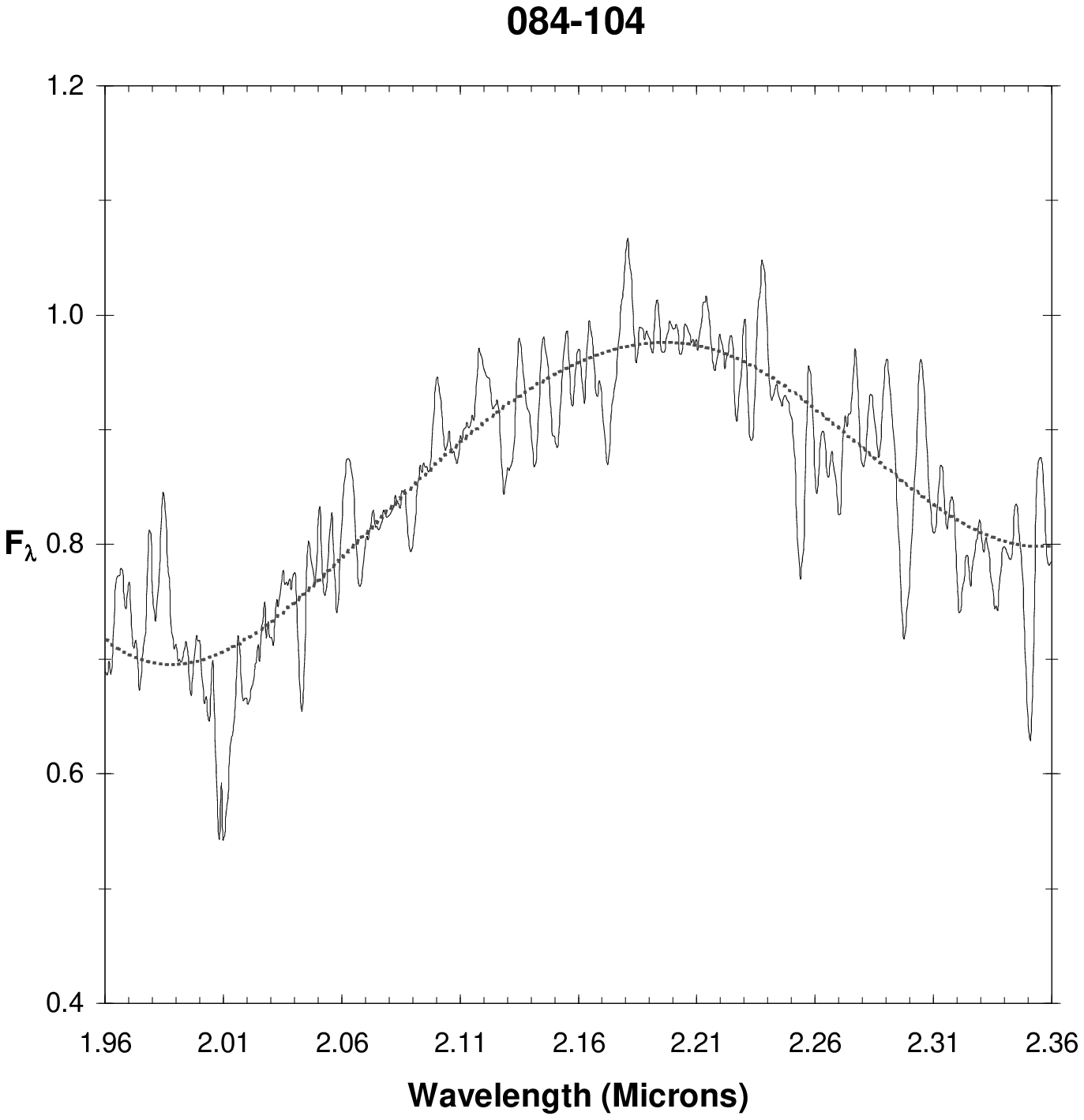,width=7.5cm,angle=-90}}
\caption{Normalised, dereddened \emph{K-}band NIRI spectra. Each
source shows strong H${_2}$O absorption, resulting in a flat-peaked,
triangular spectral profile. This provides strong evidence that the
objects are young and have low surface gravity. Several spectra show
evidence of CO absorption at 2.295 $\umu$m. The 6th order cubic
spline fit to the continua used for spectral typing has been
overplotted onto each spectrum.} \label{kspectra}
\end{figure*}

\addtocounter{figure}{-1}
\begin{figure*}
\centering{\ } 
  \hbox{
    \psfig{file=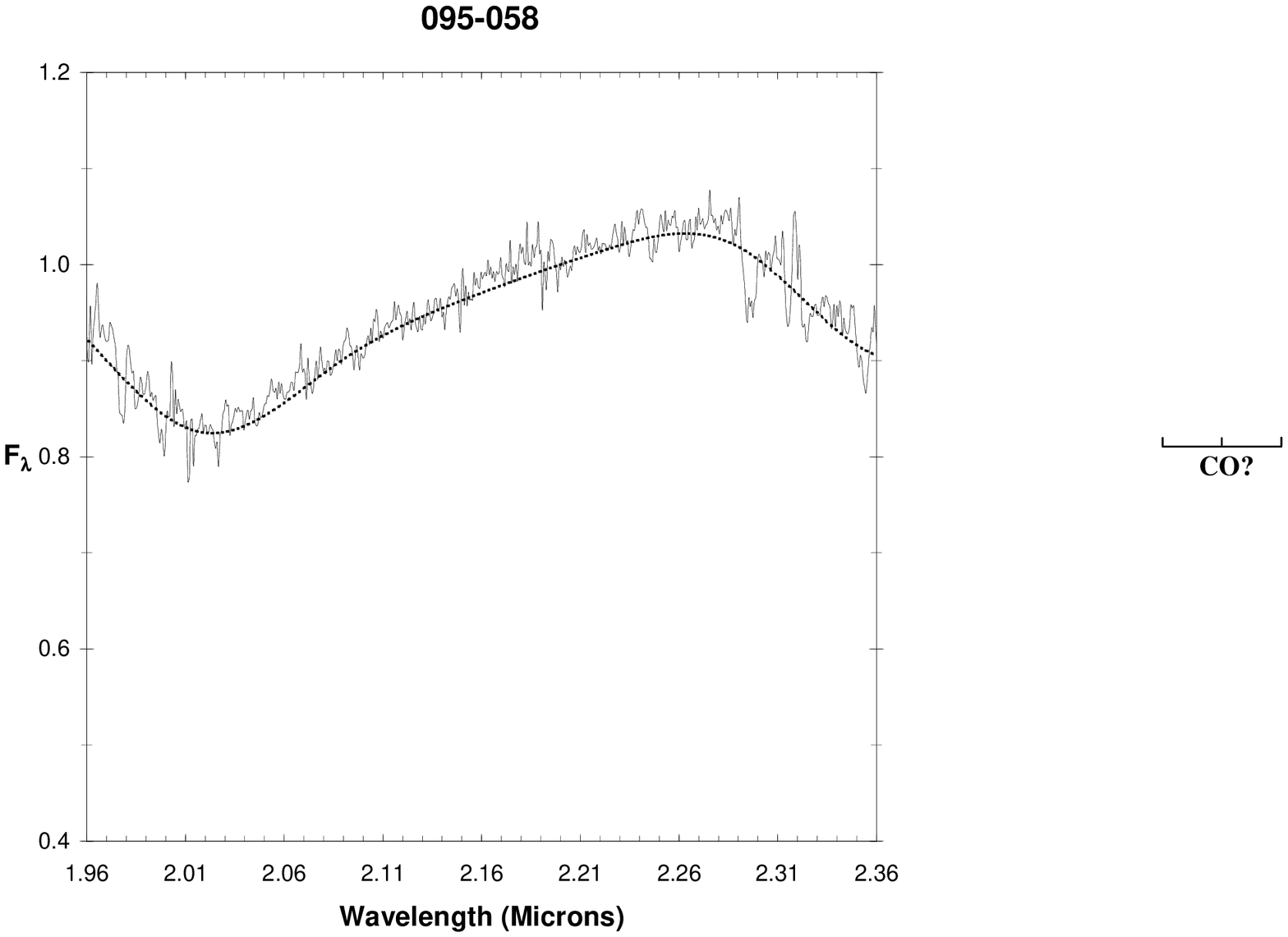,width=7.5cm,angle=-90}
    \psfig{file=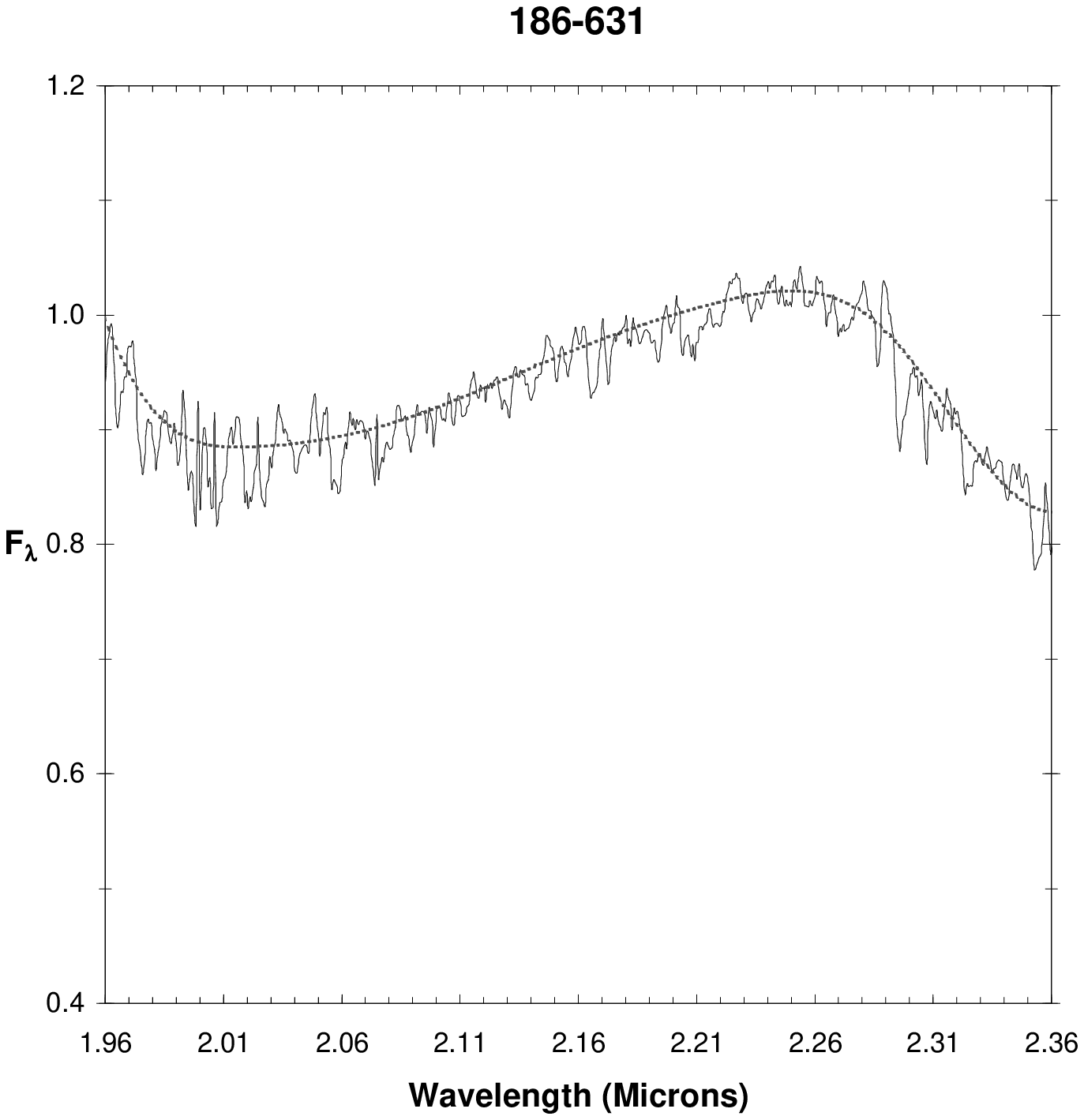,width=7.5cm,angle=-90}}
  \hbox{
    \psfig{file=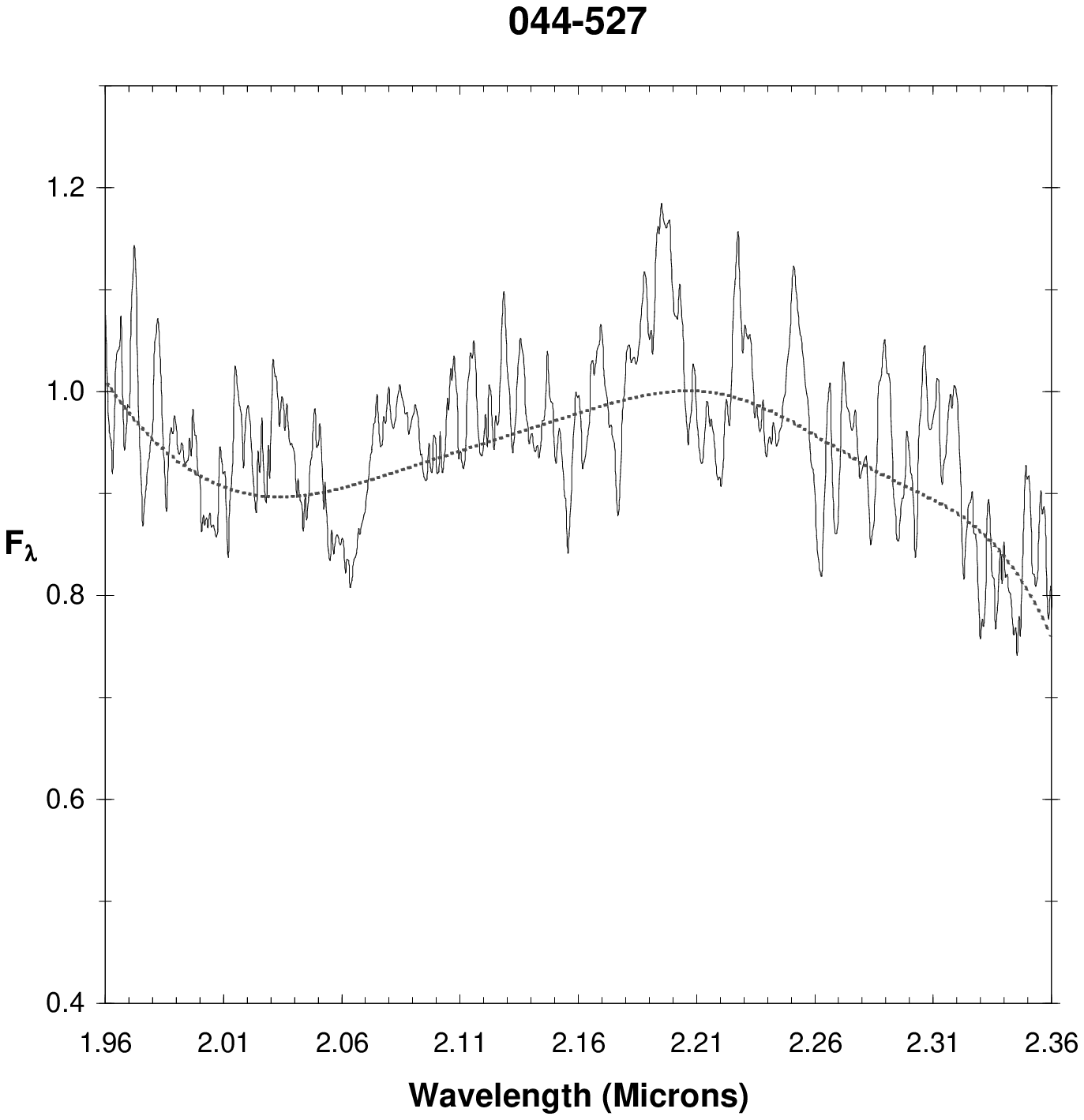,width=7.5cm,angle=-90}
    \psfig{file=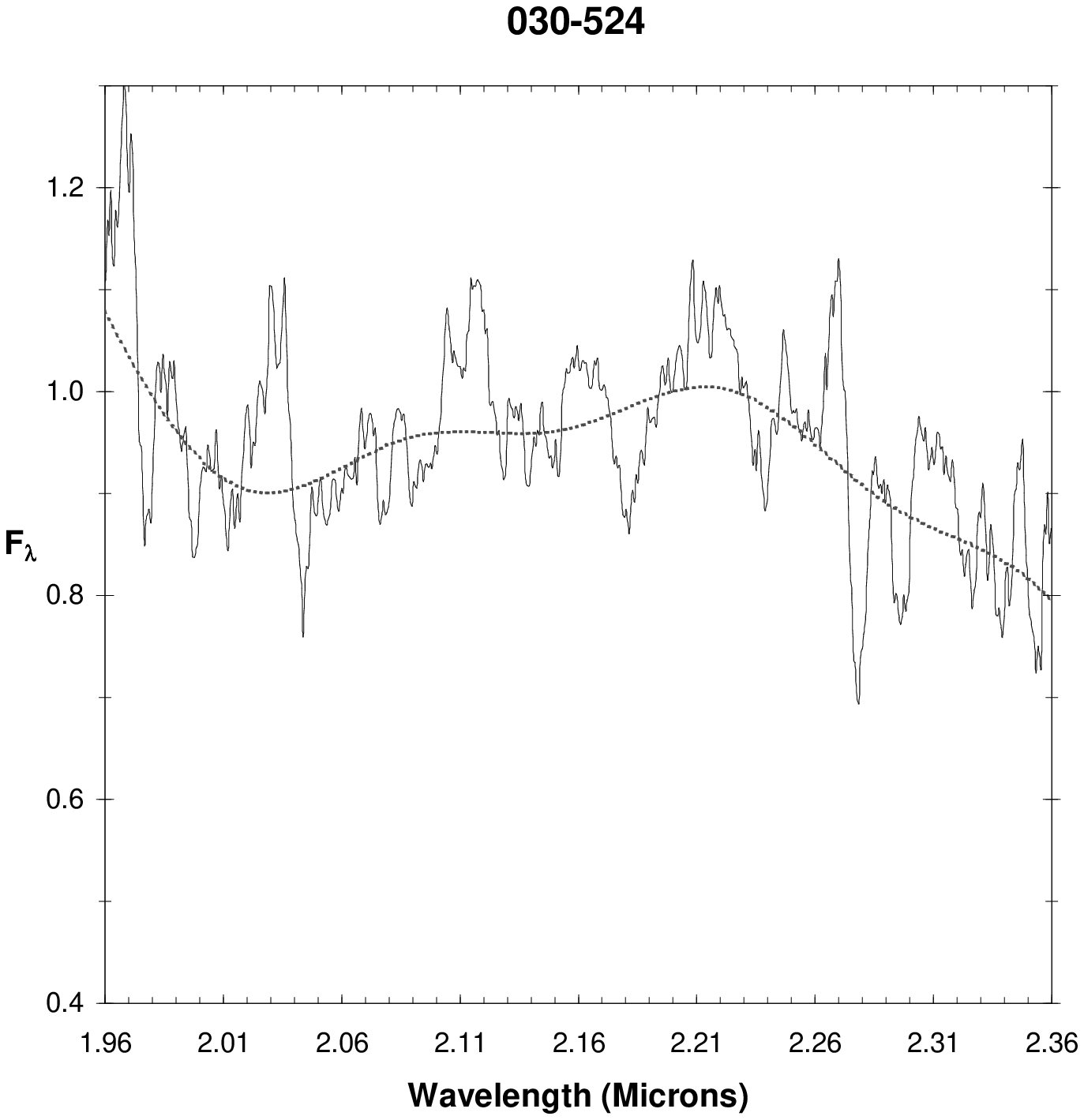,width=7.5cm,angle=-90}}
\caption{continued} \label{kspectrab}
\end{figure*}

\section{Spectral Typing}
\label{specType} Accurate spectral typing is critical if we want to
have a good understanding of the true behavior of the IMF at the
sub-stellar regime. Until recently the spectral classification of
low mass objects in young stellar nurseries was determined via the
use of spectral indices that were empirically derived from a variety
of field brown dwarf and giant spectra combinations e.g. the optical
work carried out in IC 348 by Luhman et al.(2003). Comparisons with
synthetic spectra created from detailed theoretical model
calculations (e.g. D'Antona \& Mazzitelli, 1997 (hereafter DM97);
Baraffe et al. 1998; Chabrier et al. 2000 and Allard, Hauschildt \&
Schwenke, 2000) have also been used to spectral type objects and
derive intrinsic temperatures. For the evolved population of field
dwarfs these techniques are an excellent tool for spectral
classification, yielding accurate results. However, significant
limitations exist when applying these techniques to very young
objects. Infrared spectral indices that compare relative strengths
of molecular and atomic features observed in field dwarfs, often
fail to predict a real spectral type as might be defined at optical
wavelengths (Lucas et al.2001). This is due to the fact that
near-infrared indices are predominantly assigned to gravity
sensitive features that dominate the spectra. The very young brown
dwarf and planetary mass candidates found in Orion are still at an
early stage in their evolution and thus are still contracting and
have large radii. The consequential low surface gravity results in
broad water absorption bands that are suppressed by collision
induced H$_{2}$ absorption in more evolved high surface gravity
field dwarfs (Kirkpatrick et al. 2006). The v$=$2$-$0
vibration-rotation bands of CO starting at 2.29 $\umu$m are weaker
in absorption in young, late M-type sources, seemingly due to
gravity differences. However, veiling of warm circumstellar dust may
also contribute. As a consequence of these physical differences,
spectral indices will often predict spectral types that are too
late. Theoretical models become uncertain at ages below 1 MYr and
thus should be avoided as a sole tool for accurate spectral typing.

\begin{figure}
\centering{\ } 
    \psfig{file=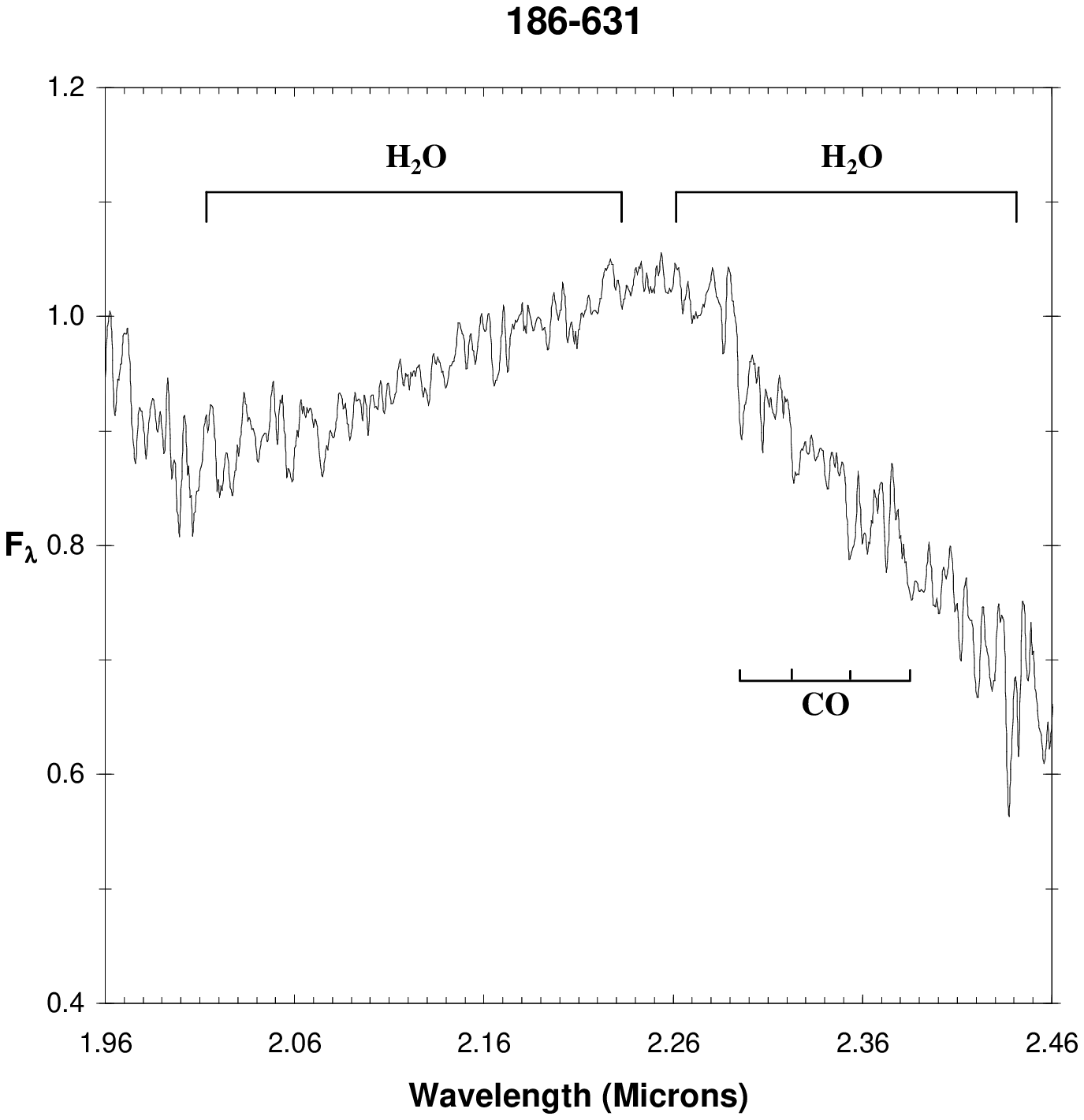,width=7.5cm,angle=-90}
\caption{Strong H$_{2}$O absorption can be seen either side of a
peak situated close to 2.26 $\umu$m. The triangular profile
indicates that the object has low surface gravity and is therefore
young. Multiple band heads of CO can be seen at 2.29, 2.32, 2.35 and
2.38 $\umu$m.} \label{186-631abs}
\end{figure}

In order to ascertain accurate spectral types of young objects with
the use of empirical measurements, indices must be derived and
calibrated from similar objects of known spectral type. Until
recently this has been difficult to do for young substellar objects
due to the lack of high quality optical spectra for near-infrared
counterparts. High quality infrared spectra of 39 young ($\sim$1 - 2
Myr) objects ranging in spectral type from M3 to M9.5 were kindly
supplied by Kevin Luhman (original spectra can be seen in
Brice\~{n}o et al. 2002; Luhman et al. 2003b; Luhman 2004). These
are located in Taurus and IC 348 star forming regions. The low
resolution (R=100) spectra were obtained with SpeX on IRTF. The
spectral types for these objects were calibrated from high quality
optical spectra and are therefore reliable for calibration. All
spectra are F$_{\lambda}$ and had been corrected for reddening. RRL
have optically determined spectral types for 17 ONC sources in our
infrared sample, 9 of which were observed at infrared wavelengths by
Lucas et al. 2001. The 8 additional sources have been covered by our
sample here. We chose to use Luhman's sample only when deriving
spectral types. This is because the quality of his template spectra
were generally of higher quality and thus more reliable than ours.

A total of four indices were constructed to spectral type each of
the Orion brown dwarf candidates. These differ slightly from those
mentioned in the previous paper (Lucas et al. 2006) as they required
careful adjustment in order for them to satisfy the spectral
coverage in the IRIS2 and NIRI spectra. Each index measures the
strength of the water absorption bands. The F$_{\lambda}$ flux value
was determined by the median flux in a 0.02$\umu$m wide interval
centered on the desired wavelength. Flux ratios were plotted as a
function of spectral type for each object in the Luhman sample. A
fit was derived for each spectral index using a cubic polynomial.
The polynomial fits are not weighted as the true error for each data
point cannot be quantified accurately. If systematic errors are
neglected we can approximate an error using the internal scatter
within the finite wavelength regions applicable to the spectral
index. Pseudo errors were generated for Luhman's data, for each
index, and then used to generate weighted fits. Due to the smooth
nature of Luhman's low resolution spectra, the weighted fits showed
an insignificant difference to the non-weighted fits, implying that
our approach is reliable, provided systematic errors are small.

The index ratios that were chosen had to satisfy two requirements.
Firstly the quality of the fit should be high (i.e. coefficient of
determination (R$^{2}>$ 0.9 where possible). R$^{2}$ is the
coefficient of determination and is shown in Equation
\ref{CoeffOfDeterm}. The numerator represents the sum of the squared
errors and the denominator represents the total sum of the squares.
$y_{i}$ represents a data value, $\hat{y_{i}}$ represents a model
value and $\bar{y}$ is the mean of the data values. Secondly the
gradient of the fit should not be too shallow so that a small error
in a given index will not result in a significant span of spectral
types. Index ratios were rejected if they produced a fit that was
too shallow regardless of how good the fit was. The four indices are
defined in Equations \ref{WH} to \ref{QK}.

\begin{equation}
\label{CoeffOfDeterm} R^{2} = 1-\frac{\displaystyle\sum_{i}
(y_{i}-\hat{y_{i}})^{2}}{\displaystyle\sum_{i}(y_{i}-\bar{y})^{2}}
\end{equation}

\begin{equation}
\label{WH} WH=\frac{F_{\lambda}(1.562\umu m)}{F_{\lambda}(1.665\umu
m)}
\end{equation}
\begin{equation}
\label{WK}WK=\frac{F_{\lambda}(2.050\umu m)}{F_{\lambda}(2.190\umu
m)}
\end{equation}
\begin{equation}
\label{QH} QH=\frac{F_{\lambda}(1.562\umu m)}{F_{\lambda}(1.665\umu
m)}\left[\frac{F_{\lambda}(1.740\umu m)}{F_{\lambda}(1.665\umu
m)}\right]^{1.581}
\end{equation}
\begin{equation}
\label{QK} QK=\frac{F_{\lambda}(2.050\umu m)}{F_{\lambda}(2.192\umu
m)}\left[\frac{F_{\lambda}(2.340\umu m)}{F_{\lambda}(2.192\umu
m)}\right]^{1.140}
\end{equation}

\begin{figure*}
\centering{\ } 
 \vbox{
  \hbox{
    \psfig{file=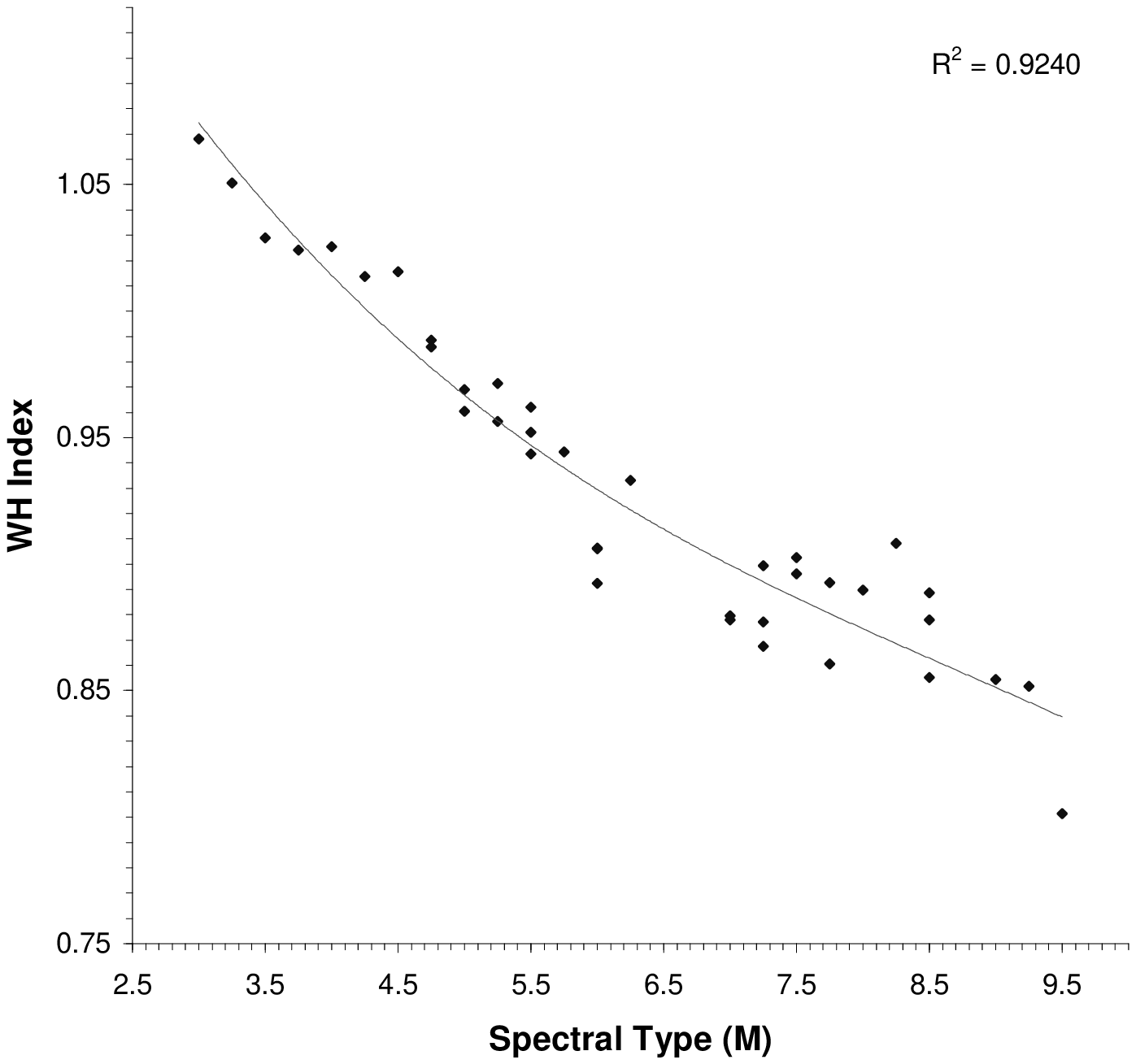,width=6.9cm,angle=-90}
    \psfig{file=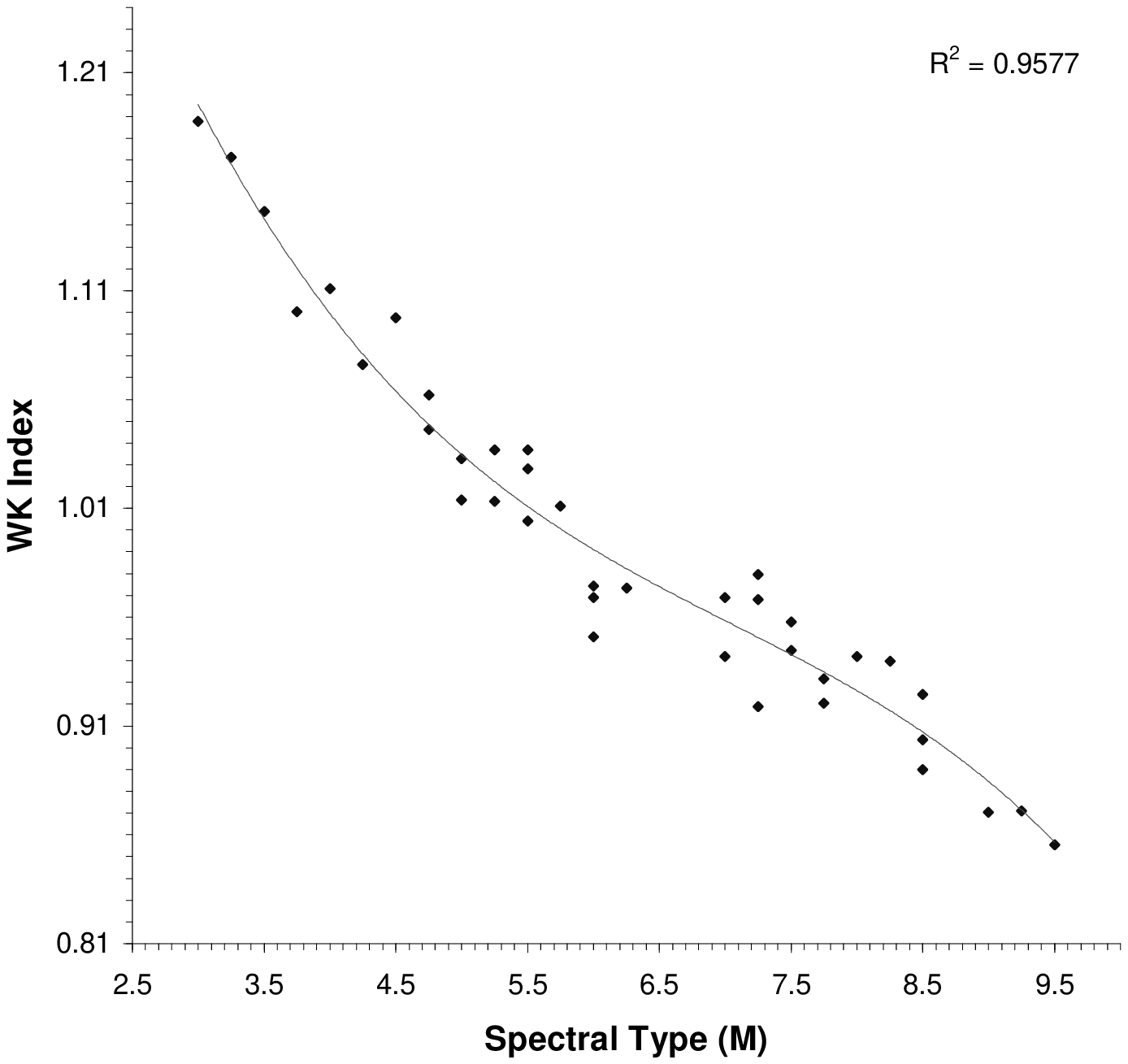,width=6.9cm,angle=-90}}
  \hbox{
    \psfig{file=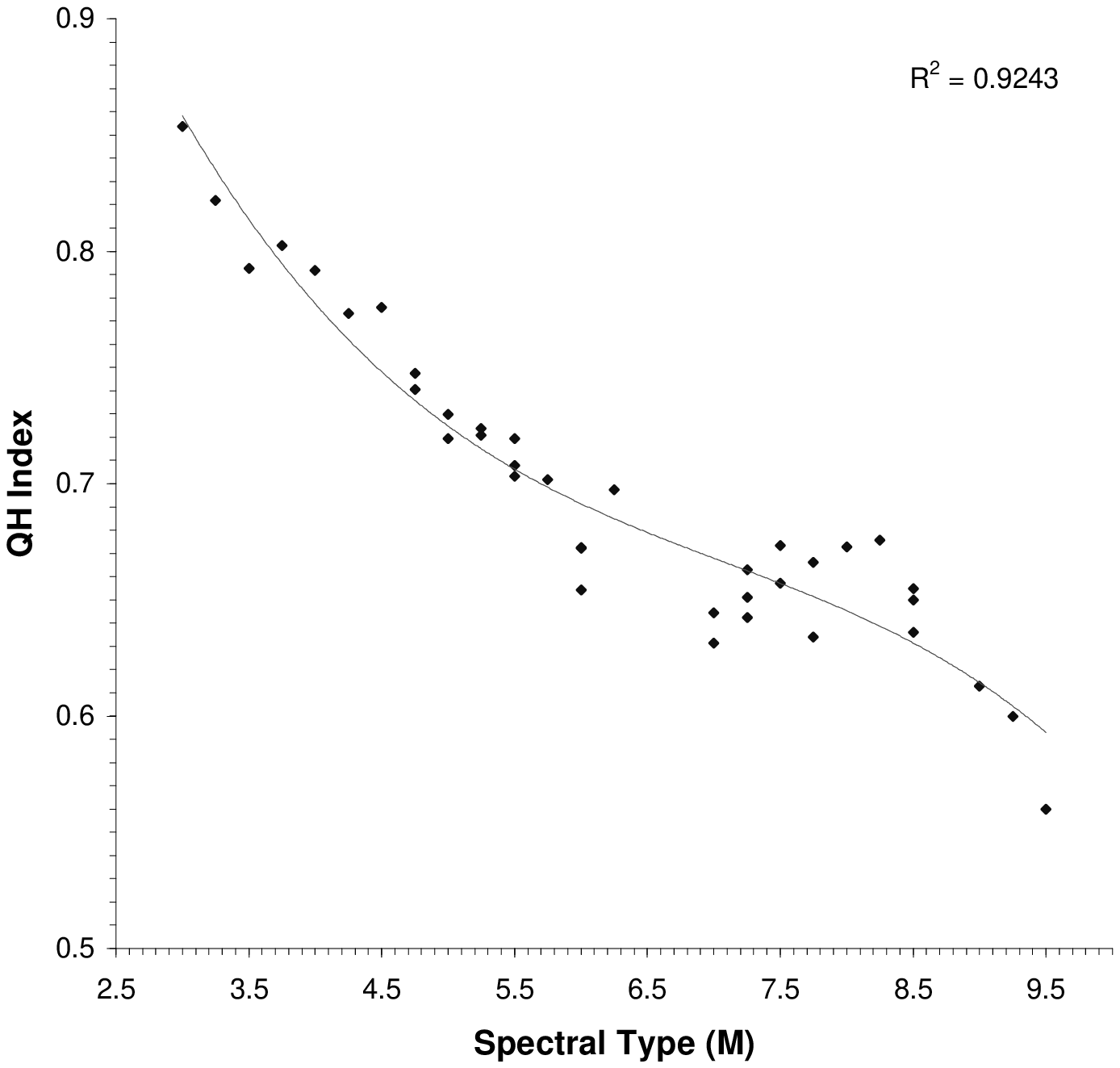,width=6.9cm,angle=-90}
    \psfig{file=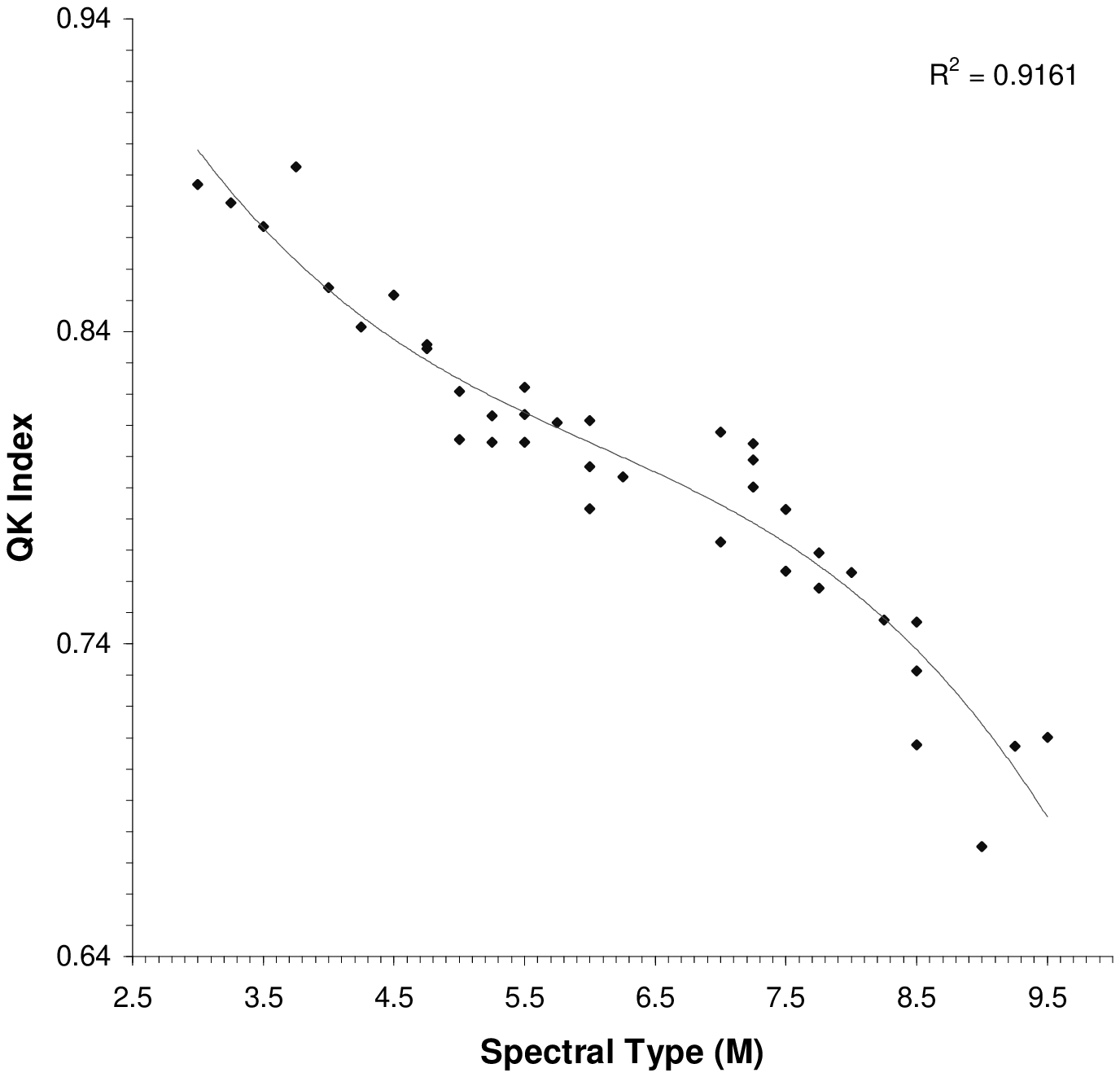,width=6.9cm,angle=-90}}}
\caption{Index strength as a function of spectral type. Each data
point represents an object from Luhman's near-infrared
spectroscopic sample of optically calibrated young ($\sim$1Myr)
brown dwarfs. For each ratio the median flux value in a 0.02$\umu$m
interval was used. The relationship between spectral type and index
are fitted by cubic polynomials. The R$^2$ values are correlation
coefficients (see Section \ref{specType} for details.).}
\label{indices}
\end{figure*}

The QH and QK indices (Equations \ref{QH} and \ref{QK}) are a
reddening independent measurement of water absorption, assuming the
Cardelli, Clayton \& Mathis (1989) reddening law of
$\lambda^{-1.61}$. To ensure errors were minimal, these indices were
constructed so that the two wavelength regions either side of the
central wavelength position were approximately equidistant, i.e.
keeping the exponents close to unity. The QH index was not used on
all the \emph{H-}band data. This was because the spectral coverage
on the right hand side of the triangular peak, seen close to 1.68
$\umu$m, was not always available. It is also important to note that
for most of the AAT spectra that had sufficient wavelength coverage,
the noise increased heavily as the wavelength approached the edge of
the \emph{H-}band atmospheric transmission window; thus rendering
these wavelength regions unreliable for spectral typing. The WH and
WK indices measure the strength of the water absorbtion on the
shorter side of the flux peak in each waveband. Due to the fact that
these indices are not reddening independent, any inaccuracies in the
derived A$_{V}$ values for an object will result in  a small error
in the determined index. However, these indices are still useful and
provide robust spectral types for each object. This is particularly
true for the WK index as it is less affected by possible veiling due
to hot circumstellar dust seen beyond 2.2 $\umu$m. We find that
altering A$_{V}$ values by $\pm$2 changes the derived spectral type
by less than 0.5 in the majority of cases. The spectral indices
described above are illustrated in Figure \ref{indices}. The fit
parameters are summarised in Table \ref{fitparamtable}. Meaningful
errors could not be assigned to the each cubic polynomial
coefficient and constant term because the errors in Luhman's data
are unquantifiable.

\begin{table*}
\caption{Parameters for cubic polynomial fits to the indices
associated with the IRIS2 and NIRI data from this work, the UKIRT
data from Lucas et al. 2001 and the GNIRS data in paper 1. The
parameters for the WH$_{(G)}$ index are not listed as they are
identical to those of the WH$_{(UK)}$. The precision of each
coefficient is supplied to allow accurate spectral typing. However,
using the 3rd and 2nd order terms to this accuracy only should be
adequate. The R$^{2}$ values represent the correlation coefficient
(see Section \ref{specType} for details.).} \label{fitparamtable}
\begin{minipage}{13cm}
 \centering
  \begin{tabular}{@{}lccccc@{}}
  \hline
  \hline
   Index Name & $\alpha$$x^{3}$ & $\beta$$x^{2}$ & $\gamma$$x$ & Const. & R$^{2}$\\
 \hline
 WH & $-$0.0004533128 & 0.0118652078 & $-$0.1265391170 & 1.3595251240 & 0.9240\\
 WK & $-$0.0017114928 & 0.0362490654 & $-$0.2866080191 & 1.7754365826 & 0.9577\\
 QH & $-$0.0015340095 & 0.0326200036 & $-$0.2526304540 & 1.3626402514 & 0.9243\\
 QK & $-$0.0013174804 & 0.0239071298 & $-$0.1633687914 & 1.2085945875 & 0.9161\\
\hline
 WH$_{(UK)}$ & $-$0.0007842812 & 0.0177687976 & $-$0.1770229758 & 1.5162233490 & 0.9453\\
 WK$_{(UK)}$ & $-$0.0012965703 & 0.0279422070 & $-$0.2225218246 & 1.5852276527 & 0.9449\\
 QH$_{(UK)}$ & $-$0.0013136355 & 0.0309180885 & $-$0.2698736864 & 1.3226247704 & 0.9276\\
 QK$_{(UK)}$ & $-$0.0012109998 & 0.0229457232 & $-$0.1598483183 & 1.2099851645 & 0.8974\\
\hline
 WK$_{(G)}$ & $-$0.0019817649 & 0.0442701944 & $-$0.3617322968 & 1.9892775060 & 0.9520\\
 QH$_{(G)}$ & $-$0.0011389261 & 0.0260481258 & $-$0.2218064454 & 1.2761433125 & 0.9137\\
 QK$_{(G)}$ & $-$0.0009214824 & 0.0159583585 & $-$0.1151253304 & 1.1080547649 & 0.9261\\
\hline
\end{tabular}
\end{minipage}
\end{table*}

A further set of spectral indices were created to re-analyse an
existing \emph{H-} and \emph{K-}band data set of young ONC objects.
The data were obtained at UKIRT in November 1999, and previously
published in Lucas et al. (2001). At the time of publication few
optically calibrated young brown dwarf spectra existed. As a result
of this, spectral types were calibrated from indices applied to
evolved field dwarfs, consequently leading to over-estimated
spectral types.

Both W and Q indices were constructed for the \emph{H-} and
\emph{K-}band UKIRT data. Due to the fact that the UKIRT
\emph{H-}band data had greater wavelength coverage to the IRIS2
data, the preferential WH index described in paper 1 (Lucas et al.
2006) was selected for analysis. Minor adjustments were required for
the WK, QH and QK indices described in paper 1, and the WK and QK
indices used with the IRIS2 and NIRI data in this paper. Each new
index was created based on the criteria described earlier in this
section. The indices that were used for spectral typing the UKIRT
data can be seen in Equations \ref{WHUK} to \ref{QKUK}, and are
represented graphically in Figure \ref{UKindices} in Appendix
\ref{UKPlotappendix}. For clarity, the names of the indices
associated with the UKIRT data have been appended with the
subscripted letters `UK' in parenthesis. The fit parameters for of
the UKIRT indices are summarised in Table \ref{fitparamtable}. The
QH and QH$_{(UK)}$ indices are reasonably insensitive beyond M 7.5
due to the cubic polynomial fits being fairly flat beyond this
point.

\begin{equation}
\label{WHUK} WH_{(UK)}=\frac{F_{\lambda}(1.525\umu
m)}{F_{\lambda}(1.675\umu m)}
\end{equation}
\begin{equation}
\label{WKUK}WK_{(UK)}=\frac{F_{\lambda}(2.080\umu
m)}{F_{\lambda}(2.190\umu m)}
\end{equation}
\begin{equation}
\label{QHUK} QH_{(UK)}=\frac{F_{\lambda}(1.530\umu
m)}{F_{\lambda}(1.670\umu m)}\left[\frac{F_{\lambda}(1.780\umu
m)}{F_{\lambda}(1.670\umu m)}\right]^{1.551}
\end{equation}
\begin{equation}
\label{QKUK} QK_{(UK)}=\frac{F_{\lambda}(2.080\umu
m)}{F_{\lambda}(2.192\umu m)}\left[\frac{F_{\lambda}(2.345\umu
m)}{F_{\lambda}(2.192\umu m)}\right]^{0.830}
\end{equation}

The GNIRS indices (except WH$_{G}$) from paper 1 are represented in
Figure \ref{Gindices} in Appendix \ref{GPlotappendix}. The fit
parameters associated with these objects are presented in Table
\ref{fitparamtable}. These data were not included in the original
publication.

Indices that were applicable to the spectral data were used with the
solutions to their corresponding cubic polynomial fits, to determine
a spectral type. Indices were run on the cubic spline fits to take
into account the poor signal to noise inherent to most of the data.
The final spectral typing results were derived using the solutions
from all applicable indices in most cases. If a pseudo continuum fit
did not represent the data well in a wavelength region required for
a particular index, the result was discarded. When data quality was
clearly superior at one wavelength, more weight was given to the
spectral types derived from indices using the same passband.
Spectral types were rounded to the nearest 0.5 sub-type.

All results from the spectral typing procedure were checked by
over-plotting template spectra from Luhman's sample. Several objects
observed at \emph{H-} and \emph{K-}band had spectra that appeared to
be later in one of the observed wavebands (see Section
\ref{specTypeResults} for details). Over-plotting template spectra
confirmed these measurements, demonstrating that the discrepancy was
not due to a weakness in spectral index. In the cases where the
difference in spectral type was not obviously due to poor data
quality, the final type was derived using an average.

011-027 has poor spectral coverage at the blue end of the
\emph{H-}band spectrum because of its location on the array. Due to
the fact that there is no spectral information at 1.562 $\umu$m we
could not use our indices to type this object. Instead we
over-plotted a selection of template spectra from Luhman's sample
and calibrated the object by visual inspection. 011-027 is not
plotted in Figure \ref{hspectraa} as it has a Rayleigh-Jeans type
continuum. The spectrum of 031-536 has adequate spectral coverage
but was not spectral typed using our indices. This is because the
spectrum rises, becoming more noisy in the wavelength region around
1.562 $\umu$m that is required for spectral typing. The apparent
rise in flux is due to increasing noise towards the edge of array.
The \emph{H-}band profile beyond 1.60 $\umu$m is triangular in
shape, showing strong water absorption either side of the peak seen
close to 1.68 $\umu$m. The pseudo continuum fit follows the noise in
this region and therefore results in a generated spectral type that
is too early if indices are used. To overcome this problem we have
visually typed this object by over plotting a selection of template
spectra. Due to the fact the spectrum of 031-536 is noisy and
unreliable at $\lambda$ $>$ 1.60 $\umu$m we assign a large error to
the spectral type.

\subsection{Spectral Typing Results}
\label{specTypeResults} Table \ref{types} shows the final spectral
types that were calculated for each object and summarises the
indices used to determine these values. The GNIRS data from Lucas et
al. 2006 has also been included in the Table \ref{types}. 12 of the
sources had H$_{2}$O absorption that was stronger than that of any
object in Luhman's sample. These objects were assigned spectral
types `$>$ M9.5' because we cannot trust that the cubic polynomial
fits are valid for objects later than this. These 12 sources,
together with 5 objects that have derived spectral types of M9.5 and
$\geq$ M9.5 are shown below the dotted line in Table \ref{types}.
Five sources have a steep Rayleigh-Jeans type continuum and appear
hotter than the warmest object (M 3.0) in Luhman's sample of
template spectra. A further 7 sources had spectral profiles
representative of objects ranging from M3.0 to M4.5 in the template
sample. All 12 early M-type sources are likely to be foreground
stars or reddened background stars. We confirmed this by plotting
the sources on an H-R diagram combined with pre-main-sequence
isochrones (see section \ref{HRsection} for details). All sources
lie below the 50 Myr isochrone, demonstrating that they do not
belong to the cluster. Due to the fact that these objects are not
$\sim$1Myr old cluster members, the derived spectral types are
invalid. In addition to this, the A$_{\emph{v}}$ values derived from
the colour-magnitude diagram will be incorrect. Non cluster members
are separated in Table \ref{types} by a solid line. We loosely
classify these objects as `early M'.

For the highest quality data the results generated from the
different indices agreed well, typically to within 1 spectral type.
However, in a few cases results differed by up to 2 spectral types.
The significant anomalies in determined spectral type (i.e.
discrepancies $>$1.5) have been carefully scrutinised. The anomalies
are discussed in Appendix A.


The spectral typing method that we have used makes it hard to
determine a robust error for each of the indices. This is because we
cannot accurately determine the error in the cubic spline fit at the
wavelength intervals used for each index. In addition, we cannot
quantify the error in the spectral types assigned to the template
spectra. A visual inspection indicated the cubic spline fits
represented the data well at wavelengths critical for spectral
typing (a spectral index was avoided if this was not the case). The
error in spectral type was subsequently derived using the scatter in
the results. This was then increased to account for systematic
errors. The standard error of the mean (standard deviation $/$
$\sqrt{n}$) was used to determine the error in spectral type. This
value was then increased by half a sub-type to take into account
further errors. The errors in Table \ref{types} have been rounded to
the nearest half sub-type. For sources with spectral types based on
a single measurement the error was estimated based on the signal to
noise of the data. Errors have been neglected for objects that
appear later than M9.5 and for reddened background stars.

The effective temperatures in Table \ref{types} were determined
using the revised temperature scale of Luhman et al. (2003). The
temperature scale was constructed for young sources ($\sim$ 2 Myr)
in the IC 348 star forming region. It is therefore a suitable
approximation for the youthful objects in the ONC. The errors in
effective temperature are somewhat unquantifiable due to the absence
of uncertainties in the spectral type to effective temperature
scale. We have assigned errors to the derived effective temperatures
based on the error in spectral type. Because the temperature scale
is not defined beyond M9, upper limits have been assigned to objects
later than M9. Because the Luhman temperature scale is not defined
beyond M9, upper limits have not been assigned to objects $>$M9. We
have not attempted to derive effective temperatures for sources we
believe to be background stars. This is because our spectral typing
scheme is calibrated from very young objects and is not designed for
evolved main sequence stars.

\begin{table*}
\centering
 \begin{minipage}{17cm}
 \centering
 \caption{Final spectral types for all sources with sufficient signal
to noise to be spectral typed (including GNIRS and NIRI data from
paper 1). Spectral types derived from UKIRT and GNIRS indices are
subscripted by (UK) and (G), respectively. 17 sources have water
absorption that is stronger than or equal to that of the coolest
object in Luhman's template data set. These appear below the dotted
line in the table. 13 sources have spectra that show a smoothly
declining Rayleigh-Jeans tail. These are probably reddened
background stars and are shown below the solid line in the bottom
section of the table. Source 047-436 is a proplyd and has therefore
been excluded from this table.}
 \label{types}
  \begin{tabular}{@{}ccclllllll@{}}
  \hline
  \hline
   Source$^{a}$ & Data Set & H$_{dr}$ & WH$^{b}$ & WK$^{b}$ & QH$^{b}$ & QK$^{b}$ & Final Type & T$_{eff}$(K)$^{c}$ & $M/M_{\odot}$$^{d}$\\
 \hline
   010-109 & NIRI & 15.55 & ----- & 7.7 & ----- & 8.7 & M8.0$\pm$0.75 & 2710$^{+127}_{-232}$ & 0.030$^{+0.020}_{-0.017}$\\
   013-306 & NIRI \& UKIRT & 14.37 & 9.1$_{(UK)}$ & 9.0 & 7.7$_{(UK)}$ & 9.9 & M9.0$\pm$0.75 & 2400$^{+232}$ & 0.014$^{+0.010}$\\
   014-413 & UKIRT & 14.16 & 8.1$_{(UK)}$ & 8.4$_{(UK)}$ & 5.9$_{(UK)}$ & 6.1$_{(UK)}$$^{e}$ & M7.5$\pm$0.75 & 2795$^{+112}_{-317}$ & 0.040$^{+0.025}_{-0.018}$\\
   016-410 & UKIRT & 12.23 & 4.2$_{(UK)}$ & ----- & 5.8$_{(UK)}$ & ----- & M5.0$\pm$1.00 & 3125$^{+145}_{-135}$ & 0.150$^{+0.025}_{-0.050}$\\
   016-430$^{g}$ & GNIRS & 18.64 & 11.5 & 10.7$_{(G)}$ & 9.3$_{(G)}$ & 6.4$_{(G)}$ & M9.0$\pm$2.00 & 2400$^{+480}$ & 0.014$^{+0.046}$\\
   019-354 & UKIRT & 13.81 & 5.7$_{(UK)}$ & 7.5$_{(UK)}$ & 5.2$_{(UK)}$ & 8.0$_{(UK)}$ & M6.5$\pm$1.00 & 2935$^{+122}_{-140}$ & 0.075$^{+0.045}_{-0.035}$\\
   030-524 & NIRI & 17.40 & ----- & 7.8 & ----- & 7.9 & M8.0$\pm$0.75 & 2710$^{+127}_{-233}$ & 0.040$^{+0.025}_{-0.025}$\\
   043-014 & UKIRT & 13.76 & 8.0$_{(UK)}$ & ----- & 7.6$_{(UK)}$ & ----- & M8.0$\pm$0.25 & 2710$^{+42}_{-77}$ & 0.030$^{+0.005}_{-0.005}$\\
   044-527 & NIRI & 16.88 & ----- & 8.6 & ----- & 8.5 & M8.5$\pm$0.50 & 2555$^{+155}_{-155}$ & 0.024$^{+0.004}_{-0.005}$\\
   053-503 & UKIRT & 13.57 & 7.5$_{(UK)}$ & 7.9$_{(UK)}$ & ----- & 6.6$_{(UK)}$ & M7.5$\pm$0.50 & 2795$^{+85}_{-85}$ & 0.040$^{+0.030}_{-0.020}$\\
   067-651 & IRIS2 & 14.46 & 6.1 & ----- & ----- & ----- & M6.0$\pm$0.75 & 2990$^{+101}_{-82}$ & 0.082$^{+0.030}_{-0.017}$\\
   068-019 & UKIRT & 13.73 & 5.7$_{(UK)}$ & 7.9$_{(UK)}$ & 5.7$_{(UK)}$ & 7.7$_{(UK)}$ & M7.0$\pm$0.75 & 2880$^{+82}_{-127}$ & 0.060$^{+0.030}_{-0.025}$\\
   069-209 & UKIRT & 13.67 & 5.7$_{(UK)}$ & 9.8$_{(UK)}$$^{e}$ & 5.8$_{(UK)}$ & 7.6$_{(UK)}$$^{e}$ & M6.0$\pm$0.50 & 2990$^{+67.5}_{-55}$ & 0.090$^{+0.022}_{-0.025}$\\
   077-453 & IRIS2 & 12.76 & 6.2 & ----- & ----- & ----- & M6.0$\pm$0.75 & 2990$^{+101}_{-82}$ &0.082$^{+0.030}_{-0.030}$\\
   084-1939 & UKIRT & 14.78 & 9.8$_{(UK)}$ & 9.3$_{(UK)}$ & 8.8$_{(UK)}$ & 8.9$_{(UK)}$ & M9.0$\pm$0.25 & 2400$^{+77}$ & 0.014$^{+0.002}$\\
   091-017$^{f}$ & UKIRT & 13.68 & 4.3$_{(UK)}$ & 4.5$_{(UK)}$ & 4.5$_{(UK)}$ & 3.4$_{(UK)}$ & M4.5$\pm$0.50 & 3198$^{+72}_{-72}$ & 0.157$^{+0.060}_{-0.032}$\\
   092-532$^{g}$ & GNIRS & 17.50 & 9.8 & 7.5$_{(G)}$ & 5.0$_{(G)}$ & 3.9$_{(G)}$ & M7.5$\pm$2.00 & 2795$^{+263}$ & 0.040$^{+0.028}$\\
   095-058 & NIRI \& UKIRT & 13.25 & 7.4$_{(UK)}$ & 9.6 & 6.5$_{(UK)}$ & 6.9$^{e}$ & M7.5$\pm$0.75 & 2795$^{+112}_{-162}$ & 0.040$^{+0.025}_{-0.015}$\\
   096-1943 & UKIRT & 14.75 & 8.5$_{(UK)}$ & 9.4$_{(UK)}$ & ----- & ----- & M9.0$\pm$0.50 & 2400$^{+155}$ & 0.014$^{+0.006}$\\
   107-453$^{g}$ & GNIRS & 18.66 & 8.6 & 4.7$_{(G)}$ & 9.6$_{(G)}$ & 5.6$_{(G)}$ & M8.0$\pm$2.00 & 2710$^{+280}$ & 0.030$^{+0.052}$\\
   121-434 & IRIS2 & 12.66 & 6.9 & ----- & 7.0 & ----- & M7.0$\pm$0.25 & 2880$^{+27}_{-42}$ & 0.060$^{+0.010}_{-0.010}$\\
   186-631 & NIRI & 15.07 & 5.9 & 8.7 & 5.8 & 8.2 & M7.0$\pm$1.00 & 2880$^{+110}_{-170}$ & 0.058$^{+0.030}_{-0.030}$\\
   255-512 & IRIS2 & 15.31 & 6.3 & ----- & 6.8 & ----- & M6.5$\pm$0.50 & 2935$^{+55}_{-55}$ & 0.065$^{+0.015}_{-0.023}$\\
   \multicolumn{9}{c}{\dotfill} \\
   015-319 & NIRI \& UKIRT & 16.78 & 11.5$_{(UK)}$ & 10.2 & 7.8$_{(UK)}$ & 10.9 & $> $M9.5 & $<$2400 & $<$0.014\\
   024-124 & IRIS2 & 12.70 & 10.7 & ----- & ----- & ----- & $> $M9.5$^{f}$ & $<$2400 & $<$0.014\\
   031-536 & IRIS2 & 16.23 & ----- & ----- & ----- & ----- & M9.5$\pm$1.50$^{f}$ & $<$2400 & $<$0.014\\
   037-628 & UKIRT & 17.18 & 9.4$_{(UK)}$ & ----- & 10.4$_{(UK)}$ & ----- & $> $M9.5 & $<$2400 & $<$0.014\\
   055-230 & UKIRT & 13.11 & 9.4$_{(UK)}$ & 10.1$_{(UK)}$ & 9.1$_{(UK)}$ & 10.0$_{(UK)}$ & M9.5$\pm$0.5 & $<$2400 & $<$0.014\\
   056-141 & IRIS2 & 15.54 & 12.5 & ----- & ----- & ----- & $>$M9.5 & $<$2400 & $<$0.014\\
   057-247$^{g}$ & GNIRS & 18.49 & 8.8 & 9.3$_{(G)}$ & 10.1$_{(G)}$ & 10.0$_{(G)}$ & $\geq $M9.5$^{f}$ & $<$2400 & $<$0.014\\
   061-400 & UKIRT & 17.71 & 10.5$_{(UK)}$ & ----- & 10.0$_{(UK)}$ & ----- & $> $M9.5 & $<$2400 & $<$0.014\\
   084-104 & NIRI \& UKIRT & 16.88 & 11.1$_{(UK)}$ & 10.6 & 10.6$_{(UK)}$ & 10.1 & $> $M9.5& $<$2400 & $<$0.014\\
   087-024 & IRIS2 & 14.74 & 9.7 & ----- & ----- & ----- & M9.5$\pm$0.75 & $<$2400 & $<$0.014\\
   130-053 & IRIS2 & 15.12 & 11.2 & ----- & ----- & ----- & $>$M9.5 & $<$2400 & $<$0.014\\
   137-532$^{g}$ & GNIRS & 17.14 & 10.9 & 9.2$_{(G)}$ & 10.0$_{(G)}$ & 9.2$_{(G)}$ & $>$M9.5 & $<$2400 & $<$0.014\\
   152-717$^{g}$ & GNIRS & 17.74 & 9.3 & 10.3$_{(G)}$ & 10.2$_{(G)}$ & 9.2$_{(G)}$ & $>$M9.5 & $<$2400 & $<$0.014\\
   183-729$^{g}$ & GNIRS & 17.15 & 7.6 & 10.6$_{(G)}$ & 10.0$_{(G)}$ & 10.2$_{(G)}$ & $\geq $M9.5$^{f}$ & $<$2400 & $<$0.014\\
   188-658$^{g}$ & GNIRS & 18.60 & 9.7 & 9.5$_{(G)}$ & 10.7$_{(G)}$ & 10.5$_{(G)}$ & $>$M9.5 & $<$2400 & $<$0.014\\
   192-723 & NIRI & 15.27 & 9.3 & 9.4$_{(G)}$ & ----- & 10.7$_{(G)}$ & $>$M9.5 & $<$2400 & $<$0.014\\
   196-659 & NIRI & 16.44 & ----- & 10.2$_{(G)}$ & ----- & 10.2$_{(G)}$ & $>$M9.5 & $<$2400 & $<$0.014\\
   \hline
   011-027 & IRIS2 & 14.95 & ----- & ----- & ----- & ----- & Early M$^{f}$ & ----- & -----\\
   019-108 & UKIRT & 13.56 & ----- & ----- & ----- & ----- & Early M & ----- & -----\\
   020-1946 & IRIS2 \& NIRI & 14.39 & ----- & ----- & ----- & ----- & Early M & ----- & -----\\
   022-115 & NIRI & 18.07 & ----- & ----- & ----- & ----- & Early M & ----- & ----- \\
   023-1939$^{g}$ & NIRI & 17.46 & ----- & ----- & ----- & ----- & Early M & ----- & -----\\
   047-245 & IRIS2 & 16.41 & ----- & ----- & ----- & ----- & Early M & ----- & -----\\
   053-323 & IRIS2 & 15.39 & ----- & ----- & ----- & ----- & Early M & ----- & -----\\
   057-305 & IRIS2 & 13.75 & ----- & ----- & ----- & ----- & Early M & ----- & -----\\
   127-044 & IRIS2 & 13.19 & ----- & ----- & ----- & ----- & Early M & ----- & -----\\
   199-617$^{g}$ & NIRI & 18.04 & ----- & ----- & ----- & ----- & Early M & ----- & -----\\
   205-610$^{g}$ & NIRI & 17.62 & ----- & ----- & ----- & ----- & Early M & ----- & -----\\
   235-454 & IRIS2 & 13.11 & ----- & ----- & ----- & ----- & Early M & ----- & -----\\
 \hline
\end{tabular}
\end{minipage}
\end{table*}

\begin{table*}
\begin{minipage}{17cm}
\begin{tabular}{@{}cclllllll@{}}
\end{tabular}

Notes:\\
 $^{a}$  Source names are coordinate
based, following O'Dell \& Wong (1996).\\
 $^{b}$  Spectral types that were derived using the UKIRT
indices have been appended by the subscripted letters `UK' in
parenthesis.\\
 $^{c}$  These quantities were derived using spectral type to effective temperature scale of Luhman (2003) \\
 $^{d}$  Masses were determined from an H-R diagram (see Section
\ref{massAndAgeOfCluster} for details).\\
 $^{e}$  This result was not used to derive the final spectral
type.\\
 $^{f}$  Spectral type was determined by over-plotting template
spectra.\\
 $^{g}$  Source from paper 1 (Lucas et al. 2006)\\
\end{minipage}
\end{table*}

\section{H-R Diagram}
\label{HRsection} 
\begin{figure*}
\centering{\ } 
 \vbox{
  \hbox{
    \psfig{file=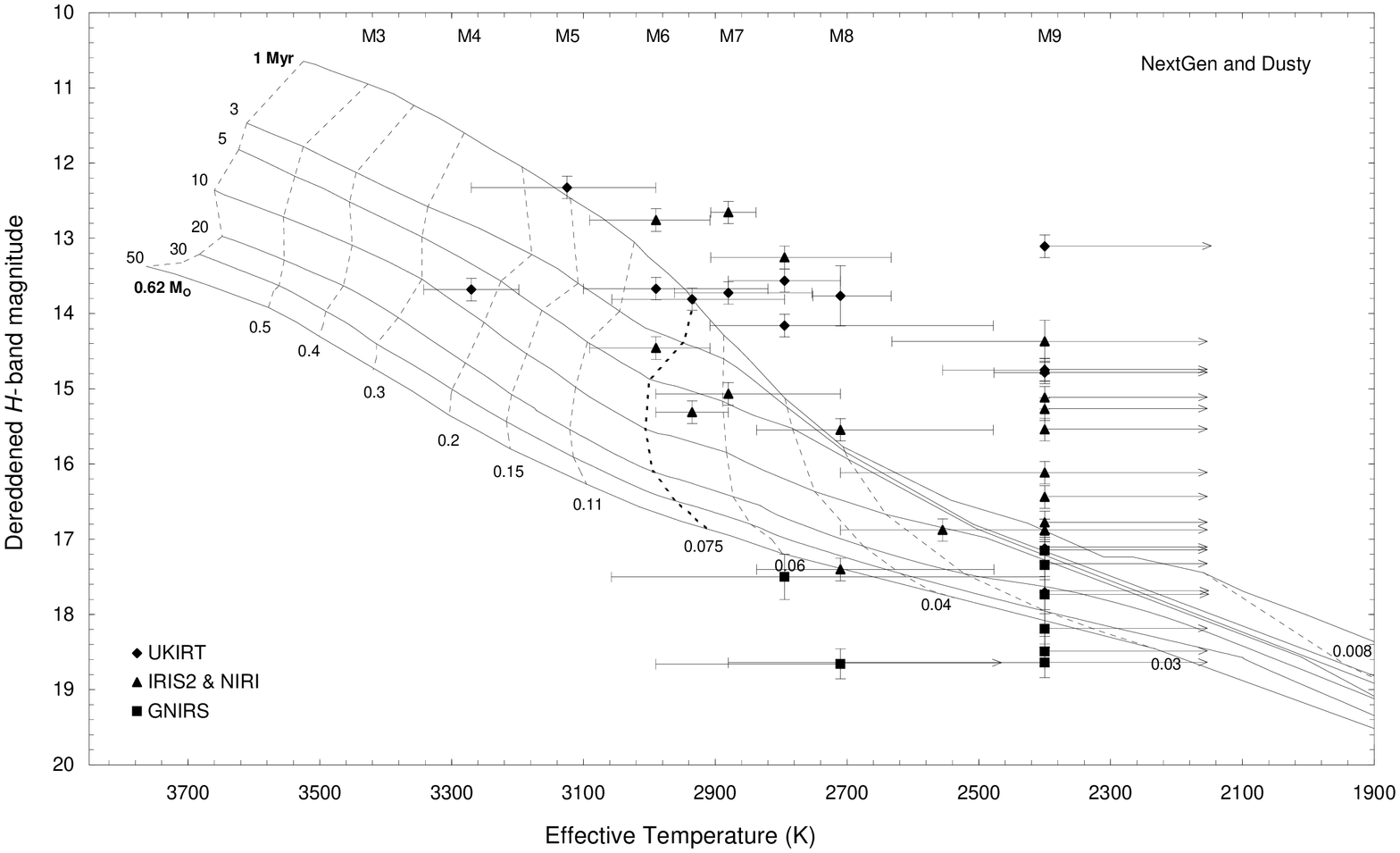,width=14cm,angle=-90}}
  \hbox{
    \psfig{file=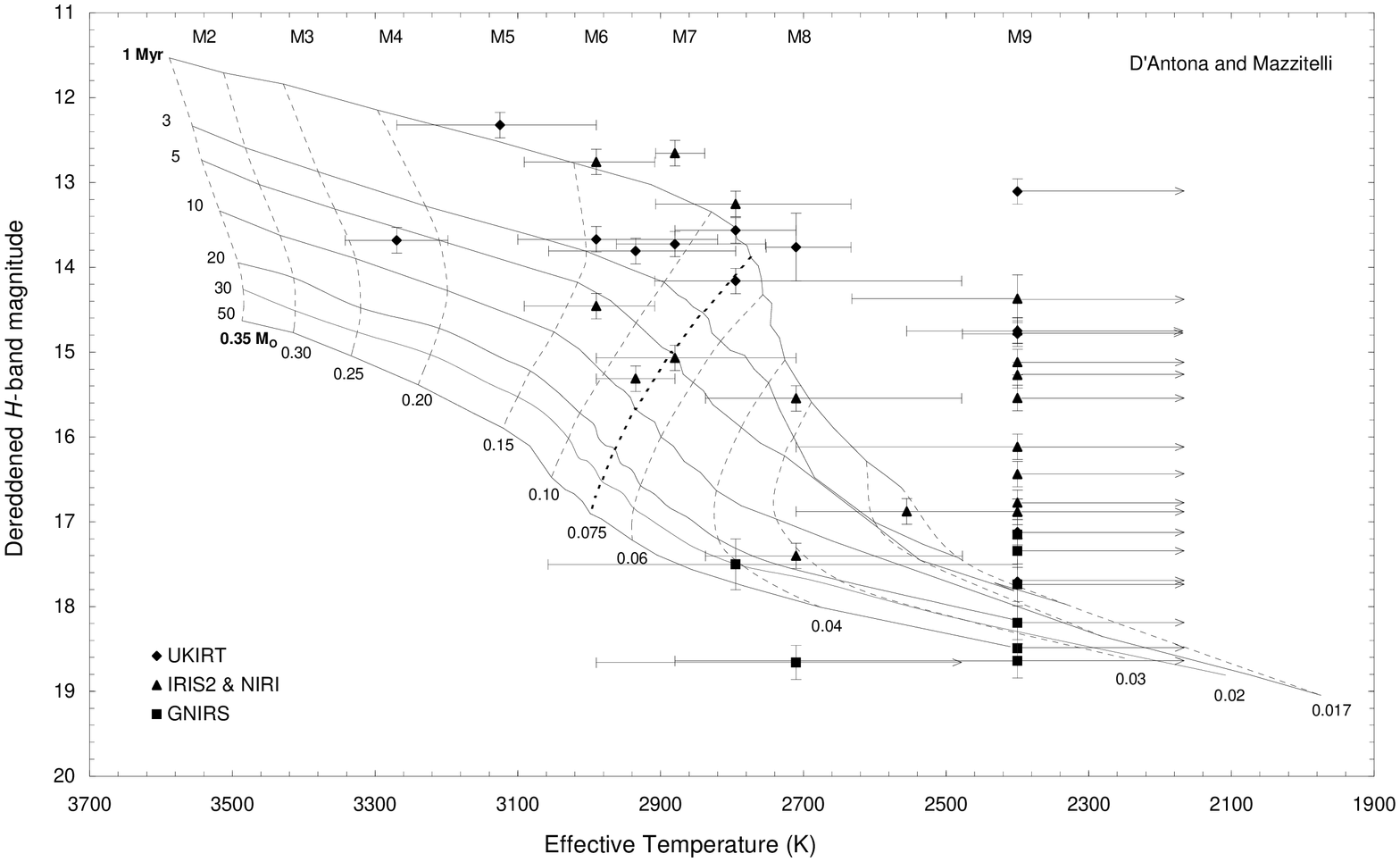,width=14cm,angle=-90}}}
\caption{H-R Diagrams for the ONC are shown with the combined
NextGen and Dusty isochrones of Baraffe et al. 1998; Chabrier et al.
2000 and Allard et al. 2001 (top), and the pre-main-sequence models
of DM97 (bottom). Arrows represent upper limits for effective
temperature.} \label{HRDiagrams}
\end{figure*}

In Figure \ref{HRDiagrams} we present H-R diagrams
for the full sample of sources with infrared spectra from this work
and from Lucas et al.(2001) and Lucas et al.(2006). Classically, an
H-R diagram is plotted with bolometric luminosity on the y-axis. We
have plotted dereddened \emph{H-}band magnitude rather than
bolometric luminosity in order to avoid the uncertainties associated
with converting to a bolometric magnitude. The upper panel of Figure
\ref{HRDiagrams} has overlaid theoretical models of the Lyon group,
while the lower panel has overlaid models from DM97. The plots in
the upper panel use different Lyon group models in different
temperature ranges. The DUSTY models (Allard et al.2001) have been
used at $T_{eff}$$<$~2500~K (the temperature at which dust begins to
become important) and the NEXTGen models (Baraffe et al.1998;
Chabrier et al.2000) have been used at $T_{eff}$$>2500$~K. For the
coolest objects (T$_{eff}$$<$~1800K) we use the magnitudes predicted
by the DUSTY models but adopt the colours of field dwarfs, as the
DUSTY models are known to be too red at such low temperatures
(Chabrier et al. 2000). The highest mass sources in the 1 and 3 Myr
dusty isochrones required a 0.15 mag shift in the \emph{H-}band to
ensure a seamless join to the NextGen isochrones.

The fluxes in the NextGen and Dusty models are presented in the CIT
format. They were converted to the MKO system using appropriate
filter transformation
equations\footnote{http://www.astro.caltech.edu/\~{}jmc/2mass/v3/transformations/}.
The absolute magnitudes of each object were converted to apparent
magnitudes using the distance modulus equation with an assumed
distance to Orion of 450 pc. The DM97 model data provided bolometric
luminosity only. To account for this we derived our own bolometric
correction at \emph{H-}band based on the models by the Lyon Group.
Two approaches were experimented with, each giving results that were
similar. The first approach generated bolometric corrections by
comparing relative masses between the models. For objects of the
same mass, \emph{H-}band magnitudes in the Lyon models were
subtracted from the bolometric luminosities presented in the DM97
models to give a bolometric correction. Where similar masses were
not available linear interpolation was used to derive a correction.
The second corrections were calculated from temperature comparisons
using a similar technique. Although the final discrepancies in the
results from the two procedures were small, the method based on
temperature comparisons was used. This is because bolometric
corrections directly relate to temperature and are therefore a more
sensible choice.

In Figure \ref{HRDiagrams} the UKIRT data from Lucas et al. 2001
appear as diamonds, the IRIS2 and NIRI data are shown as triangles
and the GNIRS data from Lucas et al. 2006 are displayed as squares.
Sources that had derived spectral types later than M9.0 are plotted
as as upper limits. Objects whose spectra are represented by obvious
Rayleigh-Jeans tails are not likely to be cluster members and have
been excluded from the plots. The errors in dereddened \emph{H-}band
magnitude are dominated by photometric uncertainties. The derivation
of errors in effective temperature is described in Section
\ref{specTypeResults}. In general the error in effective temperature
increases as spectral type becomes later. The IRIS2 object 013-306,
GNIRS object 107-453 and UKIRT objects 084-1939 \& 096-1943 have
constrained spectral types within the optically classified range of
the Luhman sample. Due to the magnitude of the errors for these
sources and an absence in temperature data beyond M9.0, error bars
could only be included for higher effective temperatures. Upper
limits were used for lower effective temperatures. In each plot the
mass track of the hydrogen burning limit (0.075M$_{\odot}$) has been
highlighted as a bold dotted line.

\subsection{Analysis}\label{analysisSection}
For objects with effective temperatures $>$ 2500K the 1Myr isochrone
of the DM97 model appears to trace the observed data quite well.
However, Luhman et al. (2003) used empirical isochrones to show that
the pre main sequence NextGen and Ames Dusty models agree well with
observational constraints for low mass sources. The DM97 models have
less detailed physics and assume gray atmospheres which, are thought
to be an inappropriate approximation for stars whose effective
temperatures fall below 4500K (Baraffe et al. 1998). Thus, the
NextGen and Ames Dusty models are regarded as more reliable and
robust than those of DM97. Both H-R Diagrams in Figure
\ref{HRDiagrams} show a considerable number of sources that lie
above the 1 Myr isochrone. This is more noticeable with the NextGen
and Dusty models. Although the error bars can account for the
scatter of most objects above the youngest isochrones, both models
suggest a population of objects that are younger than 1 Myr. This is
a particularly interesting result as it is in agreement with results
by Slesnick, Hillenbrand \& Carpenter (2004, hereafter SHC04) and
RRL. Some of these objects may not necessarily be younger than 1 Myr
in age and may in fact represent unresolved binary systems. If the
binary systems comprise objects of similar luminosity, the
components in question will be $\sim$0.75 magnitudes fainter, moving
them significantly closer to the 1 Myr isochrones. The true binary
fraction in Orion is unknown. Recent studies (e.g. Reipurth et al.
2007) suggest the overall binary fraction is 8.8$\%$ $\pm$1.1$\%$, a
factor of $\sim$1.5 less than seen in the field. The binary fraction
of the low mass population in the central part of the ONC is
comparable to that seen in the field (e.g. Padgett, Strom \& Ghez
1997; Simon, Close \& Beck 1999; K\"{o}ler et al. 2006). Ongoing
research is helping to refine this measurement (e.g Irwin et al.
2007; Cargile, Stassun \& Mathieu 2008). Binarity may therefore
explain some of the observed scatter.

Interestingly, several objects lie below the 5 Myr isochrone and
close to the 10 Myr isochrone in both diagrams suggesting they may
be associated with an earlier burst of star-formation. Evidence of a
broad and possibly bimodal age distribution was observed by SHC04.
However, RRL do not see a dichotomy in the H-R diagram. It is
uncertain whether the different positions of individual sources in
the H-R diagram are really due to different ages. As discussed by
Siess et al. (1999) and Tout et al. (1999), sources on the Hayashi
track, still contracting and$/$or accreting, are subject to scatter
on the H-R diagram. In the Hayashi region accretion generally leads
to bluer colours which results in an overestimate of the age. The
evolution of an accreting star is also accelerated which can cause
it to appear older in the H-R diagram. For this reason the H-R
diagrams for star-forming regions have significant scatter in
comparison to those based on stable main-sequence stars. Brice\~{n}o
et al. (2002) and SHC04 suggest objects with optically thick disks
observed virtually edge-on may be observed in scattered light. The
consequence of this is an underestimation in luminosity for a given
effective temperature on the H-R diagram, resulting in objects that
appear older than they really are. Due to the fact that a very small
number of our objects lie close to the 10 Myr isochrone, this
scenario may be a reasonable interpretation. The detection of
objects in scattered light may provide an explanation for the
location of two GNIRS objects (107-453 and 092-532) with effective
temperatures $>$ 2500K that appear to have ages $\geqslant$30 Myr in
each diagram, and NIRI source (030-524) that appears $\sim$50 Myr in
age in the diagram that uses the NextGen and Dusty isochrones.
Although the error bars are large for these objects, visual
comparisons with the template spectra in the Luhman sample suggest
that the spectral types have been constrained well. Under-estimation
of source flux due to scattering is therefore another possible
explanation for the abnormal location of these objects on the H-R
diagram.

The H-R diagram is most interesting below 2500K. 12 sources are
located a considerable distance above the 1~Myr isochrone in both
plots. The reduction and spectral typing of these sources was
thoroughly checked and photometric measurements taken at two
different epochs (Lucas et al. 2000 using UFTI on UKIRT and Lucas et
al. 2005 using Flamingos on Gemini South) were compared for
differences. Significant changes in flux, attributed to source
variability, were found for four objects in our sample. These were
$<$0.3 magnitudes at \emph{H-}band for two of the sources. The
remaining two sources showed larger discrepancies of 0.5 and 1.16
magnitudes at \emph{H-}band. For the photometrically discrepant
objects \emph{H-}band photometric data was used that was obtained at
an epoch closest to the spectroscopic observation. The dereddened
\emph{H-}band error bars were increased in size for these objects to
account for photometric scatter. The effect of variability in young
stellar objects is most significant at \emph{K-}band (Kaas, 1999).
For the majority of our sample photometric uncertainties due to
variability appear to be small ($\sim$0.15 mags at \emph{H-}band)
and do not have a significant effect on our H-R diagrams.

The extreme scatter of some objects above the 1~Myr isochrones can
not be explained by a population of unresolved binary sources. Even
an unresolved system of three similar objects still only accounts
for $\sim$1.2 magnitudes and cannot explain the excess flux of up to
3 magnitudes. The presence of unresolved systems is clearly an
unsatisfactory interpretation of the unusual location of this group
of objects. There is potential for a discrepancy in the assumed
distance to Orion (450 pc). However, the likelihood of this being
much greater than 50 pc is small based on the scatter present in
many independent distance measurements (e.g. Walker (1969); Breger,
Gehrz \& Hackwell (1981); Stassun et al. (2004); Jeffries (2007)).
If we assume Orion is situated at a closer proximity of 400 pc, the
isochrones become marginally brighter, with a minor shift of
$\sim$0.26 magnitudes. This shift makes an insignificant difference
to the H-R Diagram. In order to bring these objects to the 1~Myr
isochrone a minimal shift of 3 magnitudes is required in some cases.
Such a large shift is unreasonable for several obvious reasons. Eg.
it would require that most of the sources are $\sim$20-50 Myr old
which would be inconsistent with the spectroscopic observations. It
is unlikely that these objects are nearby foreground sources because
the spectral profiles are indicative of young, low surface gravity
objects.

A possible interpretation of these discrepant sources is that they
represent an extremely young population of objects. The NextGen and
Dusty models do not extend to ages below 1~Myr so a visual
comparison to reliable model data is not possible. The DM97 models
generate data that extend below the age of 1~Myr and can be used to
give an indication of the physical properties of extremely young
objects. Isochrones younger than 1~Myr have not been plotted in
Figure \ref{HRDiagrams} due to the fact that there is no calibration
available from the NextGen and Dusty models. Levine et al.(2006)
plot an H-R diagram for NGC 2024 (Orion B molecular cloud). This
includes DM97 pre-main-sequence isochrones that extend to an age of
0.1 Myr. The data for the 0.1~Myr isochrone is limited to
$\sim$2500K. At 2500K the isochrone is nearly vertical and fainter
than the majority of the outlying objects in our sample. It
therefore fails to explain the outlying sources that lie above the
isochrone. The reliability of model isochrones tails off
significantly at ages less than 1 Myr. This is partly to do with the
fact that there is variability in the initial conditions. In
addition to this the DM97 models are relatively imprecise compared
to the NextGen and Dusty models. For these reasons we cannot rule
out that the location of these objects in the H-R diagram is due to
extreme youth. Furthermore the spectral types for these objects were
calibrated from template spectra of objects 1-2~Myr in age. Like
other authors (e.g. Levine et al. 2006) we have assumed that
extremely young objects will have spectral profiles very similar to
those a few Myr in age. However, if we are looking at a population
of objects $\ll$1~Myr in age with very low surface gravities, the
water absorption bands may be deeper than in objects with similar
temperatures that are several million years older. This will lead to
spectral types that are too late. With small age differences the
extent of the effect is not likely to be large.

A possible interpretation of these bright objects is that they are
spectrum variables. It is possible that the spectra of these objects
are varying on relatively short timescales due to active accretion
(see eg. Rodgers et al. 2002). Further near-infrared spectroscopic
and photometric measurements are required to determine whether the
spectra are varying over time.

A total of 17 sources from this sample have been observed
spectroscopically at optical wavelengths by RRL. 9 of which are
included in the UKIRT sample. The optical sample also includes 5
IRIS2 and 3 NIRI objects. 12 of the 17 sources that have been
observed spectroscopically in both data sets agree to within 1
spectral subtype. A further two objects agree to within 1.25
subtypes. The remaining three objects have differences of 1.50, 2.25
and 2.75 subtypes. The two largest discrepancies are for the UKIRT
objects 096-1943 and 055-230, respectively. This is particularly
interesting as both of these sources appear late at near-infrared
wavelengths (096-1943 $=$ M9.0$\pm$0.5; 055-230 $=$ M9.5$\pm$0.5).
096-1943 and 055-230 have optical spectral types of M6.75$\pm$0.5
and M6.75$\pm$1.5, respectively. The error bars for these two
objects cannot account for the large discrepancy observed. 096-1943
and 055-230 are both cool objects that are significantly offset from
the 1~Myr isochrone in the H-R diagram. The optical observations of
these objects were obtained at different epochs to the infrared
UKIRT data. This may be evidence for two spectrum variable objects.
The difference is unlikely to be due to data quality as the
brightness of these sources ensured good data quality in the optical
and infrared. The only other M9 to M9.5 object observed at optical
wavelengths is 031-536. For this object RRL found an optical
spectral type only 0.75 subtypes earlier than us. In addition to
this 031-536 is much closer to the 1~Myr isochrone.

In all but 3 cases, our spectral types are slightly later than those
derived at optical wavelengths. The observed offset is $\leqslant$ 1
spectral sub-type, typically in the order of 0.5-0.75 sub-types. If
the optical spectral types are more reliable then a significant
number of our objects will move closer to the 1 Myr isochrone in the
H-R diagram. However, the average cluster age will remain $<1$~Myr.
The H-R diagrams published by RRL \& Peterson et al. 2008 are in
very good agreement with ours, showing a distinct population of
objects above the 1~Myr isochrone. Below 2500K RRL and Peterson et
al. 2008 also see a population of bright sources that are
significantly offset from the 1~Myr isochrone, thus confirming our
results are not due to a systematic error.

\begin{figure}
\centering{\ } 
    \psfig{file=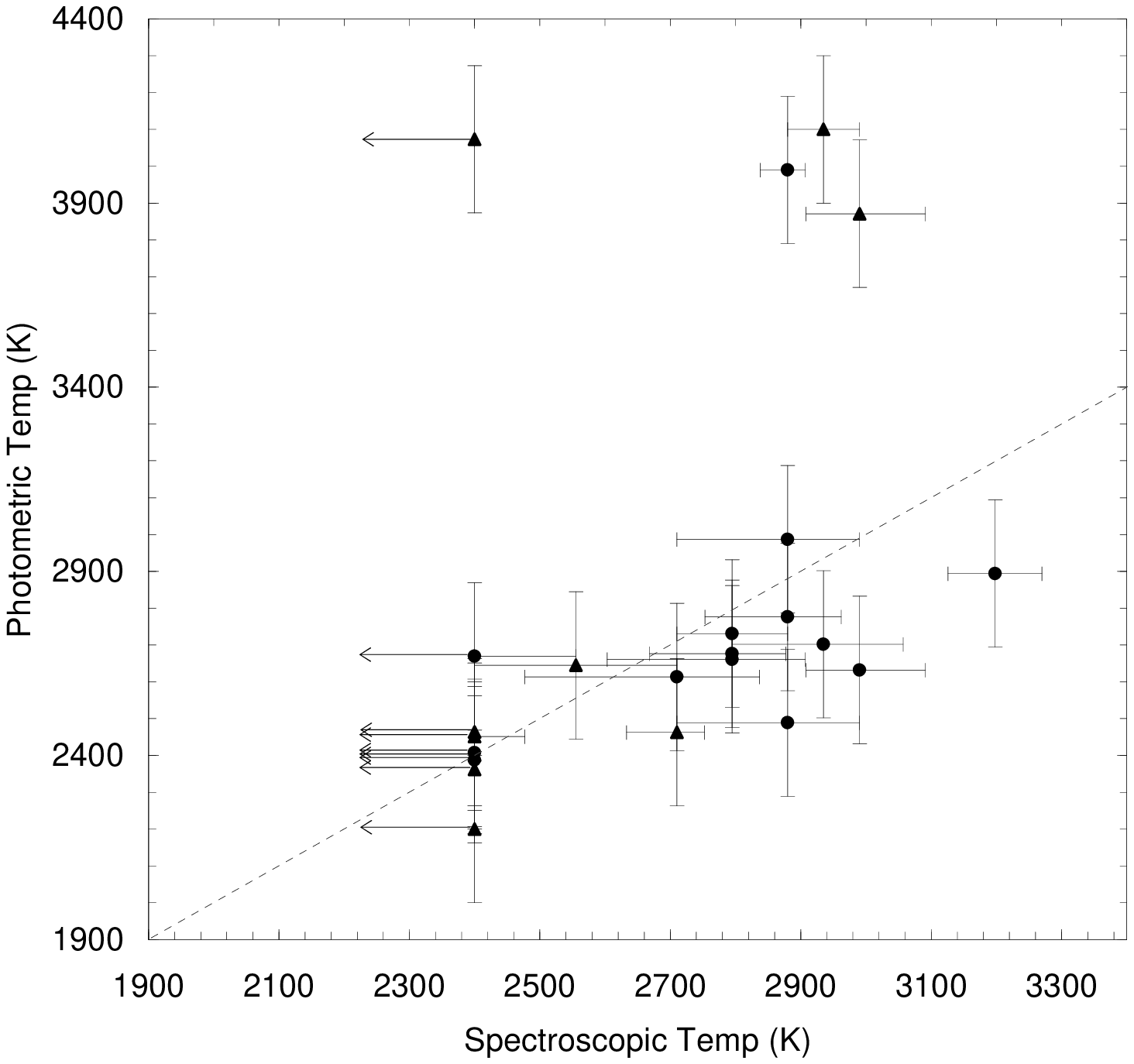,width=7.5cm,angle=-90}
\caption{Comparison of photometric and spectroscopic temperature.
Photometric temperatures are derived from dereddened (I-J) colours
by comparison with the NextGen 1~Myr isochrone. Spectroscopic
temperatures are from this work. Sources in our infrared sample
covered by RRL at optical wavelengths are depicted by circles.
Objects observed in our data set alone are displayed as triangles.
Cluster members with spectral types $>$M9.5 are plotted as upper
limits.} \label{SpecVSphot}
\end{figure}

We compare our spectroscopically determined temperatures to the
photometrically derived temperatures determined from dereddened
(\emph{I}-\emph{J}) colours and the 1~Myr NextGen isochrone (Figure
\ref{SpecVSphot}). Some objects in our sample are not detected at
\emph{I-}band because they are too faint. These have been excluded
from the plot. Sources covered at optical wavelengths by RRL are
plotted as circles, extra sources from our sample are plotted as
triangles. Sources with spectral types $>$ M9.5 are marked as upper
limits. The photometric and spectroscopic temperatures agree
reasonably well despite some scatter in the results. In general the
photometric measurements slightly underestimate effective
temperature, typically by $\sim$100~K. The trend is similar to that
seen by RRL. However, the spectroscopic temperatures derived at
near-infrared wavelengths are significantly closer compared to the
optical results of RRL. 4 sources in Figure \ref{SpecVSphot} have
photometric temperatures that are much hotter than measured from
spectroscopic data. 3 of the 4 outlying sources are from our data
alone and have anomalously blue (I-J) colours, as defined by Lucas
\& Roche (2000). This suggests that they have associated
circumstellar matter and modified colours due to scattered light.
The remaining source was also observed at optical wavelengths and
has a derived spectroscopic temperature that is consistent with
ours.

RRL found that optical sources with spectroscopic temperatures $>$
3200 K had photometrically derived temperatures that were much
hotter. We only have one cluster member in our sample hotter than
3200 K (091-017). This source was also observed at optical
wavelengths but calculated to be cooler than our spectrally derived
value. The spectroscopically derived temperatures are probably more
reliable than those derived from (I-J) photometry, in part because
the spectral types are derived from more than one index. The
\emph{I-}band fluxes are more likely to be influenced by scattered
light or even free-free emission from circumstellar matter (for
sources located in the ONC HII region).

\begin{figure}
\centering{\ }
    \psfig{file=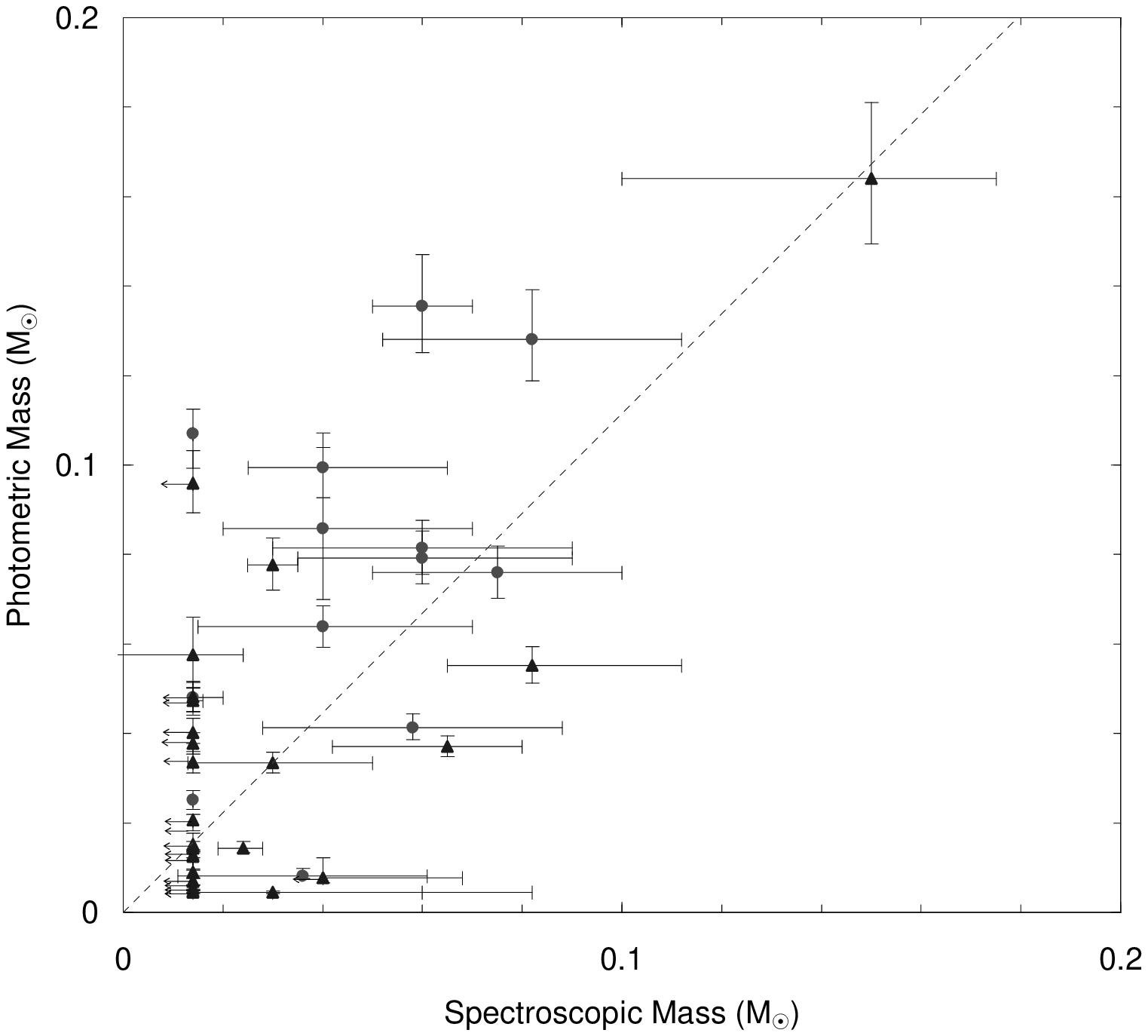,width=7.5cm,angle=-90}
\caption{Comparison of photometric and spectroscopic mass.
Photometric masses are calculated from dereddened \emph{H-}band
magnitude and the 1 Myr NextGen and Dusty isochrone. Sources in our
infrared sample covered by RRL at optical wavelengths are depicted
by circles. Objects observed in our data set alone are displayed as
triangles. Cluster members with spectral types $>$M9.5 are plotted
as upper limits.} \label{specMassVsPhotMass}
\end{figure}

In Figure \ref{specMassVsPhotMass} we plot photometric mass against
spectroscopic mass. Sources are identified by the same symbols used
in Figure \ref{SpecVSphot}. Photometric masses are determined using
the 1 Myr NextGen and Dusty isochrone together with the dereddened
\emph{H-}band magnitude for each source. Mass tracks are
approximately vertical between isochrones of different ages on an
H-R diagram (see Figure \ref{HRDiagrams}). This means that minor
errors in assumed age that may exist will have a negligible impact
on mass. Spectroscopic masses and their errors are taken directly
from this work. The agreement between photometric mass and
spectroscopic mass is reasonable if we exclude sources that are
plotted as upper limits. However, when we include every source the
overall agreement is poor. The results show significant scatter,
with the majority of the sources having photometric masses that are
higher than those derived from our spectra. The overall agreement
between photometric and spectroscopic mass is better at optical
wavelengths (see RRL), although significant scatter still exists.
One source from our sample that is also covered in the optical
sample of RRL (091-017) has a photometric mass that is significantly
lower than our derived spectroscopic mass. The water absorption in
the spectrum of this M4.5 source is weak in comparison to sources
that have masses close to the hydrogen burning limit, demonstrating
that photometric data alone was insufficient to accurately determine
the mass. The outlying object is possibly surrounded by
circumstellar material and subsequently observed in scattered light.
Only 1 of the 4 significantly outlying sources (121-434) in Figure
\ref{SpecVSphot} shows a reasonably large difference between
photometric and spectroscopic mass in Figure
\ref{specMassVsPhotMass}. 121-434 was observed at optical
wavelengths by RRL. They derive a spectral type of M 5.5 for this
object which differs from our derived value of M 7.0. The earlier
spectral type gives better agreement with the photometric mass (0.1
M$_{\odot}$ difference). However, the spectroscopic temperature
still differs from the photometric temperature by $\sim$900 K.

The comparisons made in Figures \ref{SpecVSphot} and
\ref{specMassVsPhotMass} demonstrate that we can not rely on
photometric data alone to determine the physical properties of
sources. SHC04 also observe inconsistencies between photometric and
spectroscopic masses, and come to the same conclusion. We find
photometric temperatures are more reliable than photometric masses.

\subsection{Source Mass and Age of Cluster}
\label{massAndAgeOfCluster} Masses have been determined for the ONC
sources using the NextGen and Dusty pre-main-sequence isochrones.
Source masses were derived by interpolating between isochrones and
mass tracks. For sources less than 1~Myr in age, masses were
determined by lowering each object down to the 1~Myr isochrone. The
effective temperature was kept constant in all cases. Due to the
fact that the mass tracks are approximately vertical in the regions
where this procedure was carried out, the approach is acceptable.
The quoted errors in derived mass are based on the uncertainties in
effective temperature and therefore do not take into account the
uncertainty in the spectral type to effective temperature
conversion.

Based on the pre-main-sequence NextGen and Dusty isochrones, 36 of
the 40 sources plotted in the H-R diagrams lie below the hydrogen
burning limit (0.075 M$_{\odot}$). The remaining sources are low
mass stars. 17 of the 36 sources from this work have masses that are
potentially below the deuterium burning limit (The status of these
objects are discussed in more detail in Section \ref{PMO}). However,
the template spectra must be extended to include cooler objects
before we can calibrate these sources accurately. To account for
this we assign these objects with a mass that is an upper limit
(0.014 M$_{\odot}$). If we adopt the pre-main-sequence models of
DM97 the number of sources that lie below the hydrogen burning limit
declines from 36 to 29. Source masses are presented in Table
\ref{types}.

We adopt an age of 20~Myr for the NIRI source 030-524 as it is
unlikely to be 50~Myr in age. 3 GNIRS sources appear older than
10~Myr in age in both H-R diagrams. Two of these (016-430 \&
092-532) have large errors in effective temperature which can
account for their age discrepancy. The remaining GNIRS source
(107-453) could feasibly be greater than 10~Myr in age. Masses were
determined for the GNIRS sources in paper 1. 3 more GNIRS sources
have upper limits to their temperatures and so may shift to younger
ages.

Comparison of the position of objects in the H-R diagram with the
Lyon group isochrones suggests an average cluster age that is less
than 1~Myr. We find no strong evidence that supports a significant
population of mid M-type objects $>$10~Myr in age, but have excluded
objects older than 50 Myr. This contrasts with results from SHC04,
where a bifurcation of the H-R diagram can be seen in their data,
for sources that were located in a more central region of the ONC.
RRL have an H-R diagram that is in agreement with ours (and covers
the same region of the ONC), showing few objects close to the 10~Myr
isochrone and the majority above or close to the 1~Myr isochrone. A
small bias may exist in the RRL data due to the fact that the
fainter, older sources are difficult targets for optical
spectroscopy, (although RRL showed that they would have detected a
large population of 10~Myr sources.) However, the luminosity bias in
our sample is no more severe than that in SHC04, so we should
definitely see 10~Myr objects even if they are only a small fraction
of the population.

Based on HST data, RRL provide evidence that showed 4 of the 14
apparently old sources (29\%) in SHC04 are proplyds (see O'Dell \&
Wen 1994) and are therefore not old sources. It is possible that
more of these apparently old objects are proplyd sources or young,
embedded YSOs. However, inspection of the polarimetric dataset of
Tamura et al.(2006) finds no evidence of unusually high
polarisations, which might be expected if they had associated
circumstellar matter. The objects in our sample were primarily
selected from a location West of the brightest nebulosity. SHC04
obtained spectra of objects from the inner $5^{'}_{.}1 \times
5^{'}_{.}1$ of the ONC. The nature of the remainder of the
apparently old objects detected by SHC04 therefore remains unclear.



\subsection{Planetary mass Candidates}
\label{PMO} Formally calibrated template spectra do not exist for
pre-main sequence sources later than M9.5. However, Lodieu et al.
(2008) have provided spectral types ranging from M8-L2 for 21 brown
dwarfs in the Upper Sco association, based on near infrared
spectroscopy. In their work sources later than M9 were assigned
pseudo spectral types based primarily on the spectral profile at
\emph{J-}band and the strength of the H$_{2}$O absorption in the
\emph{H-} and \emph{K-}band. In the absence of other calibration
spectra we have used the data of Lodieu et al.(2008) to derive
spectral types for the 17 sources in this dataset that have derived
types of M9.5 or later, by over-plotting and visually inspecting
spectra from Lodieu et al.(2008). We note that the higher gravity
expected for sources in Upper Sco, which has a mean age of
$\sim$5~Myr, may cause us to derive types that are slightly too
late, owing to the fact that the water absorption bands usually
appear stronger in younger sources with low gravity. The results are
presented in Table \ref{PMOtab}. The final spectral type (pseudo
type) was determined from the average of the \emph{H} and \emph{K}
types, if data in both bandpasses was available. All sources appear
to have spectral types M9.5 or later. This finding agrees well with
the spectral typing results presented in Table 3. 2 sources have
spectral types of M9.5, 11 have types ranging from L0 to L1 and 4
have a spectral type of L1.5. The \emph{H-}band spectrum of 087-024
and \emph{K-}band spectra of 084-104, 196-659, 188-658 and 152-717
have low S/N and therefore have pseudo types which are less robust.
The reliability of this process can be seen in the \emph{H-}band
spectra of 192-723 and 024-124 in Figure \ref{LodieuPlot}.

\begin{figure}
\centering{\ }
    \psfig{file=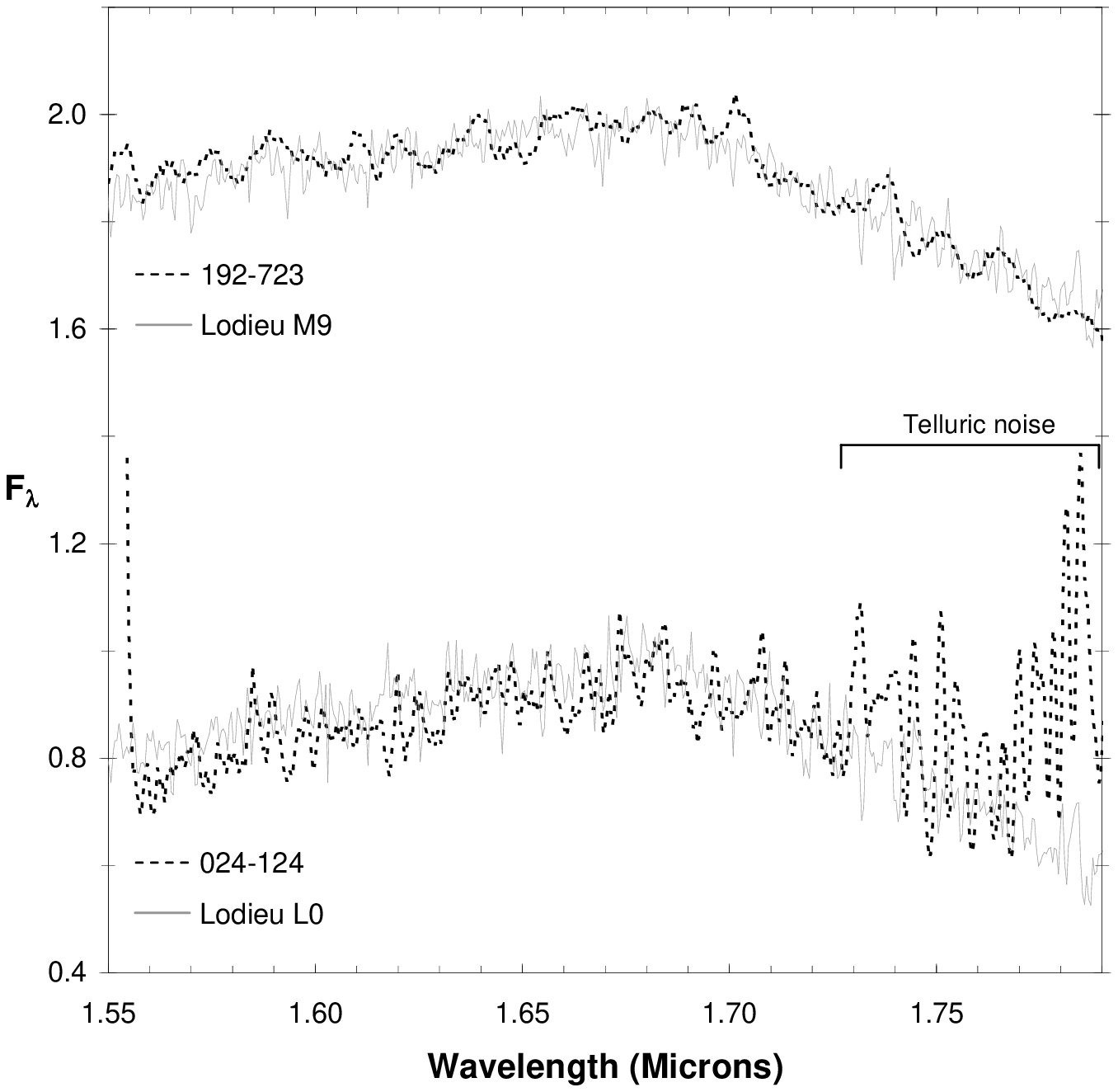,width=8.5cm,angle=-90}
\caption{The spectra of an M9 and L0 object from Lodiue's sample
over-plotted onto the \emph{H-}band spectra of 192-723 and 024-124,
respectively. Due to poor observing conditions and the relatively
low altitude of the AAO, telluric noise is present beyond 1.72
$\umu$m in the spectrum of 024-124. The spectral profile of 192-723
clearly matches that of the M9 object. For 024-124, H$_{2}$O
absorption sortward of 1.68 $\umu$m is consistent with that of an L0
source. Absorption longward if the peak is also consistent before
telluric noise dominates the spectrum.} \label{LodieuPlot}
\end{figure}

\emph{H-}band magnitudes have been compared to the 1~Myr Dusty
isochrone to determine whether these objects are likely planetary
mass objects (PMOs). This was done for the GNIRS sources in paper 1,
and so analysis has only been carried out on the remaining 12
objects. Sources that have dereddened {\it H}-band fluxes 0.5-0.75
magnitudes brighter than expected for a 13~M$_{J}$ object with an
age of 1~Myr (\emph{H}$\simeq$17 according to the Dusty isochrone)
have been assigned the status `uncertain' (see the final column of
Table \ref{PMOtab}). Sources that are even brighter than this are
marked as `BD' which stands for brown dwarf. Sources that are
fainter than expected for a 13~M$_{J}$ object at 1~Myr, or within
0.5 magnitudes, of this value, are listed as PMOs. 4 of the 12
sources (024-124, 055-230, 061-400 and 084-104) are categorised as
brown dwarfs and a further 3 sources (031-536, 056-141 and 192-723)
have uncertain status. A total of 10 out of the sample of 17 PMO
candidates are probable PMOs if they have ages of $\sim$1~Myr.

\begin{table}

 \centering
 \caption{Pseudo spectral types for the 17 sources that appear to have spectral types M9.5 or later in Table
 \ref{types}.}
 \label{PMOtab}
\begin{tabular}{@{}ccccc@{}}
  \hline
  \hline
    ID  &   \emph{H-}  &   \emph{K-}  &   Pseudo Type &   Prob. PMO$^{a}$ \\
  \hline
015-319 &   low S/N &   L1.5    &   L1.5    &   PMO \\
024-124 &   L0  &   -----   &   L0  &   BD  \\
031-536 &   L1  &   -----   &   L1  &   Uncertain   \\
037-628 &   L0  &   -----   &   L0  &   PMO \\
055-230 &   L0  &   $\sim$L0    &   L0  &   BD  \\
056-141 &   L1/L2   &   -----   &   L1.5    &   Uncertain   \\
057-247 &   $\sim$M9.5  &   $\sim$L1    &   L0  &   PMO$^{c}$   \\
061-400 &   L1  &   -----   &   L1  &   PMO \\
084-104 &   L1.5    &   L1?$^{b}$   &   L1  &   PMO \\
087-024 &   L0$^{b}$    &   -----   &   L0  &   BD  \\
130-053 &   $\sim$M9.5  &   -----   &   M9.5    &   BD  \\
137-532 &   L2  &   L1  &   L1.5    &   PMO$^{c}$   \\
152-717 &   L0  &   L0?$^{b}$   &   L0  &   PMO$^{c}$   \\
183-729 &   L0  &   L2  &   L1  &   PMO$^{c}$   \\
188-658 &   L0  &   L0/L1$^{b}$ &   L0.5    &   Probable$^{c}$  \\
192-723 &   $\sim$M9.0  &   L0.5    &   M9.5    &   Uncertain   \\
196-659 &   -----   &   L1/L2$^{b}$ &   L1.5    &   PMO \\
  \hline
\end{tabular}
\flushleft notes:\\
$^{a}$~This estimate is based on \emph{H-}band photometry.\\
$^{b}$~Spectral type may be erroneous due to low signal to noise in spectrum.\\
$^{c}$~PMO status derived in paper 1 (Lucas et al. 2006).\\
\end{table}

In Appendix \ref{hrlodapp} we plot an H-R diagram that includes the
spectral types that were derived using the template spectra of
Lodieu et al. 2008. These replace the late type objects in our
sample ($\geq$M9.5). Sources calibrated from the Lodieu spectra are
marked as open triangles. The NextGen and Dusty isochrones have been
used. Temperatures were estimated from field dwarfs using data from
Mart\'{\i}n et al. 1999. Due to the fact that these values are
estimated an error of $\pm$150 K has been used. The H-R diagram
provides a reasonable estimate of the positions of these late M $/$
early L-type objects.

\section{Surface Gravity and Cluster Membership}
The surface gravity of an object is an important parameter to
constrain as it allows us to determine whether a particular source
belongs to a cluster. The broad absorption bands seen at
near-infrared wavelengths are good indicators of surface gravity
(see Section \ref{specsec}). Surface gravity was ascertained by
studying the spectral profile of each source. In addition to visual
inspection a statistical test that makes comparisons to high and low
gravity spectra was used. This was required because of the low
signal to noise of the data. The statistical test is a useful
back-up analysis for the \emph{K-}band data, where the difference in
spectral profile is more subtle. The same computational analysis was
used to determine surface gravity in paper 1. For a description of
how the code works see Lucas et al. 2006. Both visual inspection and
computational results were used to estimate the gravity of each
object. Analysis was not carried out on sources with a
Rayleigh-Jeans type continuum. The surface gravity results are
presented in Table \ref{sGrav}. The table includes sources from
Lucas et al. 2006.

If the minimum $\chi^{2}$ result agreed with the visual
interpretation, an object was assigned a corresponding gravity
status. No formal cut-off in the difference in minimum $\chi^{2}$
was used to decide whether an object was most likely a high or low
gravity object. Differences in minimum $\chi^{2}$ that were greater
than 2 were taken to be clear indicators of surface gravity. In all
but 1 of the cases where a difference this large occurred (024-124),
the nature of the surface gravity appeared to be obvious from the
spectral profile. Limited wavelength coverage in the spectrum of
024-124 was the reason that surface gravity was uncertain from
visual scrutiny. In cases where the difference in minimum $\chi^{2}$
was less than 1, the gravity of an object was only given a gravity
measurement if the visual interpretation matched the lower
$\chi^{2}$ result. For cases where the difference in minimum
$\chi^{2}$ was very small or did not agree with the visual
interpretation the gravity status was labelled \emph{uncertain} or
\emph{probably low} depending on the individual case. Candidates
thought to have low surface gravity are assumed to be definite
cluster members and are denoted \emph{CM} in Table \ref{sGrav}.
Sources with uncertain gravity have been denoted \emph{U}. All
sources observed at optical wavelengths by RRL were shown to have
very weak absorption of the Na I doublet at 8183/8195 {\AA}. This
feature is a clear indicator of low surface gravity and thus very
strong evidence that the sources are very young and belong to the
cluster. For this reason the objects in Table \ref{sGrav} that were
observed at optical wavelengths have been classed as low
surface-gravity objects and marked as confirmed cluster members. The
membership status of 011-027 was found to be unclear by RRL. Due to
the fact that the spectrum of this object shows a smoothly declining
Rayleigh-Jeans tail we cannot determine whether or not this object
is a cluster member and have therefore excluded it from Table
\ref{sGrav}. 31 sources have spectra that are more consistent with
low-g late M or early L-type templates than high-g templates in the
H- and/or K-bands. The low gravity status is also probable for
016-410, 087-024, 156-141 and 188-658. However, the evidence is less
clear for these objects. The K-band fits are poor for 107-453. This
seems to be due to an unusually high flux at $\lambda >$ 2.2
$\umu$m. This could be due to circumstellar dust emission, which
would suggest that it is indeed a very young objects and a bona fide
cluster member. 016-430, 092-532, 044-527 and 013-306 each have
minimum $\chi^{2}$ values that are not significantly different for
the high and low-g templates. The minimum $\chi^{2}$ fit for 013-306
occurred for a high-g template, despite having a spectrum that
appears to have low surface gravity upon visual inspection. The
$\chi^{2}$ results for 192-723 indicate that the spectrum is more
consistent with a high gravity L dwarf than a low gravity field
dwarf. Photometric measurements suggest that this source is too
bright to be a 1 Myr PMO. A very young PMO at the distance of Orion
is expected to be fainter than \emph{m$_{H}$}=17.2. The apparent
magnitude of 192-723 is \emph{m$_{H}$}=16.08 and it has visual
extinction A$_{\emph{v}}$ $\simeq$4.5. A background field dwarf with
a spectral type $\geq$ M9.0 can also be ruled out based on
photometric measurements. It is possible that this object is a
foreground field dwarf. However, it is also possible that it is a
low mass cluster member that is very young ($\ll$ 1 Myr in age). A
final explanation for this result is that the template spectra for
young objects do not extend to sufficiently late spectral types. The
spectrum of 192-723 shows very strong water absorption which is
stronger than that seen in any of Luhman's calibration sample. The
lack of CO absorption in the K-band spectra also provides evidence
that this may be a young object as CO is often seen to be stronger
in absorption in late-type field dwarfs. A final observation that
may suggest the object is a foreground brown dwarf is that the
signal to noise is very good in comparison to that seen in the other
spectra that have late spectral types. The spectrum of 037-628 is
very noisy and contains a lot of structure. For this reason it was
not possible to achieve a low value of $\chi^{2}$. This object is
therefore classified as an uncertain cluster member.

\begin{table*}

 \begin{minipage}{16cm}
 \centering
 \caption{Gravity and cluster membership. Source-names of objects observed at optical
wavelengths by RRL which have been shown to have low surface gravity
from measurements of the Na I doublet at 8183/8195 {\AA} are in
italics. The table includes sources from Lucas et al. 2006. Sources
with low surface gravity are marked as cluster members CM. Objects
with uncertain gravity status are not definite cluster members and
are denoted by U. Probable cluster members are denoted CM?. The
spectrum of 037-628 contained too much structure for a reliable fit
to be determined. Objects spectral typed $\geq$9.5 are below the
dotted line.}
 \label{sGrav}
\begin{tabular}{@{}crccccccc@{}}
  \hline
  \hline
    & & & \multicolumn{4}{c}{Spectral Type : minimum $\chi^{2}$} & &\\
    & & & low-g & high-g & low-g & high-g & & \\
    Source & Type & Visual (g) & \emph{H-} & \emph{H-} & \emph{K-} & \emph{K-} & Gravity & Status \\
  \hline
\emph{091-017} &   M4.5    &   Low &   M4.5: 1.66  &   M7.0: 2.08  &   M5.5: 2.54  &   M6.0: 2.66  &   Low &   CM  \\
016-410 &   M5.0    &   Uncertain   &   M4.5: 3.07  &   M4.0: 3.19  &   ----    &   ----    &   Prob. Low   &   CM? \\
067-651 &   M6.0    &   Uncertain   &   M5.5: 15.1  &   L1.0: 16.9  &   ----    &   ----    &   Low &   CM  \\
\emph{069-209} &   M6.0    &   Low &   M5.5: 2.37  &   M8.0: 3.63  &   ----    &   ----    &   Low &   CM  \\
\emph{077-453} &   M6.0    &   Low &   M5.5: 10.6  &   M8.0: 14.7  &   ----    &   ----    &   Low &   CM  \\
\emph{019-354} &   M6.5    &   Low &   M6.5: 2.26  &   M9.0: 2.29  &   M6.5: 2.86  &   M8.0: 3.08  &   Low &   CM  \\
255-512 &   M6.5    &   Low &   M5.5: 7.34  &   M8.0: 10.7  &   ----    &   ----    &   Low &   CM  \\
\emph{068-019} &   M7.0    &   Low &   M5.5: 1.25  &   M8.0: 1.70  &   M6.5: 1.72  &   M8.0: 2.54  &   Low &   CM  \\
\emph{121-434} &   M7.0    &   Low &   M5.5: 9.50  &   M8.0: 14.1  &   ----    &   ----    &   Low &   CM  \\
\emph{186-631} &   M7.0    &   Low &   M6.0: 3.29  &   M9.0: 5.95  &   M9.5: 6.99  &   M9.0: 20.1  &   Low &   CM  \\
\emph{014-413} &   M7.5    &   Low &   M8.5: 2.89  &   L1.0: 2.92  &   M9.5: 2.56  &   M9.0: 4.89  &   Low &   CM  \\
\emph{053-503} &   M7.5    &   Low &   M7.5: 2.25  &   L1.0: 2.01  &   M7.5: 2.42  &   M8.0: 9.91  &   Low &   CM  \\
092-532 &   M7.5    &   Uncertain   &   M6.0: 2.90  &   M7.0: 3.34  &   M6.0: 2.90  &   M7.0: 3.20  &   Uncertain   &   U   \\
\emph{095-058} &   M7.5    &   Low &   ----    &   ----    &   M9.5: 12.7  &   L1.0: 59.9  &   Low &   CM  \\
010-109 &   M8.0    &   Low &   ----    &   ----    &   M7.5: 2.97  &   M8.0: 5.48  &   Low &   CM  \\
\emph{030-524} &   M8.0    &   Uncertain   &   ----    &   ----    &   M4.5: 3.01  &   M4.0: 3.13  &   Low &   CM  \\
043-014 &   M8.0    &   Low &   M8.5: 2.03  &   M3.0: 2.99  &   ----    &   ----    &   Low &   CM  \\
107-453 &   M8.0    &   Uncertain   &   M9.0: 4.10  &   L1.0: 3.80  &   M9.0: 14.3  &   L0.0: 32.1  &   Uncertain   &   U   \\
044-527 &   M8.5    &   Uncertain   &   ----    &   ----    &   M4.5: 3.60  &   M4.0: 3.76  &   Uncertain   &   U   \\
013-306 &   M9.0    &   Low &   ----    &   ----    &   M6.5: 5.20  &   M8.0: 4.62  &   Uncertain   &   U   \\
016-430 &   M9.0    &   Uncertain   &   M9.0: 2.90  &   L3.0: 2.90  &   M9.0: 4.30  &   L1.0: 4.20  &   Uncertain   &   U   \\
084-1939 &  M9.0    &   Low &   M9.5: 3.03  &   L1.0: 6.83  &   M8.5: 2.40  &   M8.0: 6.13  &   Low &   CM  \\
\emph{096-1943} &  M9.0    &   Low &   M9.0: 6.28  &   L1.0: 6.25  &   M9.5: 3.71  &   L1.0: 15.8  &   Low &   CM  \\
031-536 &   M9.5    &   Uncertain   &   M9.5: 2.25  &   L3.0: 2.32  &   ----    &   ----    &   Low &   CM  \\
055-230 &   M9.5    &   Low &   M9.0: 4.19  &   L1.0: 6.16  &   M7.5: 4.21  &   M8.0: 4.99  &   Low &   CM  \\
087-024 &   M9.5    &   Uncertain   &   M9.0: 7.82  &   L1.0: 8.38  &   ----    &   ----    &   Prob. Low   &   CM? \\
\multicolumn{9}{c}{\dotfill} \\
057-247 &   $\geq$M9.5   &   Low &   M9.0: 2.00  &   L1.0: 3.20  &   M8.0: 1.90  &   M9.0: 3.10  &   Low &   CM  \\
\emph{183-729} &   $\geq$M9.5   &   Low &   M8.0: 2.90  &   L1.0: 3.20  &   M8.0: 3.70  &   M9.0: 5.70  &   Low &   CM  \\
015-319 &   $>$M9.5   &   Low &   ----    &   ----    &   M9.0: 13.9  &   L1.0: 11.2  &   Uncertain   &   CM? \\
024-124 &   $>$M9.5   &   Uncertain   &   M9.5: 10.5  &   L1.0: 13.0  &   ----    &   ----    &   Low &   CM  \\
037-628 &   $>$M9.5   &   Uncertain   &   N/A &   N/A &   ----    &   ----    &   Uncertain   &   U   \\
056-141 &   $>$M9.5   &   Low &   M9.5: 7.84  &   L3.0: 12.7  &   ----    &   ----    &   Low &   CM  \\
061-400 &   $>$M9.5   &   Low &   M9.0: 1.42  &   M3.0: 1.53  &   ----    &   ----    &   Low &   CM  \\
084-104 &   $>$M9.5   &   Low &   M9.5: 1.52  &   M3.0: 1.53  &   M9.5: 18.2  &   L1.0: 22.0  &   Low &   CM  \\
\emph{130-053} &   $>$M9.5   &   Uncertain   &   M9.5: 56.3  &   L1.0: 82.6  &   ----    &   ----    &   Low &   CM  \\
137-532 &   $>$M9.5   &   Low &   M9.0: 1.50  &   L2.0: 1.60  &   M9.0: 5.00  &   L0.0: 9.80  &   Low &   CM  \\
152-717 &   $>$M9.5   &   Low &   M9.0: 2.00  &   L3.0: 3.20  &   M9.0: 5.70  &   L0.0: 8.70  &   Low &   CM  \\
188-658 &   $>$M9.5   &   Uncertain   &   M8.0: 5.90  &   L1.0: 6.70  &   M9.0: 4.40  &   L0.0: 5.20  &   Prob. Low   &   CM? \\
192-723 &   $>$M9.5   &   Low &   M7.5: 25.3  &   L1.0: 23.2  &   M9.0: 13.2  &   L1.0: 6.35  &   Uncertain   &   U   \\
196-659 &   $>$M9.5   &   Low &   ----    &   ----    &   M9.5: 6.16  &   L1.0: 6.45  &   Low &   CM  \\
  \hline
\end{tabular}
\end{minipage}
\end{table*}

Gorlova et al. 2003 provide evidence of gravity sensitivity of Na I
at \emph{K-}band (2.21$\umu$m). The molecular feature is shown to be
weaker in absorption in young cluster objects compared to field
dwarfs of the same spectral type. However, the Na I absorption
feature can not yet be used to accurately measure surface gravity. A
selection of high quality spectra is required to construct a
reliable spectral type to surface gravity relation for the Na I
feature at 2.21 $\umu$m. In general the K-band spectra are too noisy
to accurately identify and measure the equivalent widths of the Na I
feature. However, there is some evidence of the absorption feature
in the spectra of 192-723, 010-109, 186-631, 091-017, 084-1939,
014-413, 096-1943, 013-306 and 095-058. Although these features
cannot be used to derive definite gravity measurements, their
apparent weakness provides further evidence that they are probable
cluster members. Despite the indication of a weak Na I feature in
the spectrum of 013-306, the overall gravity status is uncertain
from visual interpretation and $\chi^{2}$ comparison.

The $\chi^{2}$ and visual analysis demonstrate that 33/40 of the low
mass candidates are probable cluster members. The majority of
uncertain cluster members in Table \ref{sGrav} have spectral types
that are M8 or later. The expected number of M9-L5V solitary field
dwarfs lying between Earth and Orion in the 26 arcmin$^{2}$ field of
view of the deep Gemini imaging data from Lucas et al.(2005), is
0.36-0.72, down to \emph{H}$=$19. This rises to 0.61-1.22 objects
when unresolved binaries are included (Lucas et al, 2006). 21
sources in Table \ref{sGrav} have spectral types $\geq$ M9.0, 14 of
which are in the field of view of the Gemini imaging. By assuming a
Poisson distribution with a mean of 1.22 for field contaminants, it
is possible to estimate the probability of different numbers of non
cluster members being present in the 26 arcmin$^{2}$ area with
$\geq$M9.0. With the above assumptions the likelihood of 7 or more
sources being foreground field dwarf contaminants is 0.16\%. The
probabilities of $>$3 and $>$2 contaminants are 3.55\% and 12.49\%
respectively. When M6 - M8 objects are included, the expected number
of contaminating objects rises by a factor of $\sim$2 (Lucas et al.
2006). It is therefore feasible that a few of the sources which have
an uncertain gravity status are foreground field dwarfs.
Contamination is most likely for the sources with the earlier
spectral types. Contamination by field dwarfs at the distance of
Orion and/or behind the cluster can be ruled out. This is because
the typical absolute magnitudes for M9 and L5 field dwarfs at K-band
are 10.29 and 11.61, respectively (Kirkpatrick et al. 2000). The
$\chi^{2}$ fits rule out contamination from any objects later than
L1.

\section{Modelling Age and Mass}
The substellar IMF and age distribution for the ONC are still poorly
constrained. This is because a very large number of high quality
optical and infrared spectra are required to confirm cluster
membership and fully characterise the population. One measurement
that is well constrained for the ONC is the luminosity function. A
Monte Carlo approach can be used to experiment with different age
distributions and a variety of different mass functions, to generate
a range of luminosity functions. The luminosity functions produced
by different models can be compared to the real (dereddened) luminosity function
and tell us which initial conditions are most likely and unlikely.
We created a program that allowed us to do this for the ONC. An
explanation of how the program works and the parameters that were
varied are described in Section \ref{MCparameters}. The results from
the Monte Carlo simulations are described and analysed in Section
\ref{MCResultsSec}.

A limitation of this approach is that we use the Lyon Group models
to generate a luminosity function. The scatter in the H-R diagram
(see plot in Appendix \ref{hrlodapp}) shows that these models cannot
fully describe what we observe. For example 055-230 (the source furthest
from the 1~Myr isochrone) in Figure \ref{HRDiagrams} is much cooler
than predicted by the Lyon Group models. However, the temperatures
of some of the brighter, outlying objects are uncertain (see
Section \ref{analysisSection}) since two of them have earlier
optical spectral types that are more consistent with the Lyon Group
models. E.g. 055-230 has an optical type of M6.75 and an
infrared type of M9.5 or L0 (the latter is the pseudo type which was
calibrated from the template spectra of the Lodieu et al. (2008)).
We do not think that this scatter will significantly affect the
results of these simulations because the proportion of outliers is
small. Also the spread extends to both warmer and cooler effective
temperatures, which tends to cancel any inaccuracy or limitations of the
theoretical models.

\subsection{Model}
\label{MCparameters} In this section we give an overview of what the
model does then go into detail regarding the parameters that can be
varied.

Our model generates a luminosity function in the following way. A
random number is scaled to generate an age based on a desired age
distribution. A second random number is then used to generate a mass
which is based on the type of mass function that has been chosen.
The age and mass values are then used to select a pre-main-sequence
isochrone of the derived age, and an \emph{H-}band luminosity
corresponding to the mass that was generated. The H-band luminosity
value is then added to two\footnote{The code generates a further 2
histograms that have smaller bins. These have not been included in
the paper due to the low number statistics.} differently binned
histograms. This is carried out for the same number of objects in
our extinction limited sample of sources (explained in Section
\ref{MCResultsSec}). This was done for convenience when normalising
the histogram. The algorithm is repeated a million times. Finally
the two histograms are normalised producing a luminosity function
that is represented in a variety of bin sizes.

The model allows us to experiment with four types of age
distribution. Variations can be made to two of the scenarios to give
a total of six different types of age distribution. In all cases the
earliest age that can be generated is 0.3 Myr, a limit chosen
because younger sources are likely to be too deeply embedded in
their natal cloud core to be present in our extinction-limited
luminosity function. The most simplistic age distribution we use is
flat. This scales a random number so that it has an equal chance of
lying anywhere between 0.3 Myr and a maximum age prescribed by the
user. The second type of distribution that can be selected is
Gaussian. In this type of scenario the user can define the mean age
and the standard deviation or spread. The third type of age
distribution is skewed. We use a Gumbel Skew for this age function.
The skew can be set to be positive or negative and also gives the
user the flexibility of setting the mean age and the spread. The
advantage of a skewed distribution is that it can simulate a sharp
burst of star formation that declines over a period of time (-ve
skew). Equally it is possible to investigate a scenario where the
star-formation rate increases before dropping off more rapidly (+ve
skew). The final type of age distribution that we investigate is
bimodal. We chose to generate bimodal age distributions using two
different methods, each of which was modelled in our simulations.
The first was to use two Gaussians and the second to use two skews.
In the case of two skews we decided the most realistic scenario was
to use two -ve skews to simulate two bursts of star-formation that
tail off over time. In the bimodal models the user was able to set
the ratio of sources to be generated in each distribution.

The nature of the mass function could be varied in three different
ways. By default a log-normal mass function as derived by Miller and
Scalo (1979) was used. The mean and standard deviation of this
function could be altered. We explored a further option that enabled
the mass function to be altered by truncating the Miller and Scalo
mass function by a user-defined power law at low masses. This enabled us to
experiment with previously published values of $\alpha$ that have
been used to describe the behaviour of the mass function as it
extends down to the planetary mass regime. The final parameter
allowed us to truncate the mass function at the deuterium burning
limit. The reason for doing this was to see whether a population of
planetary mass objects was required to produce a luminosity function
that represented that observed for the ONC. We use a rejection
algorithm to ensure that objects more massive than 1.2 M$_{\odot}$ are
not created by the mass function since this is the upper mass limit
of the NextGen model grid.

From the simulations a luminosity function that is based on the
NextGen and Dusty pre-main-sequence models or those created by DM97
could be generated. However, only the NextGen and Dusty models were
simulated. This is because the DM97 models do not extend below the
deuterium burning limit, and that these models are thought to be
less accurate. A limitation with our model is that the NextGen and
Dusty isochrones do not extend to ages less than 1 Myr. For
generated ages $<$ 1 Myr we adopt luminosities from the 1 Myr
isochrone. To some extent this will lead to a luminosity function
that lacks objects at the brightest luminosities. However, based on
our input parameters we do not believe that this will have a
significant effect on the results. We tested this by creating a 0.5
Myr isochrone based on data from the DM97 models. Although this
isochrone cannot be used for the final modelling results and
analysis, it allows us to see what happens to the luminosity
function when younger, brighter sources are added. The results
revealed an interesting difference. In general there was a slight
increase in the number of objects in the brightest bins, slightly
reducing the number of objects in the faintest bins. The overall
reduced $\chi^{2}$ values improved by approximately 0.4 in the best
fitting cases. This demonstrates that using a 1 Myr isochrone for
sources younger than 0.75 Myr in age has a noticeable effect on the
brightest and faintest end of the luminosity function. However, the
overall change is subtle and doesn't alter the best fitting age and
mass parameters. As previously mentioned, the 0.5 Myr isochrone is
only an estimate so we cannot trust that the output luminosity
function is reliable.

The histograms that represent the luminosity function are binned so
that two use full magnitude bins and two use half magnitude bins.
The two histograms that have complete magnitude bins are offset by
half a magnitude. The two histograms containing half magnitude bins
are offset by quarter of a magnitude. These different binnings were
used in order to assess the effect of random statistical
fluctuations that may be present in the observed luminosity
function.

\subsection{Calculation of $\chi^2$}
\label{MCResultsSec} 
\begin{figure*}
\centering{\ } 
  \hbox{
    \psfig{file=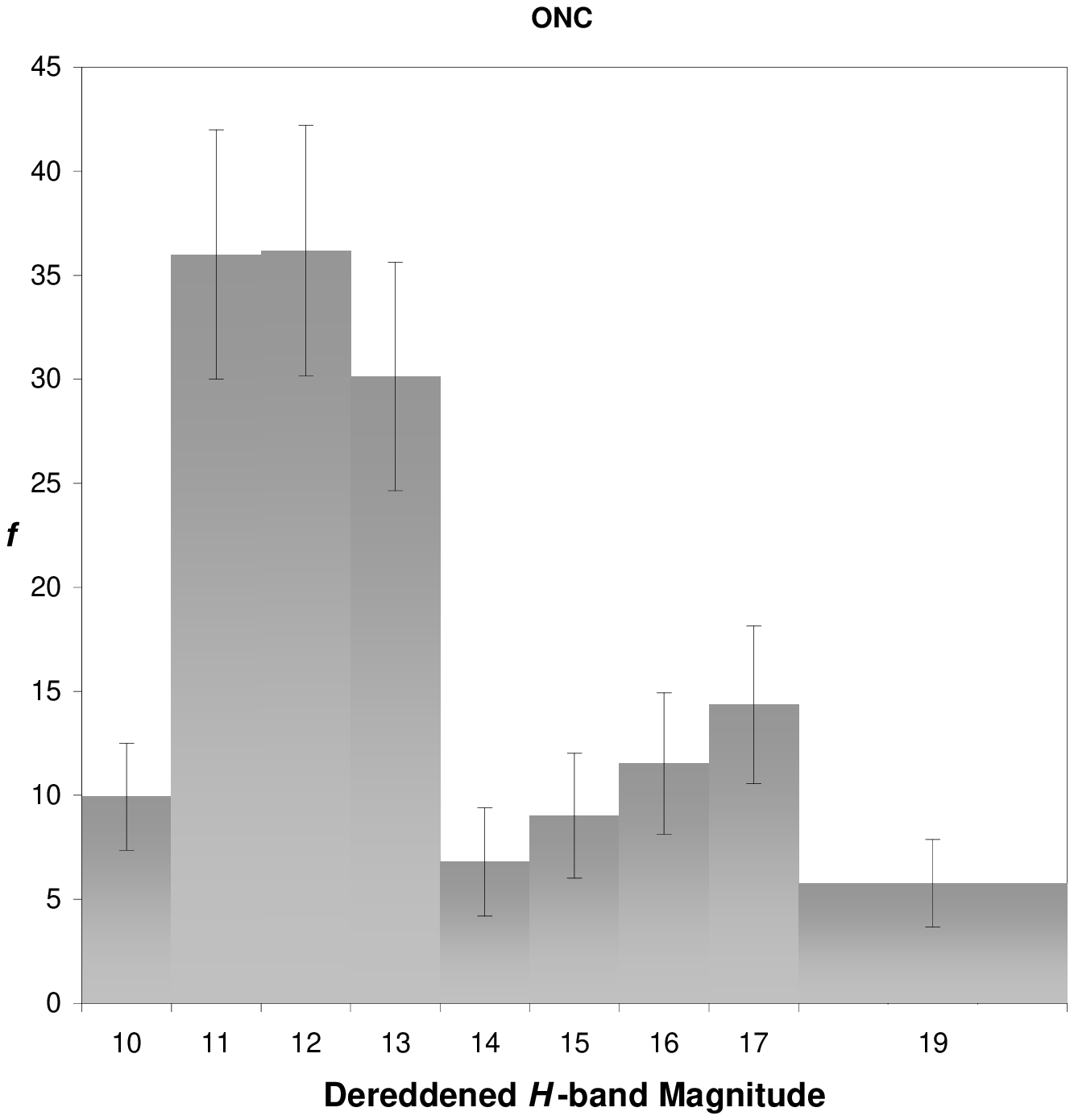,width=8cm,angle=-90}
    \psfig{file=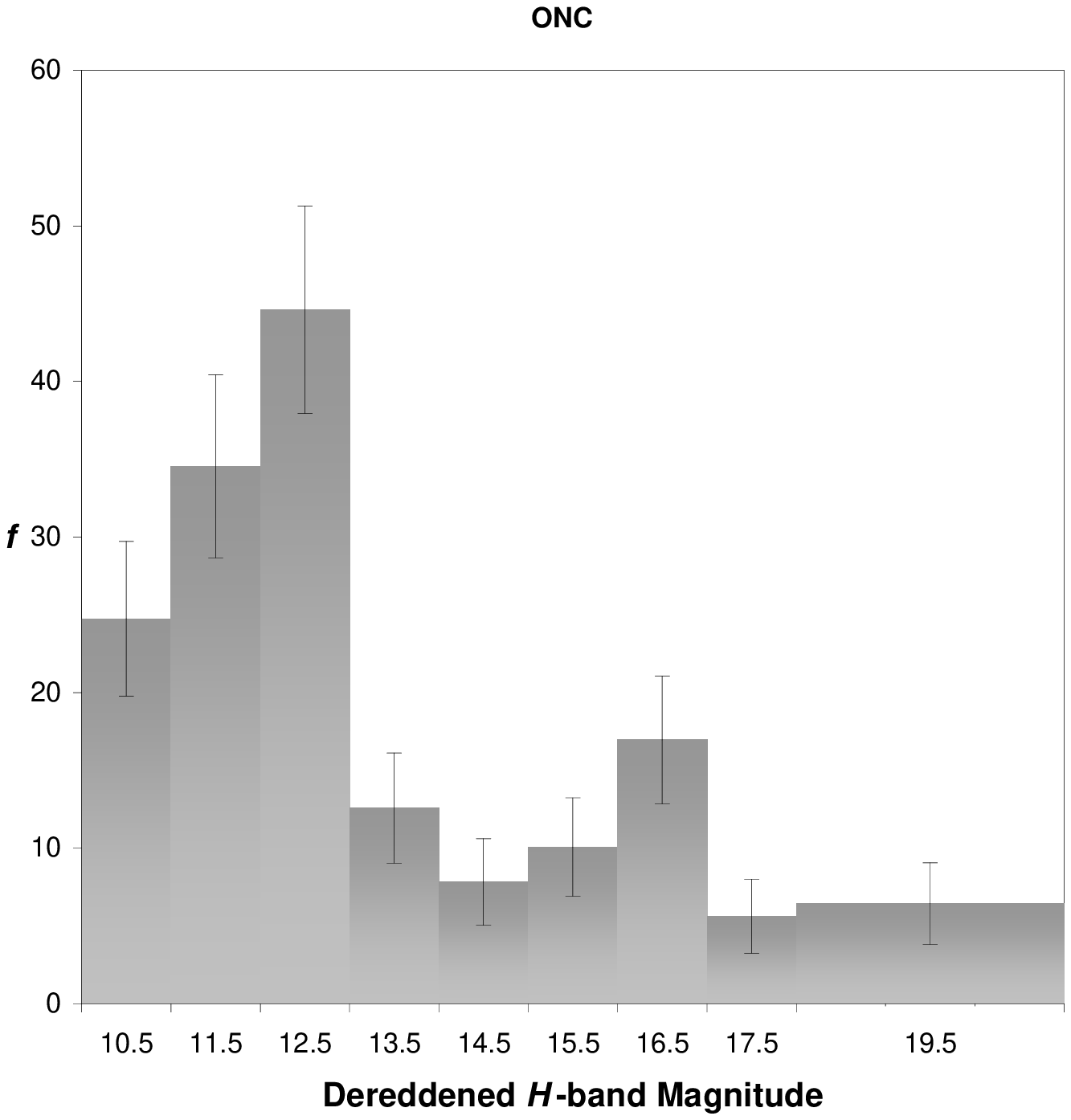,width=8cm,angle=-90}}
\caption{The measured extinction limited (A$_{\emph{v}}<5$), dereddened
\emph{H-}band luminosity function for the ONC. Two different binning
variations are presented. The broad final bin in each of the
diagrams was obtained by averaging three single magnitude
bins that had low number statistics.} \label{ONCLF}
\end{figure*}

In order to make a comparison to our model
luminosity functions we need to be sure that our observed luminosity
function suffers no contamination and is complete at all magnitudes.
To do this we use an extinction limited sample, including all
sources with A$_V<$ 5. We use deep photometric data from Lucas,
Roche and Tamura (2005) to generate our observed luminosity
function and converted it to a dereddened luminosity function
using the Rieke \& Lebofsky (1985) extinction law.
 Completeness limits were found for the faintest magnitude
bins using the ADDSTAR routine in IRAF. By randomly generating faint
sources throughout the field, the number of these detected in each
magnitude bin was able to be determined. Contamination due to
foreground and background stars and dwarfs is likely to be small.
However, we expect contamination from background stars to exist in
our sample. This will be most significant at the faintest
magnitudes. We determine the contamination in each bin based on
sources we believe to be background stars from our entire
spectroscopic sample and an optical data set from RRL.
Due to the relatively low number statistics we
see no contamination in several bins. We fitted a low order polynomial
across several bins to better determine the contamination where this was
the case. We note
that the contamination rate is consistent with a Besan\c{c}on model
calculation (Robin et al. 2003) which we used to calculate the
number of sources ranging from M0-$\sim$M6 out to 450pc over the
same area of sky covered by our observations. The calculation
predicts 9 sources in total but becomes incomplete at \emph{H}=19
where we measure greater contamination. However, we are likely to
include reddened background stars in our spectroscopic sample. The
Beasan\c{c}on calculation apparently predicts a larger number of
foreground sources than the modified Wainscoat calculation shown in
Hillenbrand \& Carpenter (2000). In Figure \ref{ONCLF} we present
the corrected luminosity function as 2 alternatively binned
histograms. In both histograms the final three bins were combined into one
large bin to avoid very low number statistics. The final bin is therefore
representative of 3 bins. The error on each bin was calculated using
standard error propagation formula. The total error takes into
account the Poisson error relating to the total number of sources
observed in the bin, the error in the completeness correction and
the error associated with the contamination correction. The
completeness errors were calculated from statistics gathered when
using the ADDSTAR routine. The dichotomy between cluster membership
and contamination in each bin was represented as a binomial
probability distribution. the fractional uncertainty was therefore
calculated using the formula presented in Equation \ref{fracuncert}

\begin{equation}
\label{fracuncert}
\frac{\sqrt{npq}}{n}
\end{equation}

\begin{figure*}
\centering{\ } 
  \hbox{
    \psfig{file=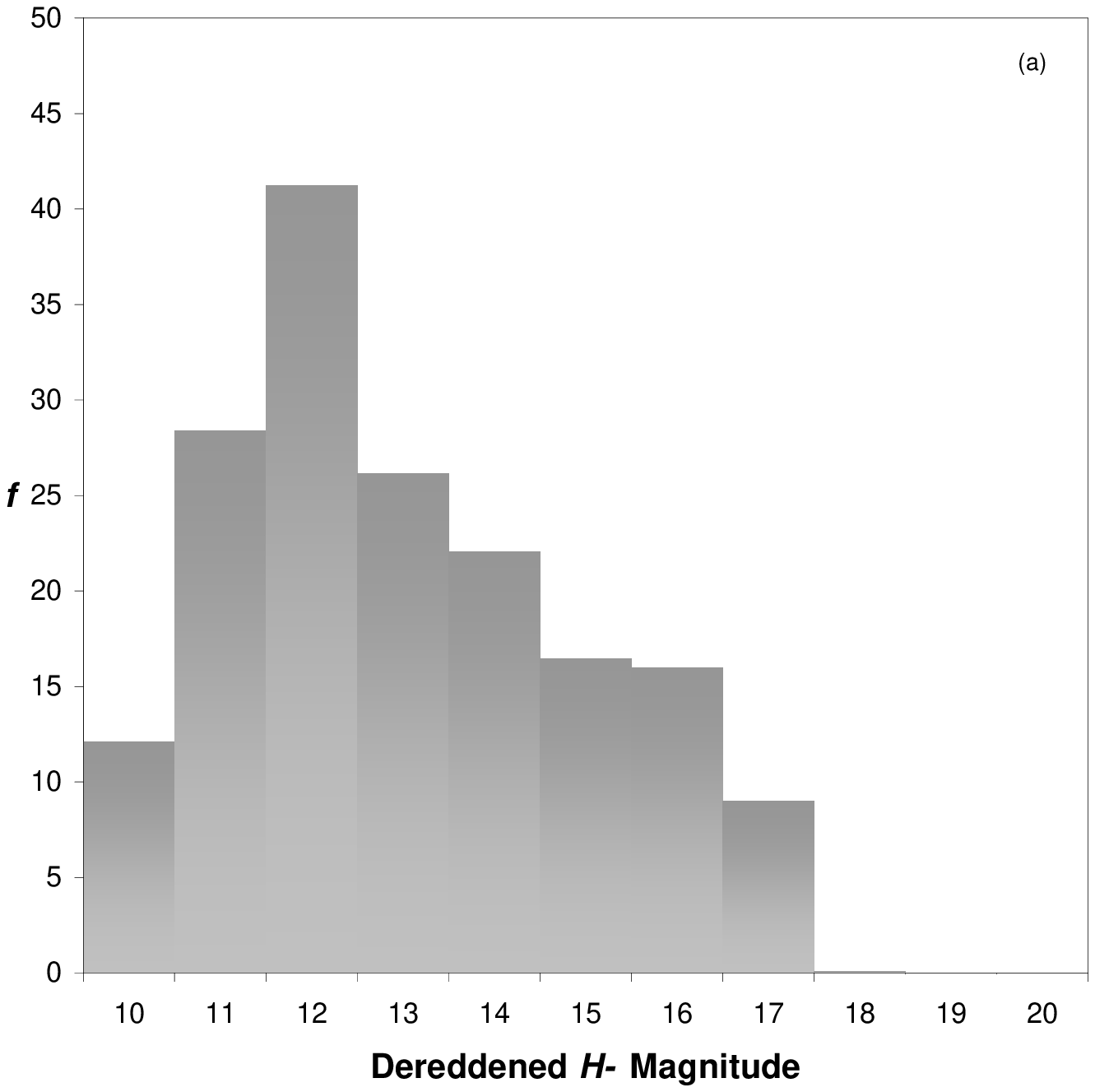,width=7cm,angle=-90}
    \psfig{file=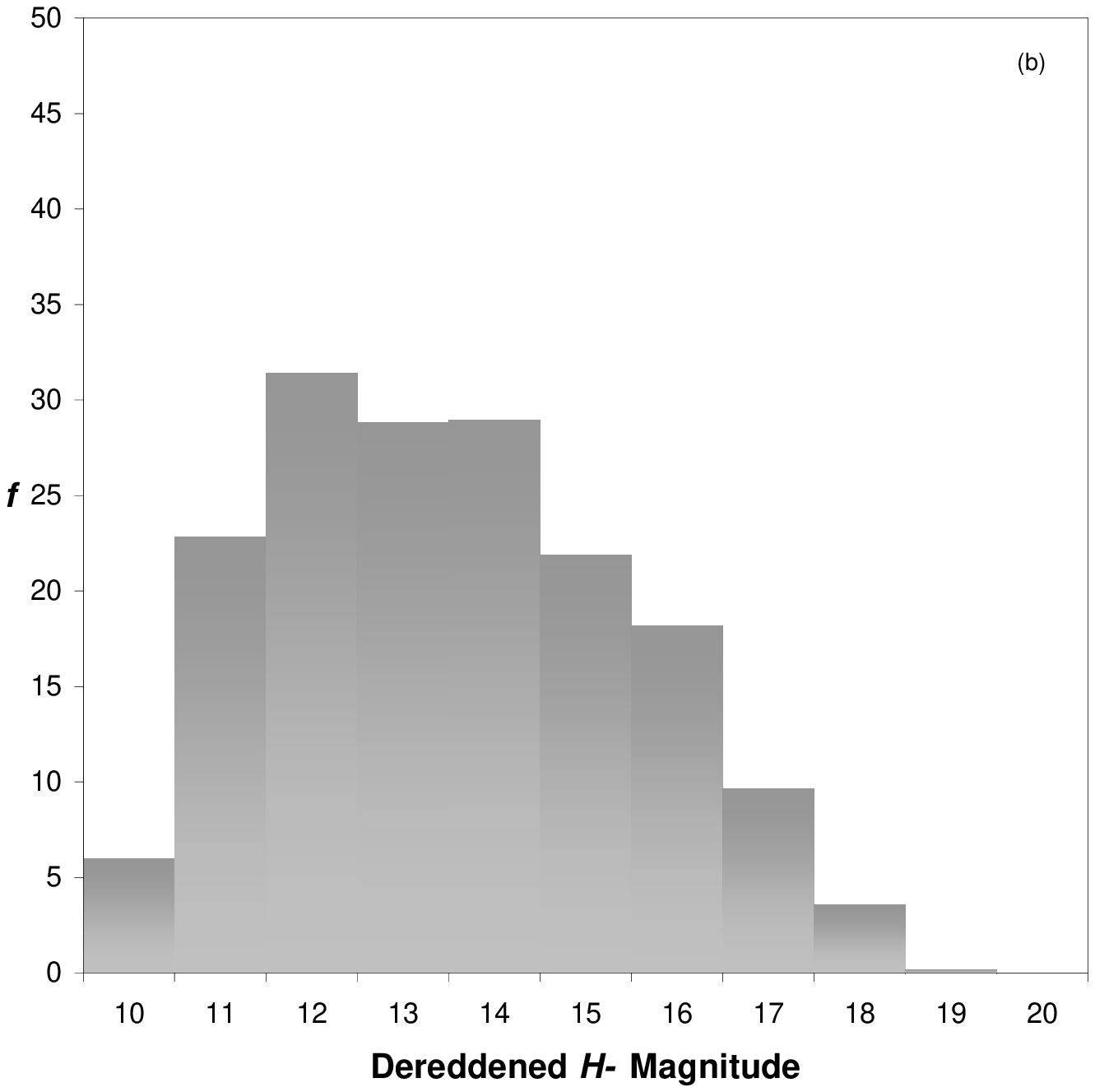,width=7cm,angle=-90}}
  \hbox{
    \psfig{file=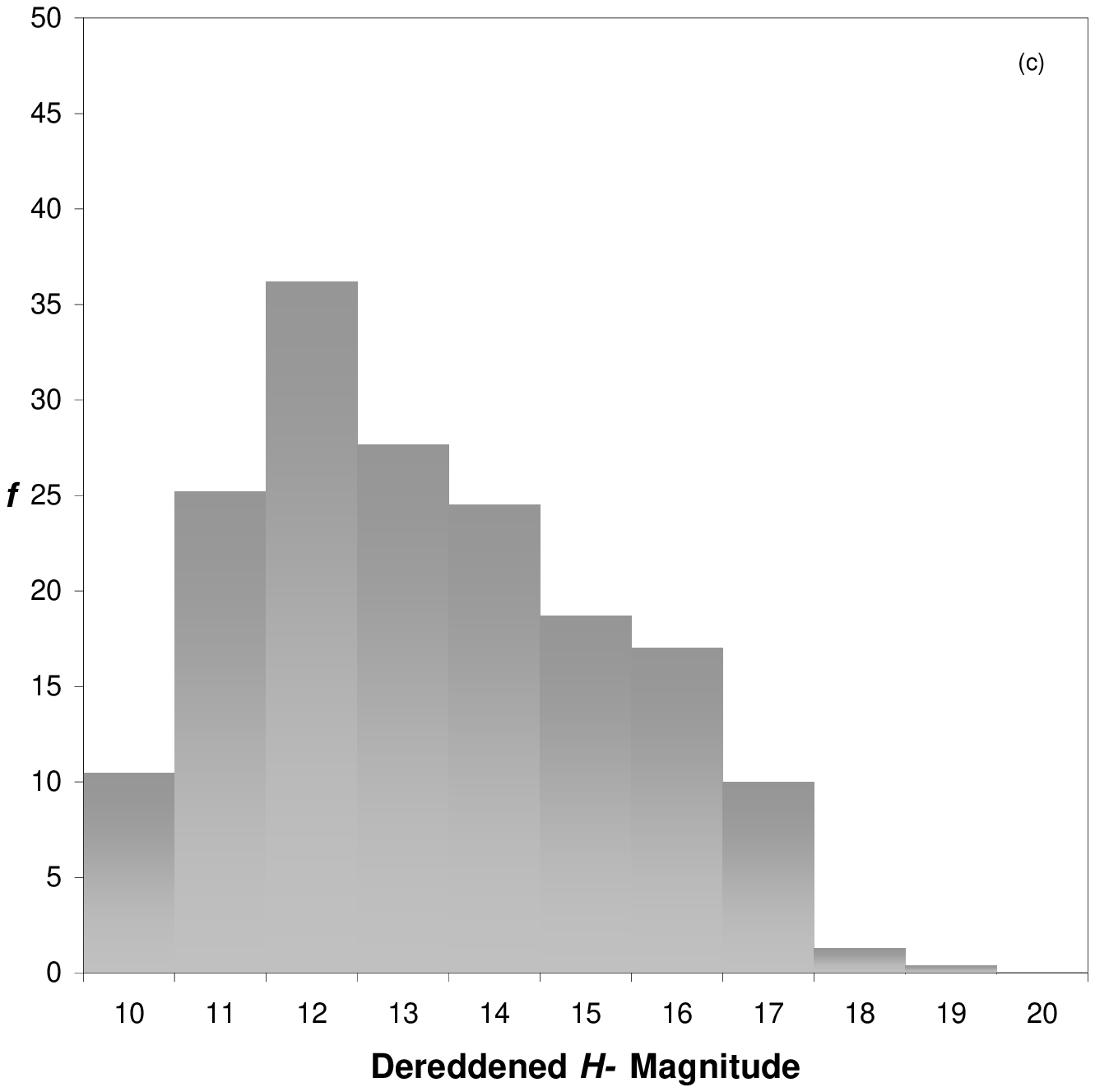,width=7cm,angle=-90}
    \psfig{file=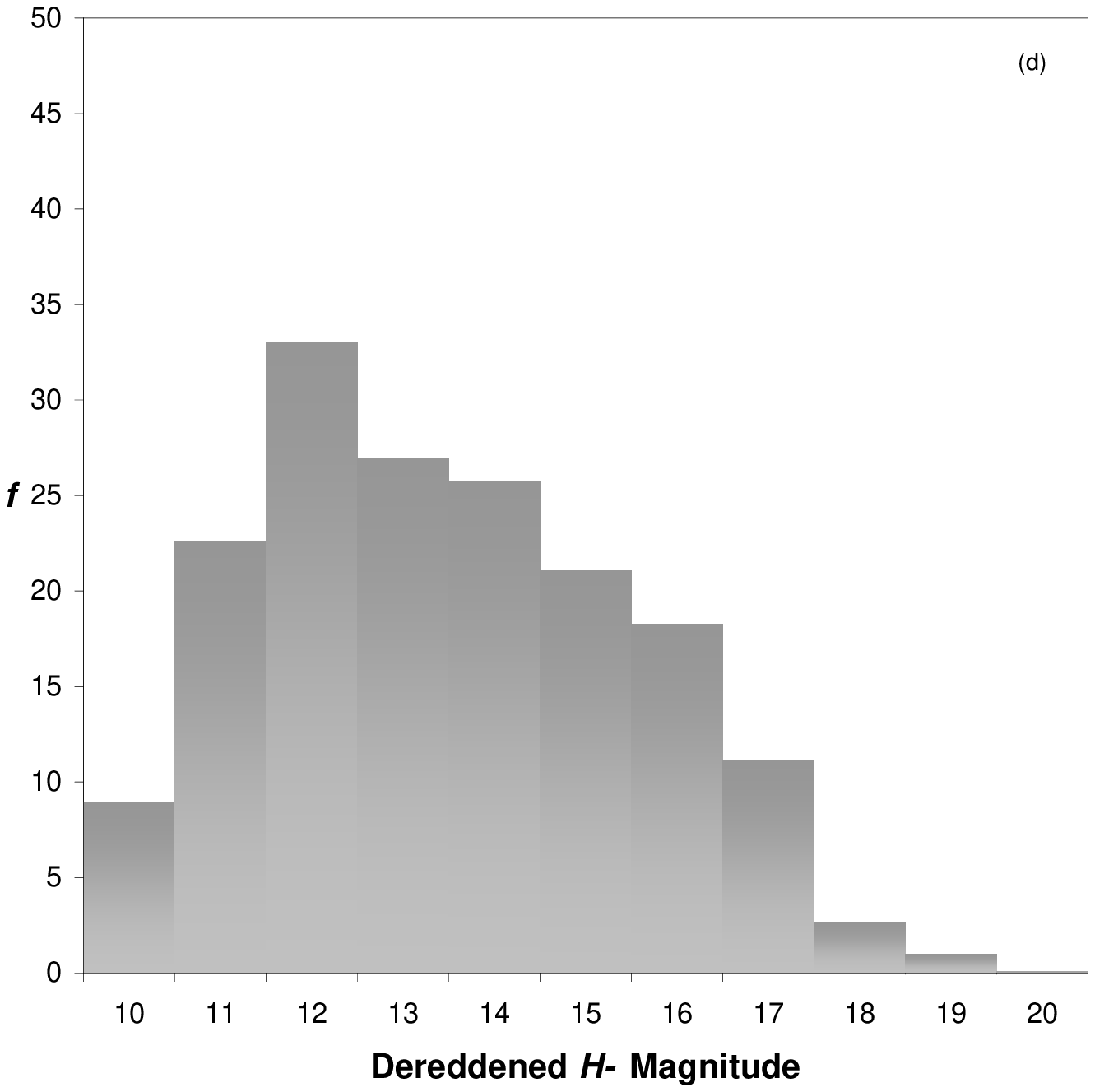,width=7cm,angle=-90}}
\caption{The effect of truncating the IMF at 0.012 M$_{\odot}$. The
IMF parameters are the same for each of the synthetic luminosity
functions displayed (Miller \& Scalo above 0.15~M$_{\odot}$ and a declining
power law with $\alpha=0.31$ at 0.012-0.15~M$_{\odot}$.)
Histogram (a) is derived from a negatively skewed 1~Myr age
population with a $1\sigma$ age spread. Histogram (b) is derived
from a Gaussian age population centered at 1~Myr, with a $3\sigma$
age spread. Histogram (c) is derived from a bimodal Gaussian age
population centered at 1 and 10~Myr, with a respective ratio of
$10:1$ sources. Histogram (d) is derived from a bimodal Gaussian age
population centered at 1 and 10~Myr, with a respective ratio of
$3.3:1$ sources. The Gaussians used for (c) and (d) had a $1\sigma$
age spread. In all cases the faintest \emph{H-}band magnitude bins
are starved of sources. To highlight this effect the faintest
magnitude bins have not been rebinned.} \label{truncplan}
\end{figure*}

\begin{figure*}
\centering{\ } 
  \hbox{
    \hspace{1.4cm}
    \psfig{file=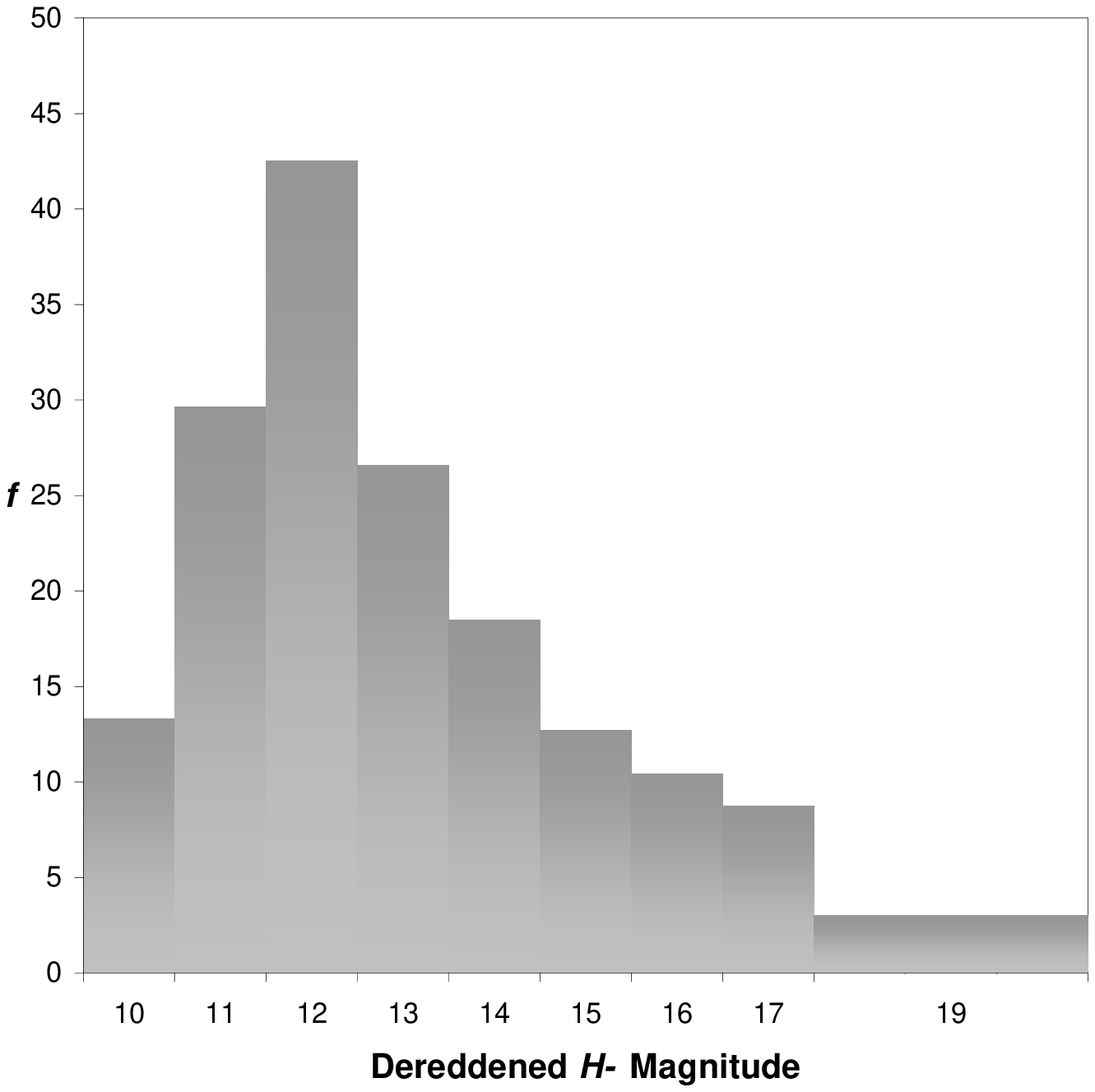,width=7cm,angle=-90}
    \psfig{file=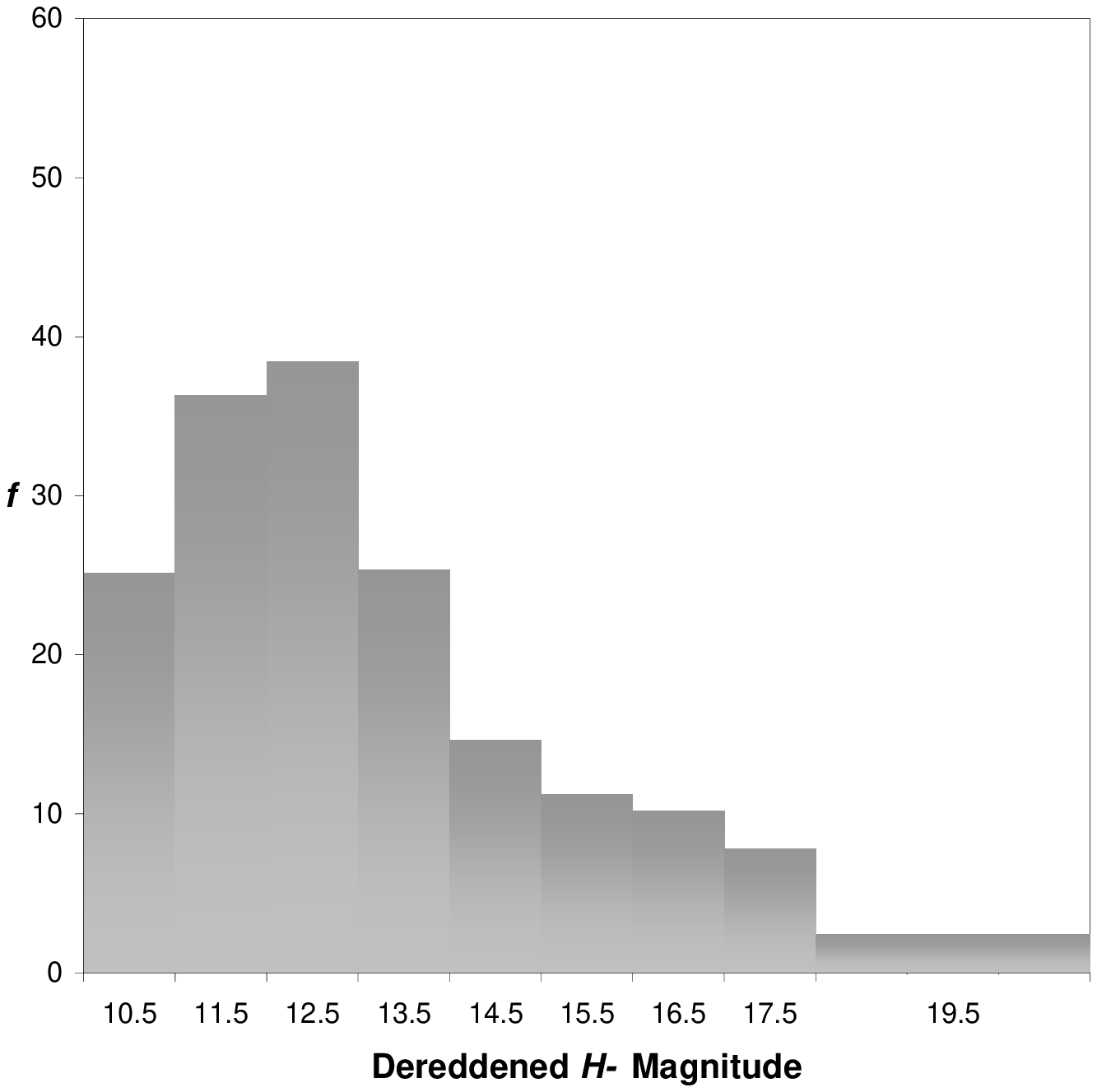,width=7cm,angle=-90}}
  \hbox{
    \hspace{1.4cm}
    \psfig{file=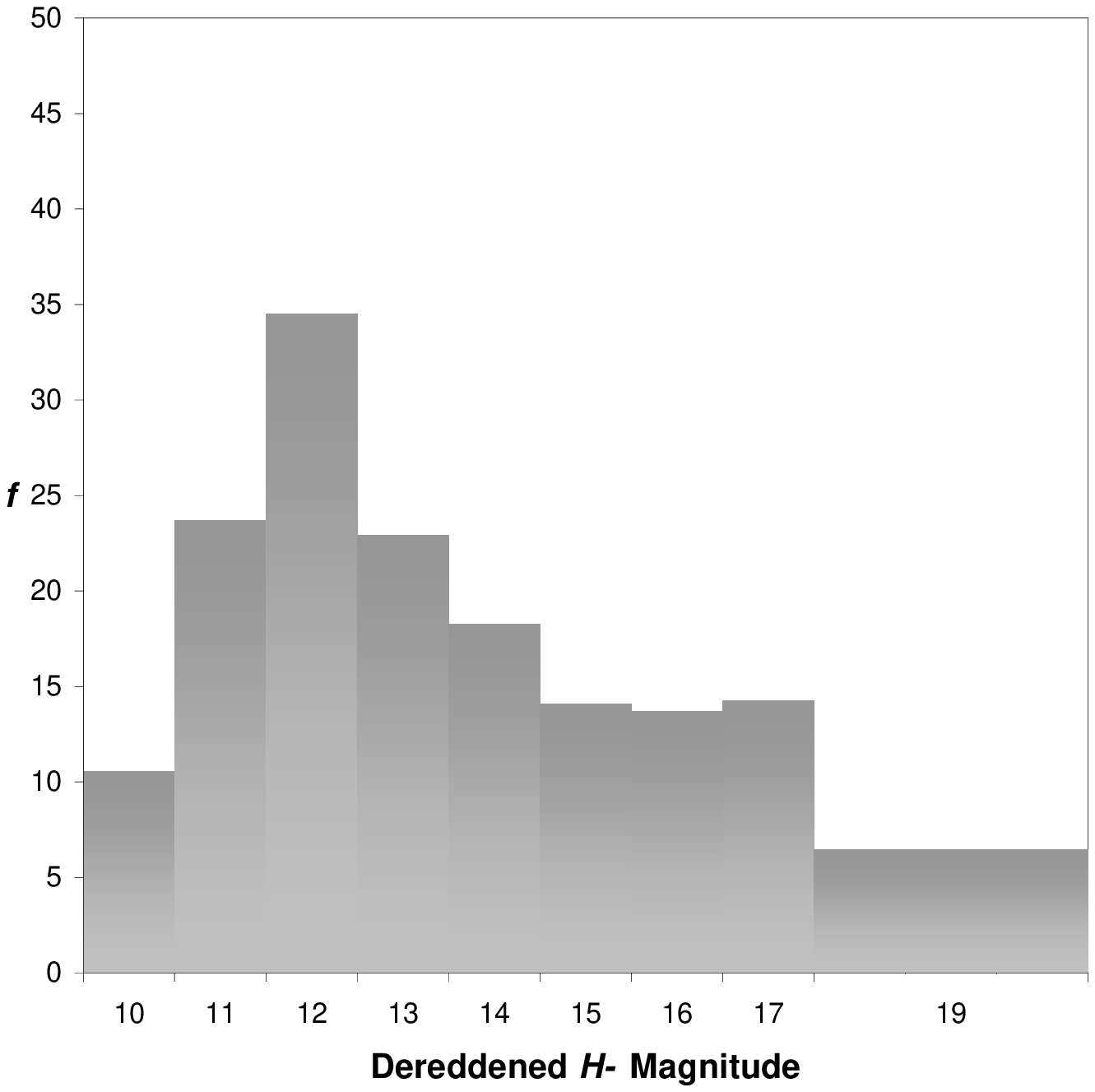,width=7cm,angle=-90}
    \psfig{file=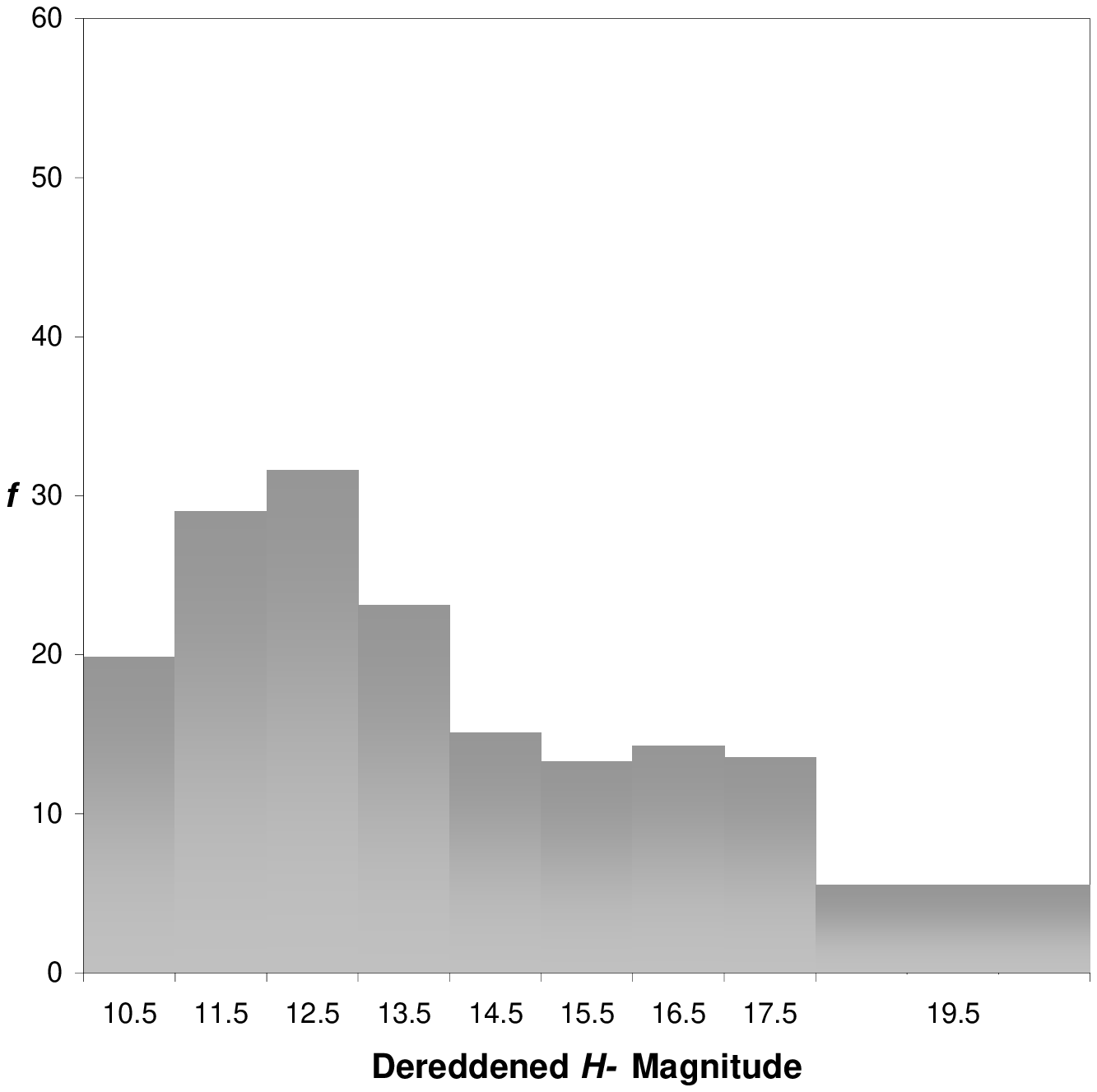,width=7cm,angle=-90}}
\caption{The luminosity functions for the two best fitting
scenarios. Both are negatively skewed age distributions with a
1$\sigma$ age spread and centred at 1~Myr. The mass distribution is
a Miller \& Scalo IMF that peaks at 0.15 M$_{\odot}$ and is
truncated by a power law in the form of $N(M) \alpha M^{\alpha}$.
$\alpha$=0.60 in the top two plots and 0.31 in the remainder. In
each case we fail recreate the sudden drop in sources that have
luminosities ranging from $\sim13 - 14.5$. $\chi^{2}$ values are
similar for both values of $\alpha$. The low mass end of the
luminosity function is best represented when $\alpha$=0.31.}
\label{bestFit}
\end{figure*}

Rather than freely explore a very large parameter space we chose to
design a model grid that tested parameters that were close to those
expected in Orion in order to reduce computational time. We made
sure that parameters in our simulations deviated sufficiently from
the most likely parameters to test whether the luminosity function
changed significantly with less likely initial conditions. In total
296 different scenarios were tested.

We calculated a reduced $\chi^{2}$ value for each scenario to
determine which parameters approximate the luminosity function best.
We calculate our degrees of freedom in each scenario as the number
of bins being fit minus the number of free-parameters in the model.
Despite the fact that the bimodal scenarios have more degrees of
freedom, reducing the $\chi^{2}$ did not change the luminosity
functions with the lowest $\chi^{2}$ quantity.

\subsection{Model Results}

Our best model fits have a reduced $\chi^{2}$ somewhat in excess of
unity. This is because the models cannot recreate the sudden drop in
source numbers in the observed luminosity function close to
\emph{H}=13.5. The steep drop observed in the dereddened luminosity
function may simply be a statistical fluctuation due to the small
size of our extinction limited sample. It is not seen in larger,
shallower surveys of the ONC (see below). All of the remaining bins
can be well fitted by our models. Our lowest reduced $\chi^{2}$
value is $\sim$5.1. (This value decreases to 3.3 when half-magnitude
bin sizes are used. As previously mentioned histograms with smaller
bin sizes are not included due to the low number of objects in each
bin, which causes the $\chi^2$ method to become invalid.) We do not expect
to be able to perfectly recreate the
luminosity function with the relatively basic IMF and age
distributions used in our model. The low number of sources ($<$200)
in the extinction limited sample mean that even if we had the age
distribution and IMF correct we would not necessarily see a low
$\chi^{2}$. In fact we would require a significantly larger number
of observed objects to represent the age population and IMF
distributions. The importance of this modelling is to see which type
of initial conditions are most likely in the ONC and which can
definitely be ruled out.

The best fit to the luminosity function, based on the scenarios we
experiment with, is a single negatively skewed age distribution. The
skew has a low age spread and is centred at 1~Myr. The mass
distribution is a Miller \& Scalo IMF that peaks at 0.15 M$_{\odot}$
and is truncated by a power law in the form of $N(M) \alpha
M^{\alpha}$, where $\alpha$=$+$0.60 (or $-$0.60 for those who
prefer the alternative definition of $\alpha$). In this scenario we do
not truncate the IMF at the deuterium burning limit
(0.012M$_{\odot}$). Using the same set of parameters with an
$\alpha$=0.31 (based on the result of Lucas et al.2005)
also gives reduced $\chi^{2}$ values that are
similar. The histograms for the two best fitting scenarios are
plotted in Figure \ref{bestFit}. $\alpha$ takes a value of 0.60 in
the top four histograms and 0.31 in the remaining four. The 0.31
power law predicts the number of fainter sources better, whereas the
0.60 power-law recreates the brighter sources better. It is
unsurprising that the single Gaussian and single positive skew
scenarios also generate relatively good fits with these parameters
as the age distributions give similar results with low age
dispersions.

\subsubsection{Requirements for Low Mass Objects}
\label{MCrequirements} The most significant result from our
simulations is that very few scenarios generate good fits to the
luminosity function if we truncate the mass function at the
deuterium burning limit ($<0.012M_{\odot}$). None of the best
fitting scenarios have a truncated mass function. After the top
eight best fitting scenarios for each type of luminosity binning,
several truncated mass functions are ranked in the reduced
$\chi^{2}$ results. However, these occur infrequently. The scenarios
that have a truncated mass function and are ranked in the top 28 best
fitting models have no sources in magnitude bins fainter than
\emph{H}=17.5. Due to the fact that we have near-infrared spectra of
very low mass objects in this magnitude range (this work and Lucas
et al. 2006), the likelihood of these model scenarios being
plausible is low. These results are ranked relatively high due to the
relatively large uncertainty in the faintest bins of the observed
data. The brightest luminosity bins have more weight when
calculating reduced $\chi^{2}$ values. This means that model
parameters that predict the number of sources in the brightest bins
to a high standard will result in reasonable fits, despite having no
sources in the faintest bins. The result would be clearer if we
could recreate the luminosity function accurately at intermediate
magnitudes. It is possible that the dearth of sources at
\emph{H}=13.5 in the observed luminosity function is a statistical
fluctuation. Muench et al. 2002 observe a significantly larger
sample ($\sim$1000 sources). The \emph{H-}band luminosity function
in their work shows a relatively smooth decline from \emph{H}$=$12
to \emph{H}$=$18, with a slight drop of in the number of sources
between \emph{H}$=$13 to \emph{H}$=$15. However, reddening
corrections have not been applied to the data.

The requirement for planetary mass candidates to recreate the mass
function is perhaps most important in the scenarios that used a
bimodal age distribution. Before running the models we anticipated
that bimodal simulations, where $\sim2/3$ of the sources are 1~Myr
in age and $\sim1/3$ belong to a population of sources $\geq10$~Myr
in age, would create a bimodal luminosity function. However, this
was not the case regardless of how we altered the mass function. We
found that too few sources appear in the brightest bins when we
sufficiently populate the faint bins. In addition to this the peak
of the luminosity function moves towards fainter sources, producing
a very poor fit. For every scenario, we experimented by truncating
the IMF at the deuterium burning limit. In all nearly all cases this
produced a worse result. We can therefore say with some confidence
that a population of planetary mass candidates is required to
reproduce the dereddened luminosity function in the ONC. Figure
\ref{truncplan} shows 4 identically binned histograms of four
different scenarios we ran. In each case the IMF was truncated at
the deuterium burning limit. Further details on the model parameters
are shown in the figure caption. The published results of planetary
mass candidates in Orion, in paper 1 complement this result. We did
not seriously investigate age distributions older than 10~Myr.
This is because observational constraints tell us that
very few ONC sources are $>$10~Myr in age. It
is likely that we could recreate the luminosity function by
truncating the mass function if we have a bimodal age distribution
that has a population of objects 15-20~Myr in age. The bulk of
sources in a simulation like this would need to be 1~Myr in age in
order to sufficiently populate the brightest bins. We do not test
such a scenario due to the fact that it is very unlikely, according
to all published H-R diagrams for the ONC.

\subsubsection{Age Distributions}
\label{MCagedist} As one may expect, a uniform age distribution is
not ideal and struggles to recreate the luminosity function well,
with no reduced $\chi^{2}$ value being better than 11.9 for both
types of luminosity binning. The best fit to the luminosity function
when using a uniform age distribution has similar IMF parameters
described above for the overall best fit (Miller \& Scalo IMF
truncated by a power law with an $\alpha$ of 0.60). The Monte Carlo
simulations show that the above IMF parameters produce the best fit
for each type of age distribution. The only parameter that varies in
the best fit is $\alpha$, which takes values of 0.31 and 0.60.

In general the bimodal scenarios produce the poorest fitting
luminosity functions. We find that a bimodal age distribution is
only suitable in the case where the ratio of 1 to 10~Myr objects is
10:1. In addition to this the spread of each age distribution has to
be small (typically $\sigma$$\leq$1~Myr). If the fraction of 10~Myr
sources becomes much larger, too few bright objects are created.
Similarly if we make the spread of the bimodal Gaussians or Skews
too broad (3-5~Myr), the sources are redistributed so that
intermediate luminosity bins become most populated. The difference
between cases running bimodal Gaussians and bimodal Skews is
moderate. We find that negatively skewed bimodal age distributions
generate better fitting luminosity functions than bimodal Gaussian
distributions. We struggle to recreate the twin peaked luminosity
function observed using a bimodal age distribution. Even with
unrealistic extreme cases using 1 and 50~Myr age populations
combined with a mass function that is truncated with a steep power
law, we fail to generate a luminosity function with two clearly
defined peaks. In bimodal scenarios where the ratio of 1 to 10~Myr
objects was $1:0.7$, the peak of the mass function was moved to
$\sim0.25 M_{\odot}$. This change in the mass function was made to
try and ensure the brightest luminosity bins were sufficiently
filled. This failed to work giving reduced $\chi^{2}$ values of
$\sim16$. Instead the intermediate magnitude bins
\emph{H}$=$12.5-15.5 were over populated.

We analysed the results for single Gaussian and skew populations and
found that negatively skewed age distributions produce higher
quality fits than Gaussian age distributions. The worst of the three
was a positive skew. The results are consistent with those utilising
bimodal age distributions. We find that single age distributions
work well, providing a mean age of 1~Myr is used and the age spread
is small. If the age spread is increased beyond a few Myr the peak
of the luminosity function migrates from the brightest bins towards
bins of intermediate magnitude, suggesting the age dispersion in the
ONC is small. Our modelling results indicate that the star formation
rate in the ONC has not been steadily increasing since an age of
10~Myr. As previously mentioned, a single negative skew produced the
best overall fit to the observed luminosity function.

We summarise the results of the Monte Carlo simulations in Table
\ref{MCResults}. Rather than tabulate all the results we take the
top three reduced $\chi$$^{2}$ values for the single magnitude
binned histogram. These histograms were offset by 0.5 magnitudes as
shown in the right-hand image in Figure \ref{ONCLF}. Table
\ref{MCResults} summarises how effective each type of age
distribution is. In some cases lower reduced $\chi$$^{2}$ values
exist for different types of luminosity binning. However, in general
the order of best fitting scenarios is essentially the same for both
types of luminosity binning.

\begin{table*}
 \begin{minipage}{16cm}
 \centering
 \caption{Top three reduced $\chi$$^{2}$ values for each type of age distribution. Results are
derived from a histogram with a bin size of 1 magnitude, using the
offset binning shown in the right-hand histogram of Figure
\ref{ONCLF}. The results indicate a clear preference for a narrow age distribution
centred near 1~Myr and an IMF that declines with decreasing mass in
the brown dwarf regime.}
 \label{MCResults}
\begin{tabular}{@{}lccccc@{}}
\hline \hline
Age Distribution    &   Reduced $\chi$$^{2}$    &   Mean Age (Myr)  &   Age Dispersion ($\sigma$/Myr)   &   $\alpha$    &   Truncated Mass Function?    \\
\hline
    &   5.165   &   1   &   1   &   0.6 &   n   \\
-ve Skew    &   5.728   &   1   &   1   &   0.31    &   n   \\
    &   6.421   &   1   &   1   &   0.9 &   n   \\
\hline
    &   6.180   &   1   &   1   &   0.6 &   n   \\
Gaussian    &   6.560   &   1   &   1   &   0.31    &   n   \\
    &   8.192   &   1   &   1   &   0.9 &   n   \\
\hline
    &   9.329   &   1   &   1   &   0.6 &   n   \\
+ve Skew    &   10.284  &   1   &   1   &   0.9 &   n   \\
    &   10.403  &   1   &   1   &   0.31    &   n   \\
\hline
    &   11.884  &   0.3-10  &   -----   &   0.6 &   n   \\
Uniform &   12.598  &   0.3-10  &   -----   &   0.31    &   n   \\
    &   13.070  &   0.3-10  &   -----   &   0.6 &   y   \\
\hline
    &   16.715  &   1 \& 10 (10:1) &   1   &   0.6 &   n   \\
Bimodal (-ve Skew)    &   18.568  &   1 \& 10 (10:1) &   1   &   0.9 &   n   \\
    &   18.644  &   1 \& 10 (8:1) &   1   &   1.2 &   n   \\
\hline
    &   20.149  &   1 \& 10 (10:1) &   1   &   0.6 &   n   \\
Bimodal (Gaussian)    &   21.950  &   1 \& 10 (10:1) &   1   &   0.9 &   n   \\
    &   23.774  &   1 \& 10 (10:1) &   1   &   0.31    &   n   \\
\hline
\end{tabular}
\end{minipage}
\end{table*}

The star forming environment of the ONC is complicated and
our Monte Carlo simulations have initial conditions that are
somewhat simplistic. Therefore we cannot confidently suggest one
particular formation history and IMF from our results. However, we can
demonstrate which scenarios produce the best and worst fitting
luminosity functions. Although the best fit to the luminosity
function comes from a single 1~Myr age distribution (see Figure
\ref{bestFit}), the best fitting bimodal scenario (80\% of the
population having a mean age of 1~Myr and 20\% of the population
aged near 10~Myr. Miller-Scalo mass function with power law where
$\alpha$=0.31 and no truncation) generates a luminosity function
that is only subtly different. We know that 10~Myr objects do exist
in Orion (e.g. Palla et al. 2005; Palla et al. 2007). The result
from the negatively skewed 1~Myr age distribution provides evidence
that the dominant age in the ONC is very young.

We have shown that the dominant age of objects in the ONC are likely
to be $\leq$1~Myr, and that objects older than this can exist
providing there are not many. This result is consistent with what we
see in our H-R diagrams and from the Li studies by Palla et al.
(2005 \& 2007). A broad age spread has detrimental effects on the
luminosity function moving the peak to intermediate mass bins. This
explains why a negatively skewed age populations, that include fewer
old sources, always give lower reduced $\chi^{2}$ values. Poor
reduced $\chi^{2}$ values for a 3 $\sigma$ negatively skewed age
distribution appears to rule out an age distribution that has been
continually rising since 10~Myr. However, our skewness is fixed. We
expect a 3 $\sigma$ age distribution that is skewed more negatively
will work better, producing fewer 5-10~Myr objects. We therefore
deduce from our results that a continually rising star formation
rate is only plausible if it is very low until $\sim$3~Myr. Age
distributions $>$1~Myr produce poor results together with a uniform
age distribution.

\subsubsection{IMF}
For all of the tested age distributions we found that the best
associated mass function was that of Miller \& Scalo (1979),
truncated at $\sim$0.15 $M_{\odot}$ by an M$^{\alpha}$ power law
with $\alpha$ taking a value of 0.31 or 0.60. The power represents
low mass sources extending towards the deuterium burning limit. This
is in agreement with measurements made by Lucas et al. 2005.
Adopting $\alpha$$<$0.3 over-populates the faintest bins
and subsequently leads to poor $\chi^{2}$ values. This remains true
if the IMF is truncated at the deuterium burning limit. We tested
the significance of altering the dispersion of the IMF. We found
that increasing the dispersion added more high mass sources and
removed faint sources. Decreasing the dispersion had the opposite
effect. We find that the default dispersion prescribed by Miller \&
Scalo 1979 works best in the scenarios we test. Most importantly we
find that altering this parameter does not help us remove sources
from intermediate bins, where we struggle to reproduce the observed
luminosity function. In most cases we find altering the mean and
spread of the mass function from the best fitting values described
above gives poorer results. In the vast majority of the scenarios
that were tested, we require the mass function to extend to
planetary masses to sufficiently populate the faintest bins. This is
our most significant result.

\subsection{Binarity}
\label{MCbinarity} The modelling fails to take into account
unresolved binary sources. Depending on the mass-binarity relation
the overall profile of the luminosity function will change.
Photometric measurements in the Pleiades and Praesepe regions by
Pinfield et al. 2003 suggest the binary fraction increases as source
mass decreases. This result is consistent with later work in the
Pleiades by Lodieu et al. 2007. Their results suggests a binary
fraction of $\sim$30-45\% in 0.075-0.030 M$_{\odot}$ mass range.
Estimates by Pinfield et al. 2003 and Lodieu at al. 2007 support the
hypothesis that the brown dwarf binary systems are near equal-mass
ratio systems. The mass range of 0.075-0.030 M$_{\odot}$ correlates
to a range of \emph{H-}band mags of 13.834-15.770 based on the Dusty
and NextGen models. Assuming the most extreme case where all sources
have identical mass and luminosity, cluster members that are part of
an unresolved binary system will be 0.75 magnitudes brighter.
Assuming the binary fraction remains constant below 0.030
M$_{\odot}$ and remains close to 30\% for masses greater than 0.075
M$_{\odot}$ the luminosity function is likely to change in the
following way. The faintest bins are likely to decrease in frequency
as the number of contributing sources from fainter bins decreases
with the majority of the model scenarios. The region of the
luminosity function that is between \emph{H}=14 and \emph{H}=12 is
likely to become steeper, with fewer sources in the fainter of these
bins. We have not included these binary measurements in the
modelling as the results rely on the interpretation of photometric
measurements being accurate. Further measurements on the binary
fraction in star-forming regions is required before this aspect
implemented in the code (e.g. the MONITOR
project\footnote{http://www.ast.cam.ac.uk/research/monitor/}, which
detects occultations in the light curves of young low-mass stars and
brown dwarfs in open clusters and star-forming regions).
Hydrodynamical simulations by Bate, Bonnell \& Brom (2002) suggest
that the binary fraction for brown dwarfs is $\sim$5\%. If binary
systems are in fact more prevalent for higher mass stars and the
more massive brown dwarfs, this will have important implications
when modelling the luminosity function. This scenario would mean
that the faintest end of the luminosity function would remain
approximately the same. The frequency of sources in the brighter of
the intermediate bins will decrease, while the frequency of sources
in the brighter bins will increase. For the best fitting functions
this may produce a luminosity function that has a profile which is
similar to that observed in Orion. However, the simulations by Bate,
Bonnell \& Brom (2002) produce only a small fraction of the number
of sources we observe in Orion. In addition to this the limiting
resolution of the model means binaries with very close separations
cannot be formed. For this reason we have not tried to implement
this result into the models.

\section{Summary and Conclusions}
\label{summaryConc} A total of 40 of the sample drawn from this
paper and Lucas et al.(2001;2006) show strong water absorption in
their \emph{H-} and$/$or \emph{K-}band spectra. Some of these
spectra also show evidence of CO absorption at \emph{K-}band. We
have developed a spectral typing scheme based on optically
calibrated, near-infrared spectra of objects in the Taurus and IC348
star forming regions. These ranged from M3.0 to M9.5 in spectral
type. Our spectral typing scheme showed that 23 of our objects had
spectral types ranging from M4.0 - M9.5. A further 17 sources had
very strong water absorption, indicating spectral types later than
or equal to M9.5. Our calculated spectral types are accurate to
within one sub-type for the majority of our objects. 12 sources in
our sample have spectra that show clear Rayleigh-Jeans type
continua. These are most likely to be reddened background stars.

The spectral type to effective temperature scale designed by Luhman
et al. 2003 was used to calculate temperatures for each of our
sources. We plotted the objects on an H-R diagram together with
pre-main-sequence NextGen and Dusty isochrones and DM97 isochrones.
The majority of the objects in our sample lie close to or above the
1~Myr isochrone suggesting an average cluster age that is
$\leq$1~Myr. Below 2500 K we see a population of objects that lie
several magnitudes above the 1 Myr isochrone. This is in agreement
with results by RRL. The position of these objects cannot be
explained by unresolved multiple body systems or inaccurate distance
measurements. This does not indicate a problem with calibrating
sources from young objects in different clusters as a similar trend
is observed at optical wavelengths. We do not observe an obvious
population of objects 10 Myr in age.

We conclude from our results that the average cluster age is less
than 1 Myr. We find that photometrically derived temperatures from the
1~Myr NextGen isochrone show fairly good agreement with spectroscopic
measurements, but typically cooler by $\sim$100~K. We find that
spectroscopic masses show a relatively poor agreement with
photometrically determined masses. These findings are consistent
with results from SCH04 and RRL.

We derive source masses from the H-R diagram that uses the combined
NextGen and Dusty isochrones. 36 of the 40 sources plotted lie below
the hydrogen burning limit (0.075 M$_{\odot}$). The remaining 4
sources are low mass stars or lie close to the threshold. 10 of the
17 PMO candidates have masses that potentially lie below the
deuterium burning limit (0.012 M$_{\odot}$). If we use the models of
DM97 the number of sources that lie below the hydrogen burning limit
declines from 36 to 29.

We used a Monte Carlo approach to model the observed extinction
limited luminosity function for the ONC. Our simulations show that
single age distributions centred at 1~Myr with a narrow age spread
fit the observed luminosity function best. A bimodal age
distribution of 1 and 10 Myr sources also works well providing the
respective ratio of sources is $\geq10:1$. Broad singular age
distributions generate too many sources in the intermediate
luminosity bins and too few in the brighter bins. We conclude from
our results that a continually rising star-formation rate as
suggested by Palla et al. (2006 \& 2007) is only plausible if it
remained very low until $\sim$3~Myr ago. We find that a uniform age
distribution produces a very poor fitting luminosity function. The best fitting
version of the mass function is a Miller Scalo form at stellar masses that joins to
an M$^{\alpha}$ power law for M$\le$0.15~M$_{\odot}$. Our most
important result from the modelling is that planetary mass
candidates are required to produce enough sources in the faintest
magnitude bins. This holds true for scenarios that have bimodal age
distributions with a moderate number of 10~Myr objects. A poorly
fitting luminosity function is generated in every scenario where we
truncate the IMF at the deuterium burning limit.

Although negatively skewed age distributions give the best results,
we cannot definitively label it as the most appropriate age
distribution. This is because the negative skew is best when it has
a low dispersion and is centered at 1~Myr. A Gaussian distribution
that has lower dispersion, centered at the same age would also
approximate this result. A negatively skewed age population is
significantly better than a Gaussian population at larger age
dispersions. This is because it populates the intermediate bins of
the luminosity function with fewer sources. The advantage of using a
negatively skewed function for age is that it mimics a sudden burst
in star formation that tails off over time. There seems to have been
a sudden burst of star formation around 1~Myr ago in Orion. In most
cases a positively skewed age distribution gave poor results. The
only exception was for a 1~Myr population that has a low age
dispersion.

\section*{Acknowledgments}
We thank Dr Kevin Luhman and Dr Nicolas Lodieu for supplying
infrared template spectra of young brown dwarfs which have been used
to calibrate our young ONC spectra. We would also like to thank Dr
James Collett for meticulously checking our probability
distributions used in the Monte Carlo simulations.

DJW acknowledges the support of an STFC (previously PPARC) doctoral
studentship.

\section{References}

Allard F., Hauschildt P.H., \& Schwenke D., 2000, ApJ 540, 1005\\
Allard F., Hauschildt P.H., Alexander D.R., Tamanai
A., Schweitzer A., 2001, ApJ, 556,357\\
Baraffe I., Chabrier G., Allard F., Hauschildt P.H.,
1998, A\&A, 337, 403\\
Bate M. R., Bonnell I. A., Bromm V., 2002, MNRAS, 332, 65\\
Brice\~{n}o C., Luhman K. L., Hartmann L., Stauffer J. R.,
Kirkpatrick
J. D., 2002, ApJ, 580, 317\\
Breger M., Gehrz R. D., Hackwell J. A., 1981, ApJ, 248, 963\\
Burrows A. et al., 1997, ApJ, 491, 856\\
Cardelli J.A., Clayton G.C., Mathis J.S., 1989, ApJ,
345, 245\\
Cargile P.A., Stassun K.G., Mathieu R.D., 2008, ApJ, 674, 329\\
Chabrier G., Baraffe I., Allard F., Hauschildt P. 2000,
ApJ, 542, 464\\
Comeron F., Rieke G.H., Burrows A., Rieke M.J., 1993, ApJ, 416,
185\\
D'Antona F., \& Mazzitelli 1997, MmSAI, 68, 807 (DM97)\\
Gorlova N. I., Meyer M. R., Rieke G. H., Liebert J., 2003, ApJ, 593,
1074\\
Hillenbrand L.A., 1997, AJ, 113, 1733\\
Hillenbrand L.A., Carpenter J.M., 2000, ApJ, 540, 236\\
Hillenbrand L.A., Hartmann L.W., 1998, ApJ, 492, 540\\
Irwin J. et al., 2007, MNRAS, 380, 541\\
Jeffries R.D., 2007, MNRAS, 376, 1109\\
Kaas A.A., 1999, AJ, 118, 558\\
Kirkpatrick J. D., Barman T. S., Burgasser A. J., McGovern M. R.,
McLean I. S., Tinney C. G., Lowrance P. J. 2006, ApJ, 639,
1120\\
K\"{o}hler R., Petr-Gotzens M. G., McCaughrean M. J., Bouvier
J.,
Duch\^{e}ne G., Quirrenbach A., Zinnecker H. 2006, A\&A, 458, 461\\
Levine J. L., Steinhauer A., Elston R. J., Lada E. A., 2006, ApJ, 646, 1215\\
Lodieu N., Dobbie P. D., Deacon N. R., Hodgkin S. T., Hambly N. C., Jameson R. F., 2007, MNRAS, 380, 712\\
Lucas P.W., Roche P.F., 2000, MNRAS, 314, 858\\
Lucas P.W., Roche P.F., Allard F., Hauschildt
P.H., 2001, MNRAS,
326, 695L\\
Lucas P.W., Roche P.F., Tamura M., 2005, MNRAS 361, 211\\
Lucas P.W., Weights D.J., Roche P.F., Riddick F.C., 2006, MNRAS, 373, L60 (Paper 1)\\
Luhman K. L., Brice\~{n}o C., Stauffer J. R., Hartmann L., Barrado y
Navascués D., Caldwell N., 2003b, ApJ, 590, 348\\
Luhman K. L., Stauffer J. R., Muench A. A., Rieke G. H.,
Lada E. A., Bouvier J., Lada C. J., 2003, ApJ, 593, 1093-1115\\
Luhman K. L., 2004, ApJ, 602, 816\\
Luhman K. L., Peterson D.E., Megeath S.T., 2004, ApJ, 617, 565\\
Luhman K.L., Adame L., D'Alessio P., Calvet N., Hartmann L., Megeath
S.T., Fazio G.G., 2005, 635, L93\\
Luhman K.L., Wilson J.C., Brandner W., Skrutskie M.F., Nelson M.J.,
Smith J.D., Peterson D.E., Cushing M.C., Young E., 2006, ApJ, 649,
894\\
Mart\'{\i}n, E. L., Brandner, W., Bouy, H., Basri, G., Davis, J.,
Deshpande, R., Montgomery, M., King, I., 2006, A\&A, 456, 253\\
Meeus G., \& McCaughrean M.J., 2005, AN, 326, 977\\
Muench A. A., Lada E.A., Lada C.J., Alves J., 2002, ApJ, 573, 366\\
O'Dell, C. R., \& Wen, Z. 1994, ApJ, 436, 194\\
O'Dell C. R., Wong K., 1996, AJ, 111, 846\\
Padgett D. L., Strom S. E., Ghez A. 1997, ApJ, 477, 705\\
Palla F., \& Stahler S.W., 1999, ApJ, 525, 772\\
Palla F., Randich S., Flaccomio E., Pallavicini R., 2005, ApJ, 626,
L49\\
Palla F., Randich S., Pavlenko Y.V., Flaccomio E., Pallavicini R.,
2007, ApJ, 659, L41\\
Peterson D. E., Megeath S.T., Luhman K. L., Pipher J, L., Stauffer J. R., Barrado y Navascu\`{e}s D., Wilson J. C., 2008, ApJ preprint doi:10.1086/'590527\\
Pinfield D. J., Dobbie P. D., Jameson R. F., Steele I. A., Jones H. R. A., Katsiyannis A. C., 2003, MNRAS, 342, 1241\\
Prosser C.F.F., Stauffer J.R., Hartmann
L.W., Soderblom D.R., Jones
B.F., Werner M.W., McCaughrean, M.J., 1994, ApJ, 421, 517\\
Reipurth B., Guimar\~{a}es M., Connelley M., Bally J. 2007, AJ,
134, 2272\\
Riddick F., Roche P.F., Lucas P.W., 2007, MNRAS, 381, 1077 (RLL)\\
Rieke G.H., \& Lebofsky M.J. 1985, ApJ, 288,618\\
Robin A.C., Reyl\'{e} C., Derri\`{e}re,  Picaud S., 2003, A\&A, 409,
523\\
Rodgers B., Wooden D. H., Grinin V., Shakhovsky D., \& Natta A.
2002, ApJ, 564, 405\\
Simon M., Close L. M., Beck T. L. 1999, AJ, 117, 1375\\
Saumon D., Hubbard W. B., Burrows A., Guillot T., Lunine J. I., Chabrier G. 1996, ApJ, 460, 993\\
Siess L., Forestini M., \& Bertout C., 1999, A\&A, 342, 480\\
Slesnick C.L., Hillenbrand L.A., Carpenter J.M., 2004, ApJ, 610,
1045 (SCH04)\\
Stassun K. G., Mathieu R. D., Vaz L. P. R., Stroud N., Vrba F. J.,
2004, ApJS, 151, 357\\
Tinney C. G., et al. 2004, Proc. SPIE, 5492, 998\\
Tout C. A., Livio M., Bonnell I. A., 1999, MNRAS 310, 360\\
Walker M. F., 1969, ApJ, 155, 447\\
Williams D.M., Comer\'{o}n F., Rieke G.H., \& Rieke M.J., 1995, ApJ,
454, 144\\

\appendix

\section{Spectral Typing Discrepancies}
\label{Descrepancies} The first anomaly in the spectral typing
results can be seen in the QH$_{UK}$ index for 013-306 (see Lucas et
al. 2001 for \emph{H-}band spectrum). The signal to noise of this
spectrum is relatively high, resulting in a cubic spline fit that
represents the pseudo continuum very well. It is therefore most
likely that the index is not steep enough in the region of late-type
objects to efficiently differentiate between spectral type. It
should be noted that no spectral index designed for this small
wavelength range can be guaranteed to generate an accurate spectral
type at all temperatures. This is because intrinsic scatter exists
in the template data. This scatter is probably not attributed to the
inherent low signal to noise of these objects, but to small
differences in atmospheric structure and composition. Further cause
for discrepancy may be attributed to dust or hot spots in the
photosphere.

There is a discrepancy in both Q indices for the spectrum of 014-413
(see Lucas et al. 2001 for \emph{H-} and \emph{K-} spectra). The
difference in the QH$_{(UK)}$ index is not obvious and is probably
due to scatter in the index sequence as the signal to noise is
relatively high. Over plotting template spectra revealed that
H$_{2}$O absorption shortward of 1.675 $\umu$m is consistent with
that of an object ranging from M8.0 - M8.5 in spectral type.
However, the H$_{2}$O absorption on the red side of the triangular
peak is weaker and more consistent with an object ranging from M4.0
- M5.0, thus skewing the QH$_{(UK)}$ result. The difference seen in
the result of the QK$_{(UK)}$ index is likely to be due to
decreasing signal to noise in the spectrum longward of the
$\sim$2.2$\umu$m peak. The noise seems to be suppressing water
absorption at the longest wavelengths, leading to a pseudo-continuum
fit that makes 014-413 appear earlier in spectral type when using
the QK$_{(UK)}$ index. This was confirmed by over-plotting the
spectrum and pseudo-continuum fit onto template spectra. The result
from the QK$_{(UK)}$ index was therefore not used to determine the
final spectral type. The result from the QH$_{(UK)}$ index was not
discarded as the quality of the spectrum is high.

The inconsistency that is present in the QH$_{(UK)}$ index for
015-319 is most likely due to low signal to noise in the
\emph{H-}band UKIRT spectrum for this object. Without sufficient
signal to noise it is difficult to certify whether this discrepancy
is purely down to the cubic spline fit poorly representing the real
pseudo continuum. The template sample used for deriving the spectral
typing scheme contains only two objects later than M9.0 in spectral
type. It is therefore reasonable to assume the discrepancy in
spectral type may be due to potential limitations in the QH index.
Plotting template spectra on top of 015-319 clearly demonstrated
that this object is later than M9.5

Results derived using the K$_{(UK)}$ indices provide less reliable
results for the spectra of 068-019 and 069-209 due to poor signal to
noise at K-band. The WK index for 095-058 produces a spectral type
that is clearly later than those derived using the indices at
\emph{H-}band. A visual inspection of the spectra and cubic spline
fits for this object reveal no obvious clue as to why there is a
large discrepancy. The difference is most likely to be due to the
profile of the \emph{K-}band spectrum. We have investigated this
possibility by superimposing template spectra of different spectral
type on top of 095-058. This revealed that 095-058 does in fact
appear later ($\sim$M9) at K-band, based on data shortward of the
triangular peak. However, the water absorption longward of the peak
is much weaker than seen in any of the late-type template spectra,
despite the presence of multiple CO band-heads. This suggests that
warm circumstellar dust may be present, resulting in a near-infrared
excess. The UKIRT spectra were observed at a different epoch to the
NIRI data. We can therefore not determine from this data whether the
apparent difference in spectral type is due to short term spectral
variability possibly caused by dust, or if the object is a spectrum
variable intrinsically different at \emph{K-}band. Due to the
strange nature of the spectrum beyond 2.26 $\umu$m we did not use
the result from the QK index to determine a spectral type. We did
use the result from the WK index as 095-058 does appear late over
this wavelength region.

A further spectral typing discrepancy that is significant can be
seen in the \emph{H-}band indices for 186-631. This may be because
the \emph{H-}band spectrum for this object is of lower quality than
the \emph{K-}band spectrum. Careful inspection of this object
indicates telluric noise may exist in the \emph{H-}band spectrum at
wavelengths $<$ 1.58 $\umu$m. Plotting template spectra over the top
of the spectra for this object demonstrated the derived spectral
types were accurate. The consistent difference seen between
\emph{H-} and \emph{K-}band is therefore not due to scatter in the
indices. 186-631 has been observed at optical wavelengths by RRL,
and assigned a spectral type of M8.0. This measurement strengthens
the argument for telluric noise suppressing water absorption in the
\emph{H-}band spectrum. We were unable to test whether this is a
reasonable diagnosis as an alternative telluric standard for 186-631
was unavailable. We can not rule out the possibility that this
object has a spectrum that appears to be dissimilar at different
wavelengths. For this reason all \emph{H-} and \emph{K-}band indices
were used to spectral type this source.

\section{Spectral Index Plots}
\subsection{UKIRT Indices}
\label{UKPlotappendix}

\begin{figure*}
\centering{\ }
 \vbox{
  \hbox{
    \psfig{file=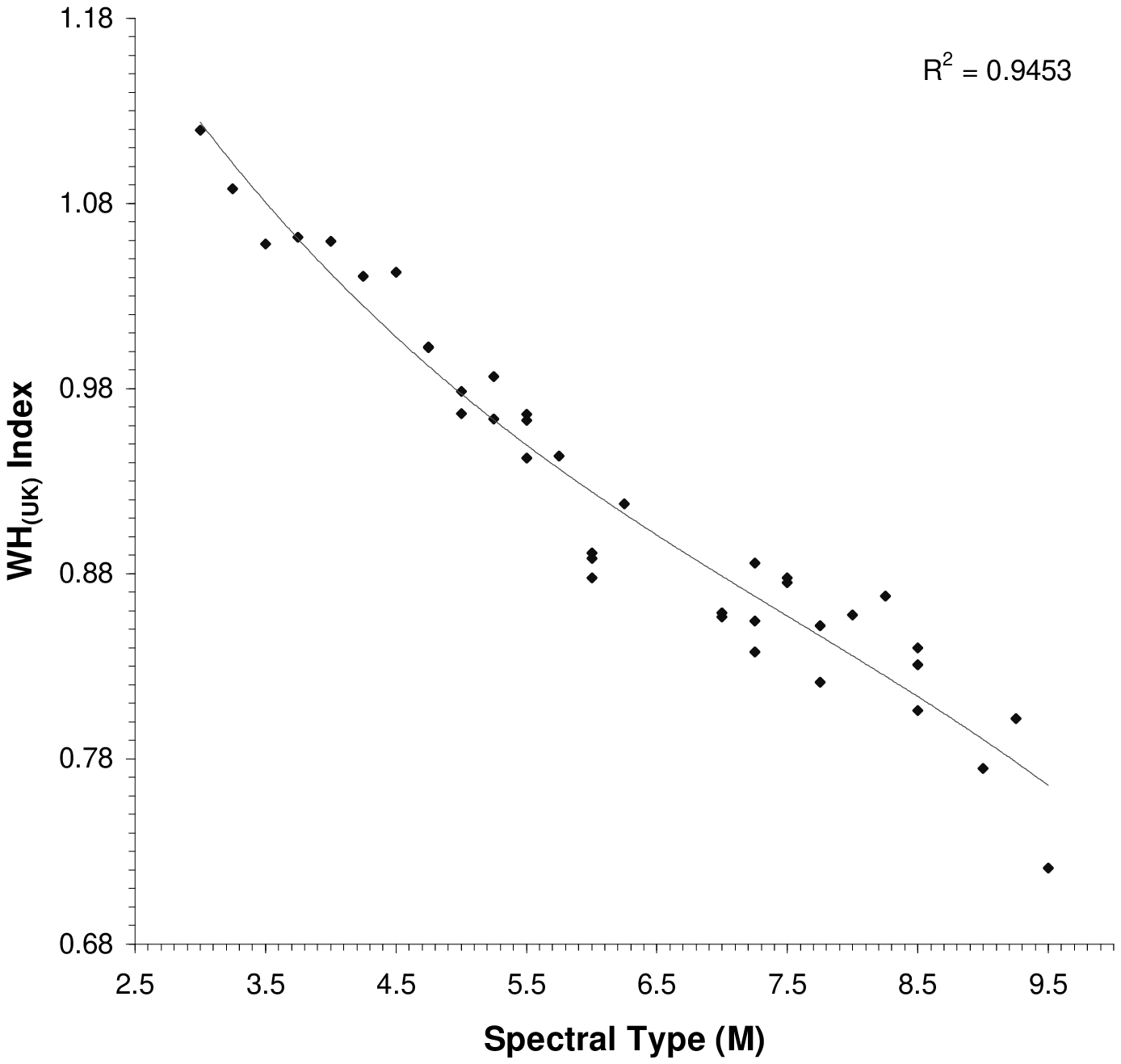,width=6.9cm,angle=-90}
    \psfig{file=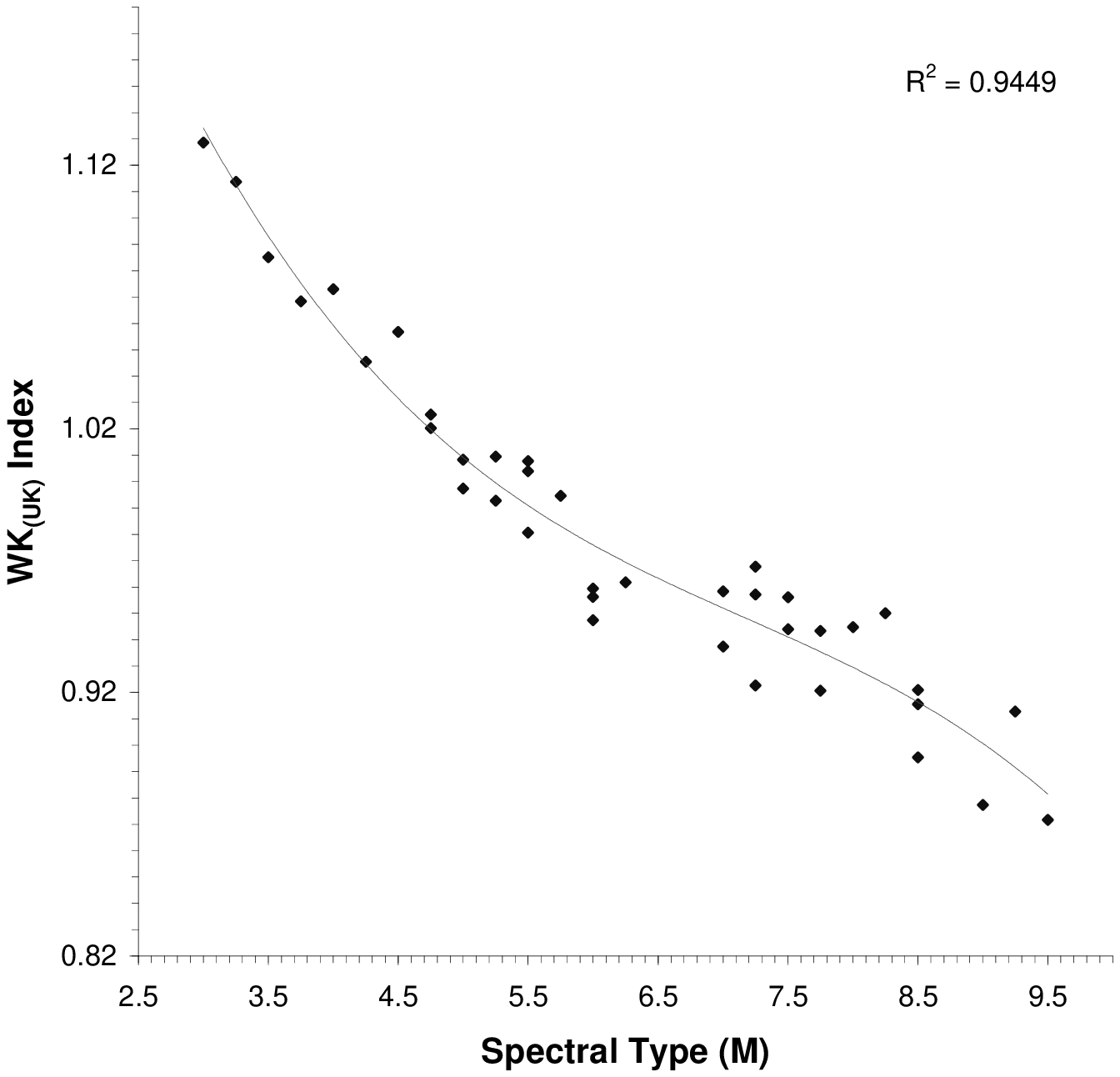,width=6.9cm,angle=-90}}
  \hbox{
    \psfig{file=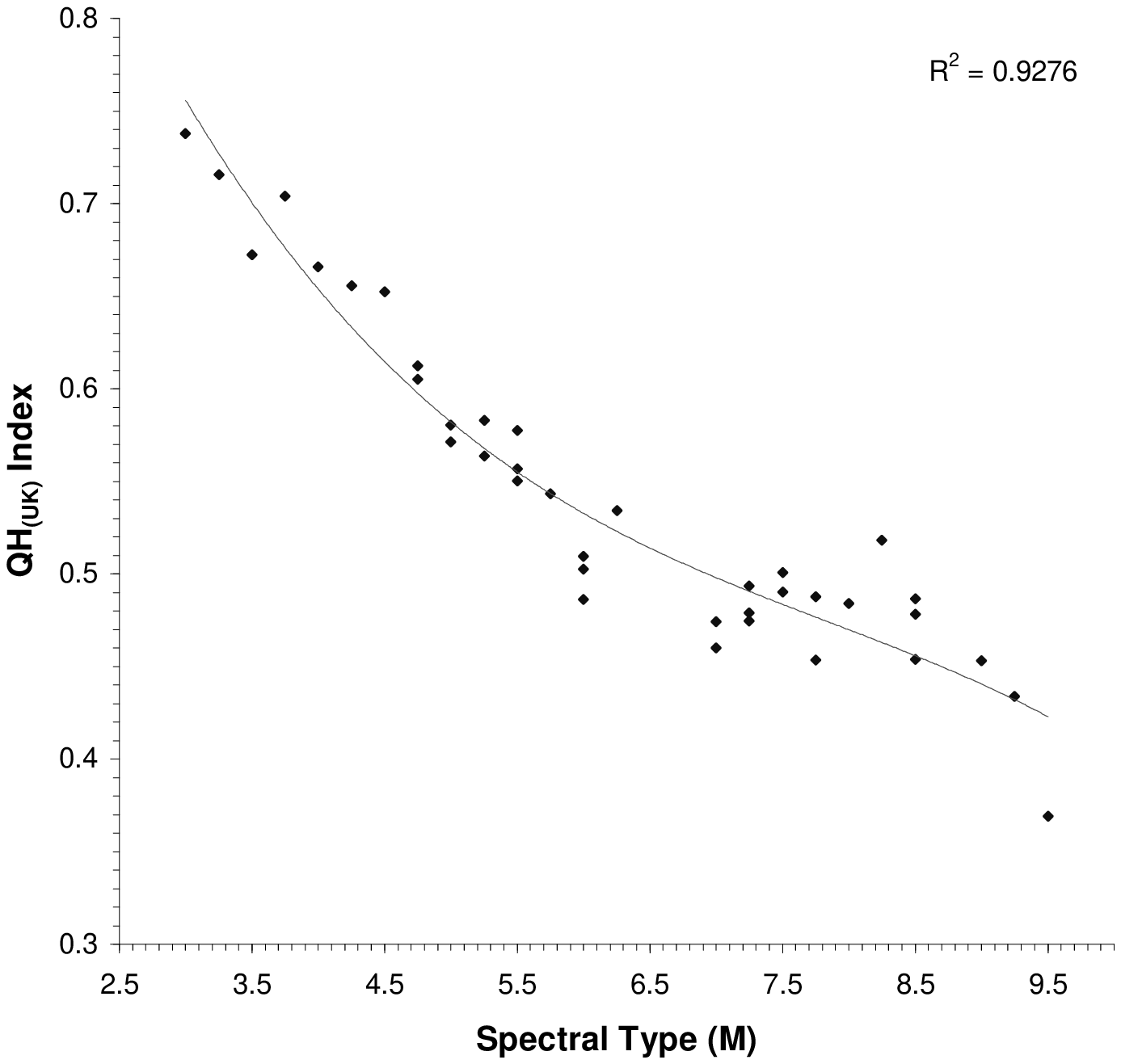,width=6.9cm,angle=-90}
    \psfig{file=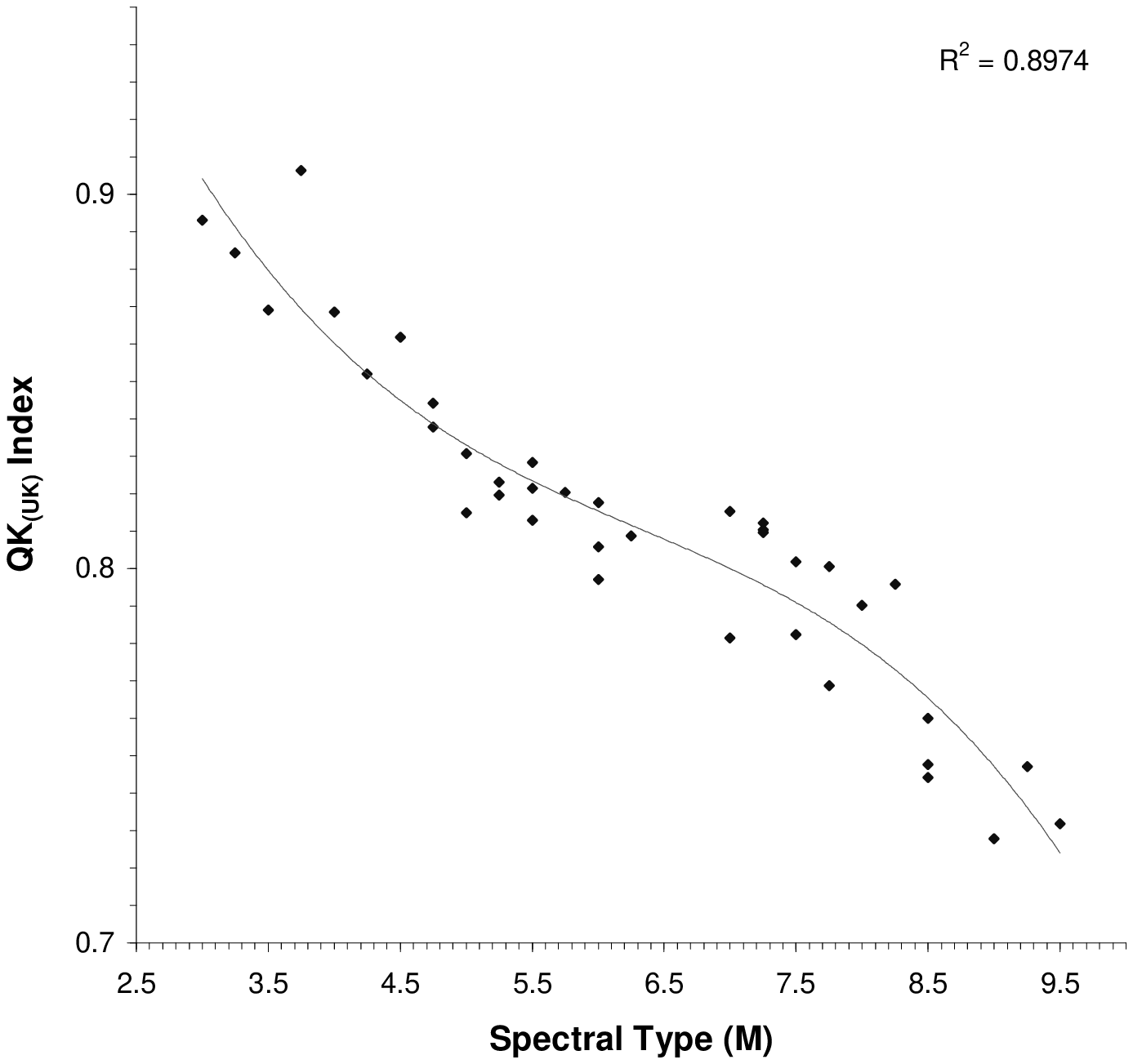,width=6.9cm,angle=-90}}}
\caption{Fits to spectral indices used to characterise the UKIRT
data. Index strength is plotted as a function of spectral type. Each
data point represents an object from Luhman's near-infrared
spectroscopic sample of optically calibrated young ($\sim$1Myr)
brown dwarfs. For each ratio the median flux value in a 0.02$\umu$m
interval was used. The relationship between spectral type and index
are fitted by cubic polynomials. The R$^2$ values are correlation
coefficients.} \label{UKindices}
\end{figure*}

\subsection{GNIRS Indices}
\label{GPlotappendix}

\begin{figure*}
\centering{\ } 
 \vbox{
  \hbox{
    \hspace{3.45cm}
    \psfig{file=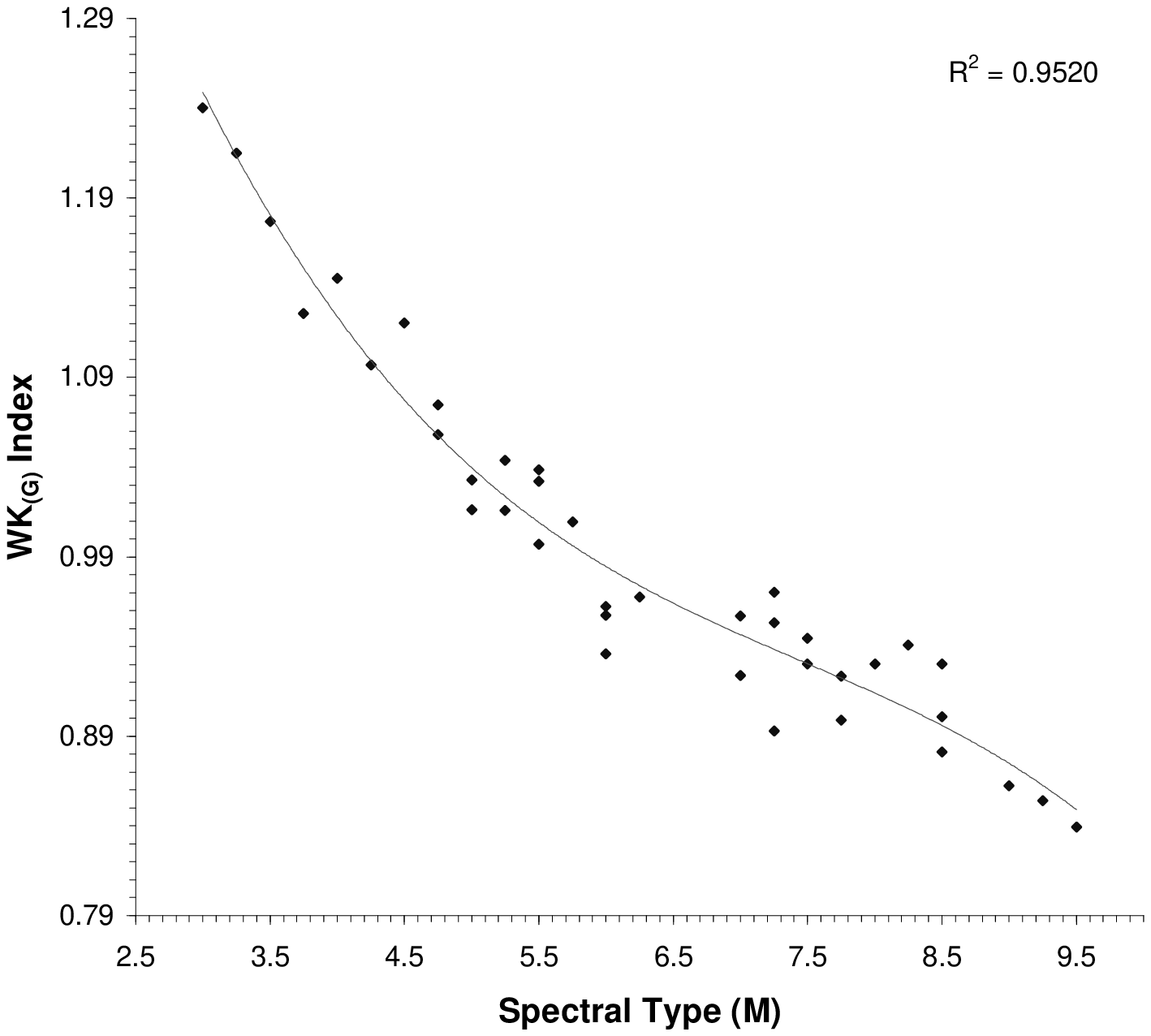,width=6.9cm,angle=-90}}
  \hbox{
    \psfig{file=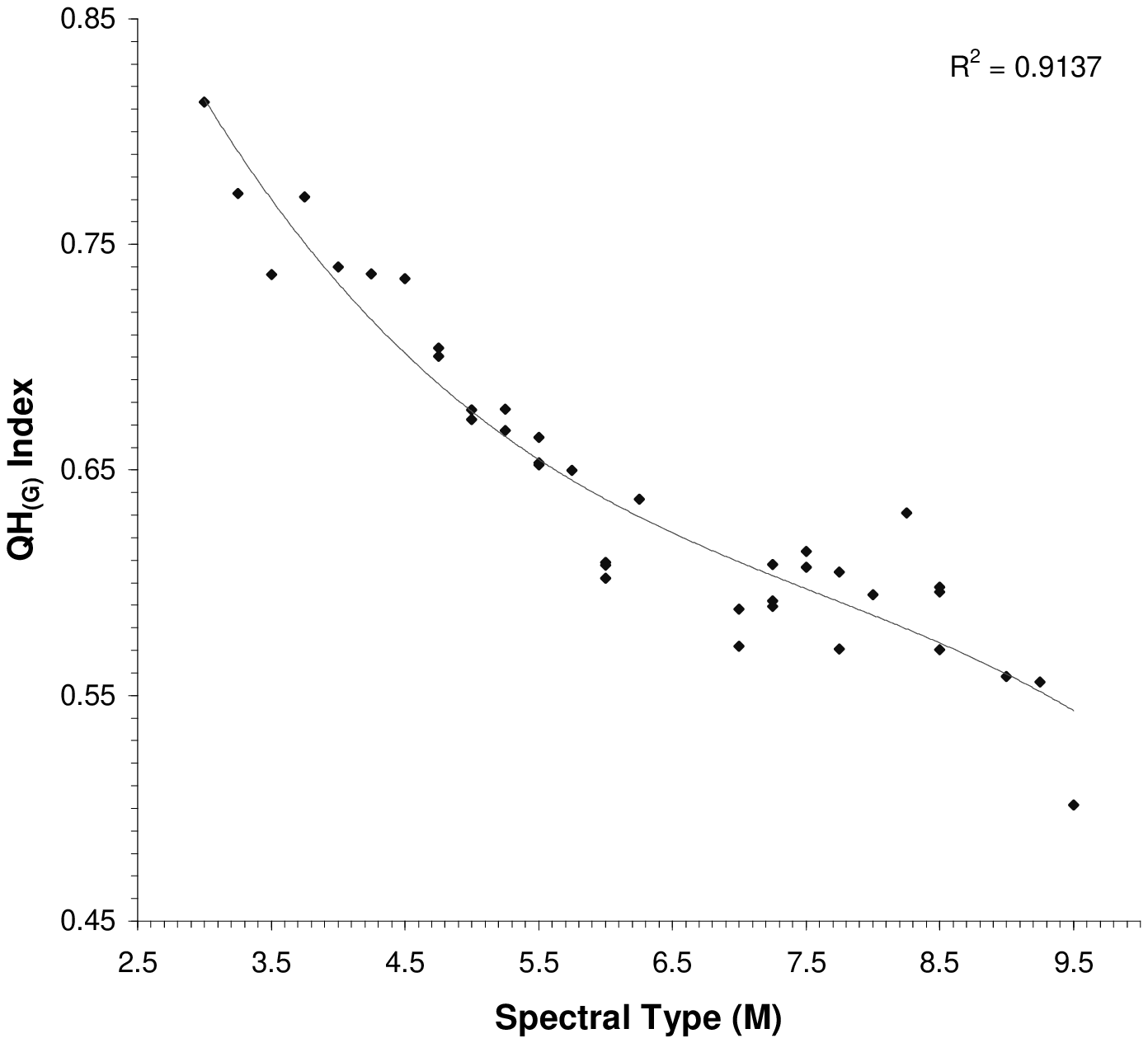,width=6.9cm,angle=-90}
    \psfig{file=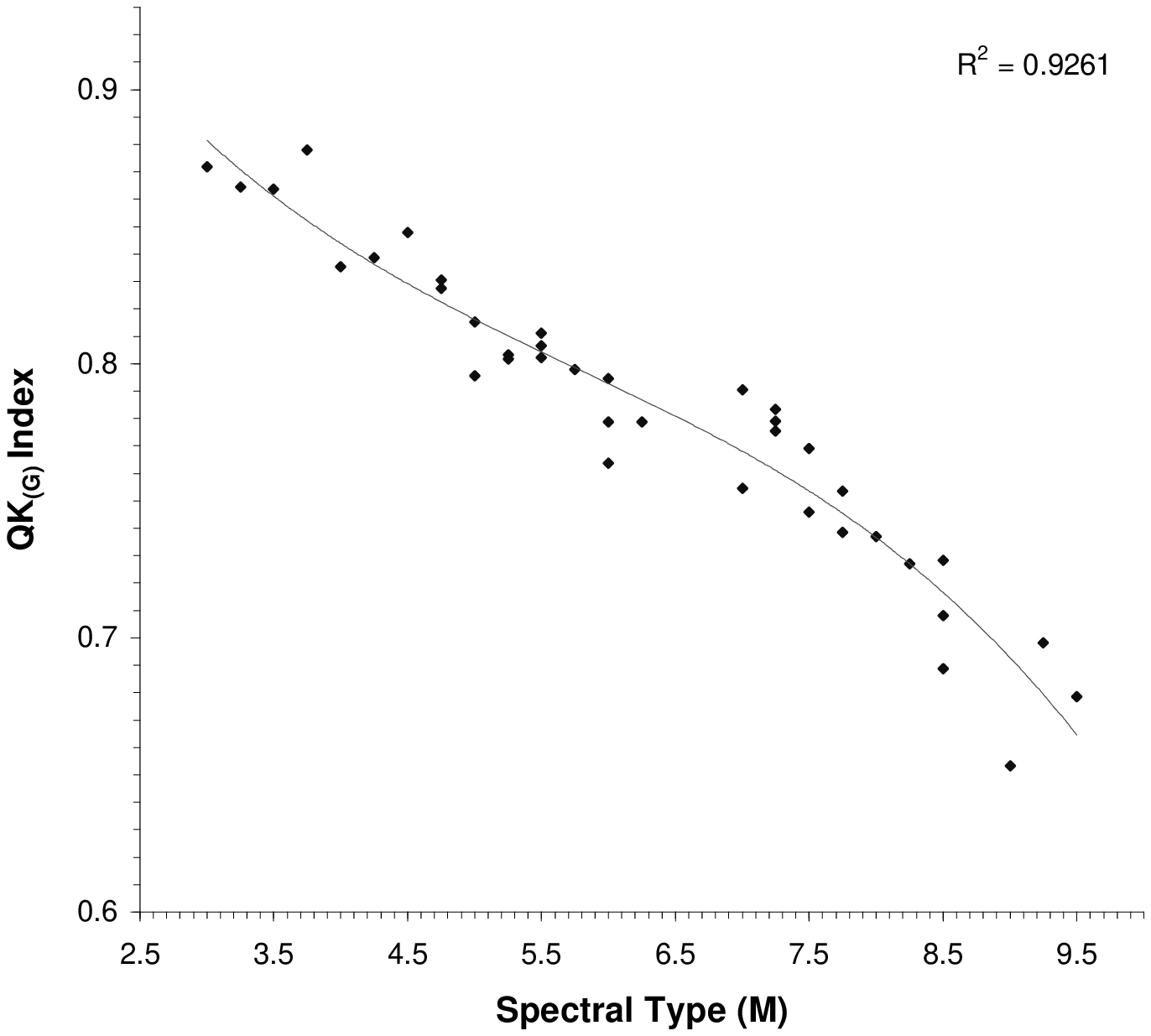,width=6.9cm,angle=-90}}}
\caption{Fits to spectral indices used to characterise the GNIRS and
NIRI data in Paper 1. Index strength is plotted as a function of
spectral type. Each data point represents an object from
Luhman's near-infrared spectroscopic sample of optically calibrated
young ($\sim$1Myr) brown dwarfs. For each ratio the median flux
value in a 0.02$\umu$m interval was used. The relationship between
spectral type and index are fitted by cubic polynomials. The R$^2$
values are correlation coefficients.} \label{Gindices}
\end{figure*}

\section{H-R Diagram Containing Objects Calibrated from the Lodieu Spectra}
\label{hrlodapp}
\begin{figure*}
\centering{\ }
    \psfig{file=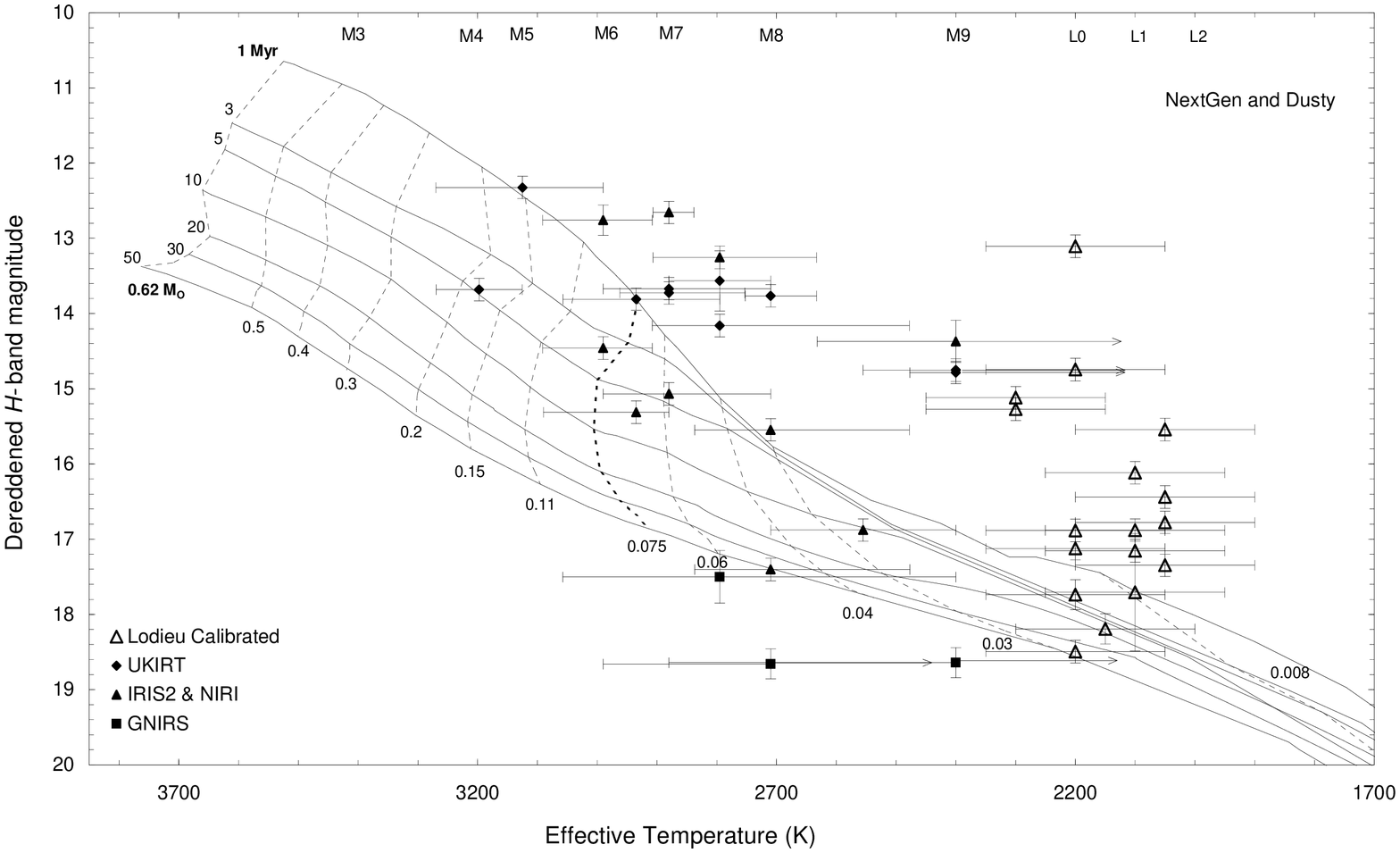,width=14cm,angle=-90}
\caption{Sources calibrated from the Lodieu spectra are marked as
open triangles. The pre-main-sequence tracks are from the NextGen
and Dusty models. Temperatures were estimated from field dwarfs
using data from Mart\'{\i}n et al. 1999. Due to the fact that these
values are estimated an error of $\pm$150 K has been used. The H-R
diagram provides a reasonable estimate of the positions of these
late M $/$ early L-type objects.} \label{LodieuPlot}
\end{figure*}

\label{lastpage}

\end{document}